\documentclass[preprint,amsmath,amssymb,floatfix]{revtex4-1}

\usepackage{graphicx}
\usepackage{dcolumn}
\usepackage{bm}
\usepackage{color}

\def\etal{\hbox{\it et al.}}

\definecolor{vk_color}{rgb}{0.858, 0.188, 0.478}

\begin{document}

\title{Uncertainty Estimates for Theoretical Atomic and Molecular Data}

\author{H.-K.~Chung}
\email[Electronic address: ]{H.Chung@iaea.org}
\affiliation{Nuclear Data Section, International Atomic Energy Agency, Vienna, A-1400, Austria}
\author{B.~J.~Braams}
\email[Electronic address: ]{B.J.Braams@iaea.org}
\affiliation{Nuclear Data Section, International Atomic Energy Agency, Vienna, A-1400,Austria}
\author{K.~Bartschat}
\email[Electronic address: ]{klaus.bartschat@drake.edu}
\affiliation{Department of Physics and Astronomy, Drake University, Des
  Moines, Iowa, 50311, USA}
\author{A.~G.~Cs\'asz\'ar}
\email[Electronic address: ]{csaszar@chem.elte.hu}
\affiliation{MTA-ELTE Complex Chemical Systems Research Group, H-1118
  Budapest, P\'azm\'any s\'et\'any 1/A, Hungary}
\author{G.~W.~F.~Drake}
\email[Electronic address: ]{gdrake@uwindsor.ca}
\affiliation{Department of Physics, University of Windsor, Windsor, Ontario
  N9B 3P4, Canada}
\author{T.~Kirchner}
\email[Electronic address: ]{tomk@yorku.ca}
\affiliation{Department of Physics and Astronomy, York University, Toronto,
  Ontario M3J 1P3, Canada}
\author{V.~Kokoouline}
\email[Electronic address: ]{slavako@mail.ucf.edu}
\affiliation{Department of Physics, University of Central Florida, Orlando, FL
  32816, USA}
\author{J.~Tennyson}
\email[Electronic address: ]{j.tennyson@ucl.ac.uk}
\affiliation{Department of Physics and Astronomy, University College London,
London WC1E 6BT, UK}

\date{\today}

\begin{abstract}

  Sources of uncertainty are reviewed for calculated atomic and
  molecular data that are important for plasma modeling: atomic and
  molecular structure and cross sections for electron-atom,
  electron-molecule, and heavy particle collisions.  We concentrate on
  model uncertainties due to approximations to the fundamental
  many-body quantum mechanical equations and we aim to provide
  guidelines to estimate uncertainties as a routine part of
  computations of data for structure and scattering.

\end{abstract}


\maketitle

\section{Introduction}\label{sec:Intro}
\label{sec:Introduction}
There is growing acceptance that benchmark atomic and molecular (A+M)
calculations should follow accepted experimental practice and include an
uncertainty estimate alongside any numerical values presented~\cite{PRA}.
Increasingly, A+M computations are also being used as the primary source of
data for input into modeling codes.  It is our assertion that these data
should, if at all possible, also be accompanied by estimated uncertainties.
However, it is not at all straightforward to assess the uncertainties
associated with A+M computations.  The aim of this work is to provide
guidelines for A+M theorists to acquire uncertainty estimates as a routine
part of their work.  We concentrate on data that are most important for 
high-temperature plasma modeling: data for A+M structure, electron-atom (or ion)
collisions, electron collisions with small molecules, and charge transfer in
ion-atom collisions.

Uncertainty Quantification (UQ) is a very active research area in
connection with simulations of complex systems arising in weather and climate
modeling, simulations of nuclear reactors, radiation hydrodynamics, materials
science, and many other applications in science and engineering.  A report
from the USA National Research Council~\cite{NRC-UQ} provides a valuable
survey.  The current state of the field is reflected in the biennial meeting
of the SIAM Activity Group on Uncertainty Quantification~\cite{SIAG-UQ}.  This
field of UQ for complex systems has a mathematical core in the description of
uncertainty propagation for chaotic deterministic and stochastic evolution
equations in many dimensions (``polynomial chaos'').  In many cases the
interest is then focused on systems for which the basic equations are not well
established and involve poorly known parameters and functional dependencies.

The present article is concerned with quantification of uncertainties in
elaborate computations, but the nature of computational A+M science for
application to high temperature plasmas is rather different from the focus
areas of present UQ science.  This A+M science is concerned with simple
physical systems and their interactions.  The underlying equations governing
the processes of interest and the ensuing dynamics are essentially known
\cite{dirac}, but except for a few special cases a true first-principles
treatment is numerically intractable: the complexity scales exponentially with
the number of electrons while for a fixed number of electrons the complexity
of the first-principles equations using a basis tends to scale polynomially in
the basis size with the number of electrons in the exponent.  A+M theory is,
therefore, about development of models that aim to approximate the exact
problem with numerically tractable procedures.  The uncertainties in these
procedures, referred to as ``model uncertainties'' in the following, are
strongly model-dependent and are often poorly understood.  The solution of any
given model is itself subject to uncertainties due to convergence and other
numerical issues associated with a grid or a basis set.  These will be
referred to as ``numerical uncertainties''.  Finally, closer to established UQ
science, uncertainties propagate through the various stages of a calculation,
e.g., from structure to collisions, in ways that are hard to quantify.

Plasma conditions in, for example, astrophysics and nuclear fusion
applications span many orders of magnitude variation in energy and in spatial
and temporal scales, and systems can be far from thermodynamic equilibrium.
Basic data may be required for quite strange-looking A+M systems; e.g., for
collision processes between neutral atoms and highly charged ions (relevant
for neutral beam heating in fusion plasma and for processes involving the
solar wind) or for neutral and low charge states of atoms in high temperature
plasma (relevant for laser-produced plasma and for plasma-wall interaction).
For applications to low-temperature industrial plasmas, similar to the case of
chemical dynamics, data are required for transient species such as molecular
radicals and molecular complexes above the dissociation threshold.
In addition, for
applications in plasma chemistry essentially always data are required for
multiple electronic states, corresponding to the possibility of charge
transfer. 
%
Very often the modeling requires data that are not accessible to direct
experiments; for example, data for atomic processes from excited initial
states, data for molecular processes resolved with respect to the rovibrational
state of the molecule, data for processes involving electronically excited
molecules, data for molecular radicals, and to some extent data involving
hazardous species such as tritium or beryllium.

To develop an effective and objective science of uncertainty assessment for
A+M applications one has to bring together physics, chemistry, computer
science, and applied mathematics communities.  The A+M and plasma modeling
communities are making the first steps in this direction, for example by
meetings such as~\cite{IAEA-ITAMP} and~\cite{StonyBrook-UQ}.  Our ultimate
goal is to develop guidelines for self-validation of computational theory for
A+M processes; i.e.~\ computational procedures by which an uncertainty estimate
is obtained along with the primary quantity of interest.  We recognize that
experimental benchmark data are sometimes available and can be used for
additional validation.  In general, this is more readily possible for
structural studies (where spectroscopic data often provide benchmark accuracy)
than for studies of collision processes.  Similarly, procedures for
uncertainty estimates are currently better developed for structure
calculations than for scattering.  This will be further elaborated below.

Energies and state-resolved cross sections are the primary data from A+M
science, but these data are normally processed further before being used in
plasma modeling codes, which tend to use effective rate coefficients for
processes in thermal plasma with explicit account of long-lived electronic
states only.  The processed data may be tabulated for interpolation or fit
functions may be used, or a combination of interpolation and function fitting.
At that stage completeness of the data (relative to processes covered and
range of collision energy) and qualitative correctness of behavior at extreme
conditions is essential; more important than pointwise accuracy.  These
processed, tabulated and fitted data are incorporated into integrated modeling
codes, and a key challenge for theory and simulation is the consistent
integration of all processes and scales together with a well-founded
assessment of uncertainties as they are generated and propagated in the
simulations.  For the propagation of uncertainties in A+M data through a
simple plasma model (no spatial dependence) we note the \hbox{HydKin} toolkit
\cite{Reiter-HydKin}, which has been developed to support fusion plasma
modeling and other applications.

The focus of the present work is on calculations based on quantum mechanics
for A+M properties and processes that are important in plasmas: atomic and
molecular structure, electron collisions with atoms and molecules (and their
ions), and charge transfer in ion-atom and ion-molecule collisions.  Processes
governed by time-dependent fields and photon-induced processes are not
considered.  Section~\ref{sec:general} contains general remarks about the need for
uncertainty estimates and about approaches for uncertainty assessment.  In
section~\ref{sec:UQstruct} we discuss uncertainty assessment for atomic and molecular
electronic structure.  Section~\ref{sec:UQcoll} is concerned with uncertainty assessment for
electron-atom and electron-molecule collisions.  In section~\ref{sec:hcoll} we consider
charge transfer in heavy particle collisions.  Section~\ref{sec:results} is concerned with
uncertainty assessment in practice, with examples from atomic and molecular
structure, electron collisions and heavy particle collisions.  In section~\ref{sec:summary}
we provide conclusions and an outlook for future work.

\section{General Considerations}\label{sec:general}
Uncertainties should be provided for observable and other physically important
intermediate quantities, such as molecular electronic excitation energies.
Quantities in structural studies for which uncertainties should routinely be
provided include:
\begin{itemize}
\setlength\itemsep{0em}
\item energy level differences, such as excitation and ionization energies and
  for molecules also dissociation energies and barrier heights;
\item configurational parameters of molecules such as bond lengths and bond
  angles at local minima and transition states;
\item properties, such as dipole moments, oscillator strengths, lifetimes, and
  polarizabilities;
\item numerical issues such as analytical representations (fits) yielding
  potential energy and dipole moment surfaces.
\end{itemize}

Quantities in collisional studies for which uncertainties should routinely be
provided include:
\begin{itemize}\setlength\itemsep{0em}
\item threshold energies;
\item cross sections and/or appropriate rates;
\item positions and widths of key resonances;
\item other observables, such as the polarization of the emitted radiation,
  branching ratios, etc..
\end{itemize}
It may also be desirable to provide uncertainties for other key computed
quantities, such as eigenphase sums or scattering lengths, which are important
for the theoretical analysis of given processes. These quantities, however,  do
not generally form input of modeling codes and therefore the provision of
uncertainties can be regarded as having lower significance.  It must be
recognized that there are difficulties in estimating uncertainties in some
cases, for example if a resonance comes out on the wrong side of a threshold.
This observation means that a computational model must have reached a sufficient
level of stability and accuracy before an uncertainty estimate is appropriate.
However it is exactly such computations that provide benchmarks and inputs to
modeling codes.

For structural studies, including computations of relative energies and
properties, the focal point analysis (FPA) technique~\cite{98CsAlSc} provides
an excellent procedure to assign uncertainties for key quantities.  Studies
building on an FPA approach can also include uncertainty estimates for effects
not explicitly computed: for example it is much easier to estimate the
magnitude of higher order electron correlation or nonadiabatic corrections to
the Born-Oppenheimer approximation than it is to compute them in specific
cases; see~\cite{jt549} for example.

In clear contrast, at present there is no well-defined general procedure for
uncertainty propagation in scattering calculations.  Notable exceptions are
the way uncertainties in dipole moments and oscillator strengths propagate
from structure to certain collisional observables.

It is important to estimate all major corrections separately, rather than
as a sum that may contain accidental cancellations, and to compare the
estimates with known values for reference ions.  Whenever possible,
calculations should be done by more than one method (such as CI and MCHF), and
the results compared for consistency.  
In the ideal scenario, for a given method of calculation, uncertainties in the
parameters of the method should be propagated towards uncertainties in the
final results (cross sections, energies, etc.).
Due to the need to approximate the many-electron Schr\"odinger equation with
a tractable model, systematic errors are in general unavoidable. The use of
different independent methods will help to reduce the influence of systematic
unknown errors.  Assuming that some uncertainty estimate is available the
results from different methods can be combined using a Bayesian approach to
produce a final probability distribution for quantities of interest, from
which a revised uncertainty can be obtained.  The use of models in combination
with experimental benchmarks to produce correlated probability distributions
for quantities of interest has become rather well established in the nuclear
data community under names such as Total Monte Carlo (TMC) or Unified Monte
Carlo (UMC); for example see Ref. ~\cite{CAPOTE20082768}.
It would be very interesting to
see such a formal and objective approach applied in the field of atomic,
molecular and optical physics as well.

Once the evaluation (comparison) between results of theoretical methods is done, the final
step is to check that the uncertainty estimates are in accord with the actual
differences between theory and experiment for known cases. 
For an objective evaluation, in an ideal situation, when experimental and
theoretical uncertainties are available, the same evaluation procedure based
on the formal statistical approach is recommended. The data (values and
uncertainties) resulted from this evaluation would account for all available
information from theory and experiment. If no systematic error in the theoretical
data is suspected (for example, if two different methods produce results
within their intervals of uncertainties), the theoretical results and their
uncertainties could be extended to cases where there are no experimental data
available. 

The above discussion demonstrates why uncertainty quantification is very
important in theoretical calculations. Unfortunately, uncertainty
quantification of the final results from a given theoretical method is often
impossible or very difficult. But it is still strongly recommended that the
authors of the produced data give an approximate estimate of uncertainty of
the produced results for the purposes discussed above. In many situations,
where the direct uncertainty propagation is not possible, sensitivity tests
could and should be performed to collect statistics and estimate uncertainties
of the final results.

\section{Uncertainty Estimates for Structure Computations}~\label{sec:UQstruct}
\vspace{-10.0truemm}
\subsection{Atoms}~\label{subsec:astruct}
The discussion of uncertainties in atomic structure computations begins with
one- and two-electron atoms and ions since these provide the traditional
testing grounds for theory in comparison with experiment.  Theoretical
uncertainties here limit the accuracy that can be achieved for more complex
atomic systems.

The highest accuracy can of course be achieved for hydrogen and other two-body
problems since the Schr\"odinger equation can be solved exactly to find the
exact nonrelativistic wave~function and energy~\cite{DrakeBandS}.
Uncertainties then come from relativistic and quantum electrodynamic (QED)
corrections, and the effects of finite nuclear size and structure (for a
general review, see ref.~\cite{Eides_2001}).  The sizes of the relativistic
and QED corrections are determined by the dual expansion parameters $\alpha$
and $\alpha Z$, where $\alpha = 1/137.035999139(31)$ is the fine structure
constant and $Z$ is the nuclear charge~\cite{PhysRevLett.95.163003}.
Beginning with the lowest order nonrelativistic energy, the relativistic
corrections can generally be represented as an expansion in powers of $(\alpha
Z)^2$.  For the case of one-electron atoms and infinite nuclear mass, the
series can be summed to infinity by solving instead the Dirac equation to
obtain the exact relativistic energies~\cite{DrakeBandS}.  However, QED
corrections (Lamb shifts) cannot be similarly summed to all orders, and so
represent a dominant source of uncertainty.  The lowest order one-loop terms
from vacuum polarization and electron self-energy are of order $\alpha^3 Z^4$
Ry.  These can be calculated essentially exactly.  Higher order terms come
from both binding energy corrections as additional powers of $\alpha Z$, and
multi-loop Feynman diagrams as additional powers of $\alpha$.  The higher order
terms are known in their entirety up to $\alpha^6 Z^6$ Ry, but the uncertainty
in the numerical coefficients gives an uncertainty of order $\alpha^6 Z^7$ Ry,
or a few kHz for the ground state of hydrogen~\cite{Pachucki_2003}.  The
uncertainty from finite nuclear size effects is about an order of magnitude
larger, and hence dominates.

For heavy hydrogenic ions up to U$^{91+}$ and beyond, considerable progress
has been made in summing the binding energy corrections (i.e.\ powers of
$\alpha Z$) to all orders for certain classes of diagrams
\cite{YerokhinShabaev_2015}, coupled with experiments for comparison (see
Gumberidze \etal~\cite{Gumberidze}, and earlier references therein).
For the ground state of U$^{91+}$, the theoretical Lamb shift is $464.26 \pm 0.5$
eV, in good agreement with the measured value $460.2 \pm 4.6$ eV.  For excited
s-states, the Lamb shifts and uncertainties scale approximately as $1/n^3$
with $n$ and $Z^6$ with $Z$.  These uncertainties place a fundamental limit on
the accuracy of atomic structure computations.

For atoms or ions containing two or more electrons, the Schr\"odinger equation
is not separable, and hence cannot be solved exactly.  Electron correlation then
enters as an important new source of uncertainty.  The correlation energy
represents the difference between the exact energy, and the Hartree-Fock (HF) approximation
arising from the use of spherically averaged potentials to obtain an
independent particle approximation.  Methods for few-electron atoms are
divided into two broad categories, depending on the relative importance of
correlation effects and relativistic corrections.  As a function of $Z$ for an
isoelectronic sequence, correlation effects are proportional to $Z^0 = 1$
(i.e.\ a constant) while the lowest order relativistic corrections are
proportional to $\alpha^2Z^3$.  There is therefore a crossover point when
$\alpha^2Z^3 = 1$, or $Z =1/\alpha^{2/3} \simeq 27$.  For $Z \le 27$,
correlation effects dominate relativistic effects.  Consequently, one should start
with the best possible solutions to the nonrelativistic Schr\"odinger
equation and treat relativistic corrections as a perturbation.  Conversely,
when $Z > 27$, one should start with exact one-electron solutions to the Dirac
equation, and treat electron correlation as a perturbation.  We will call
these two regions the low-$Z$ and high-$Z$ regions respectively.  There is a
broad region around $Z = 27$ where both methods yield useful results, and
provide interesting comparisons to assess the accuracy.


Atoms with two or three electrons provide a special case because
specialized techniques are available that yield essentially exact
solutions to the Schr{\"o}dinger equation.  This is achieved by
expanding the wave function in a Hylleraas basis set of functions
involving explicitly powers of the interelectron coordinate $r_{12} =
|{\bf r}_1 - {\bf r}_2|$, where ${\bf r}_1$ and ${\bf r}_2$ are the
position vectors of the individual electrons.  Since a Hylleraas basis
set is provably complete \cite{Klahn,Klahn2}, a variational calculation in
Hylleraas coordinates is guaranteed to converge from above to the exact
nonrelativistic energy.  The accuracy can be readily determined from
the rate of convergence as more functions are added to the basis set.
In this way, the nonrelativistic energy of the ground state of helium
has been determined to 35 or more significant figures
\cite{Schwartz,Nakashima}, and results accurate to 20 or more
significant figures can be readily obtained for the entire singly
excited spectrum of helium \cite{Drake_Yan}.  At this level of
accuracy, calculations must be done in at least quadruple precision (32
decimal digits).  Some authors go even further to use multiple 
precision arithmetic (48 or 64 decimal digits) \cite{Korobov,Bailey19,Bailey21} in order
to avoid
numerical linear dependence in the basis set and preserve numerical
stability. The record is the 101-digit arithmetic used by Schwartz 
\cite{Schwartz} for the ground state of helium. However, the standard
quadruple 
precision arithmetic provided by FORTRAN is usually sufficient, provided
that care is exercised in choosing the basis set in order to avoid
excessive numerical linear dependence. See for example Ref.\ \cite{Nistor}
for the use of triple basis sets in Hylleraas coordinates to maintain
numerical stability.
Results for lithium-like atoms with three electrons
are not as accurate because the basis sets become considerably larger
(i.e.\ 30,000 terms instead of 3,000 terms), but energies accurate to
16 figures and other atomic properties can still readily be obtained
\cite{Wang}.

At these levels of accuracy for two- and three-electron atoms, the dominant
sources of uncertainty in the low-$Z$ region are the relativistic and QED
corrections, as discussed above for hydrogen.  The Breit interaction accounts
for relativistic corrections of order $\alpha^2Z^4$ Ry, and a full
many-electron theory accounts completely for QED corrections of order
$\alpha^3Z^4$ (including the Araki-Sucher terms for QED corrections to the
electron-electron interaction)
\cite{DrakeMartin_1998,DrakeMorton_2006,DrakeYan_2008}.  Theory has also
recently been completed for all terms of order $\alpha^4Z^5$ Ry
\cite{Yerokhin_2010}, although the nonrelativistic operators become
complicated and difficult to evaluate.  The resulting uncertainty from
higher order terms is estimated to be 36 MHz for the ionization energy of the
ground state of helium, and this scales as $Z^5$ with nuclear charge and
roughly $1/n^3$ with $n$.  For a comprehensive review, and tabulation for all
states up to $n = 10$ and angular momentum $L = 7$, see ref.~\cite{Handbook}.

In the high-$Z$ region, the all-orders methods described above for hydrogenic
ions can be extended to helium-like ions and combined with $1/Z$ expansion
calculations from the low-$Z$ region (the so-called unified method) to obtain
results that are accurate over the entire range from $Z = 2$ to $Z = 100$
\cite{Drake_1987,Artemeyev_2005}.  In most cases, the theoretical accuracy is
better than the experimental.  The uncertainty from omitted terms of order
$\alpha^4Z^4$ is estimated to be $\pm$1.2$(Z/10)^4$ cm$^{-1}$ for the $n=2$
states.

Calculations of similar accuracy can also be carried out in Hylleraas
coordinates for three-electron atoms, but that is the limit to what has been
achieved to date.  Further progress is hindered by the technical difficulties
of calculating integrals involving nonseparable products of factors
containing all the interelectron coordinate of the form
$r_{12}r_{23}r_{34}\cdots$.

For many-electron atoms, one must resort instead to generally applicable
methods of atomic structure based on the HF approximation, or
its generalizations to the multi-configuration Hartree-Fock (MCHF) or
configuration interaction (CI) methods.  
The MCHF method is usually called MCSCF in quantum chemistry as
the HF approximation is called the self-consistent field (SCF) method.  
The relativistic versions of these
methods are based on the Dirac equation instead of the Schr\"odinger equation,
and are called the Dirac-Fock (DF) approximation, with generalizations to the
corresponding multi-configuration Dirac-Fock (MCDF) or relativistic
configuration interaction (RCI) methods.  The basic approximation of the HF
and DF methods is to assume that the many-electron wave~function can be
written as an anti\-symmetrized product of one-electron orbitals (a Slater
determinant).  The HF (or DF) solution is the one that minimizes the energy
over all wave~functions that can be expressed in this Slater determinant form.
The difference between the HF (or DF) energy and the exact energy is called
the electron correlation energy.  The correlation energy can be systematically taken
into account by solving a larger problem in which the mixing with other
electronic configurations is included.  The configuration mixing is induced by
the difference between the effective HF potential and the exact electrostatic
potential containing all the interelectronic repulsion terms.  The difference
between CI and MCHF revolves around whether or not the electron
orbitals are frozen (CI) or allowed to vary (MCHF) to obtain a self-consistent
solution.  The correlation energy is of key importance in chemical physics,
because it is typically the same order of magnitude (about 1 eV) as chemical
binding energies.  One might say that much of chemistry is buried in the
correlation energy.  Full spectroscopic accuracy can require correlation
energies as accurate as $\pm$$10^{-9}$ eV or better for neutral atoms.

The coupled cluster (CC) method is a variation of CI, which also starts from
the HF orbitals, but then uses Brueckner-Goldstone perturbation theory to
describe excitations from the HF reference state, organized as singles (S),
doubles (D), triples (T) etc.  The advantage is that it guarantees the
size-extensivity of the solution, but it lacks the variational character of
the CI method.

Both the CI and MCHF methods are exact in principle (within their respective
nonrelativistic or relativistic approximations) and generally applicable to
many-electron atoms and ions, but they are much more slowly convergent than
the methods based on Hylleraas basis sets for two- or three-electron atoms.
It can be shown that a CI calculation is equivalent to a Hylleraas calculation
that includes only the even powers of $r_{12}$ in the basis set, but it is the
odd powers that are most effective in reproducing the cusp at $r_{12} = 0$ in
the correlated electronic wave~function.  For this reason, a CI calculation
requires much larger numbers of configurations in the variational wave
function in order to achieve even modest levels of accuracy, and so careful
convergence studies must be carried out to assess the uncertainty in the
calculation.  Convergence uncertainties better than $\pm$$10^{-6}$~eV are
seldom achieved, even for few-electron atoms, and the convergence is typically
much worse for many-electron atoms.  For example, Chantler {\it et
al}~\cite{Chantler2010} carried out a detailed convergence study for
satellite spectra of the copper $K$-alpha photo-emission spectrum, and found
that uncertainties were of the order of $\pm$$0.01 - 0.1$ eV using the MCDF
method.  Their work includes a detailed consideration of valence-valence and
valence-core contributions to the correlation energy.  They also make use of
comparisons between the length and  velocity forms of dipole transition
integrals to assess the accuracy.  Many other similar studies have been
carried out.  A great deal of work has now been done by many authors to
develop systematic procedures to assess the theoretical/computational
uncertainties, and to assign reasonable uncertainty estimates
\cite{atoms2010001,atoms2010015,atoms2020086,atoms2020215,atoms2040382,Safronova,Safronova2,Sahoo}.  
Uncertainties of transition parameters can be
evaluated by investigating differences between results calculated in the length and
the velocity gauges for $LS$-allowed transitions~\cite{fischer2009evaluating},
and the analysis can be also extended for $LS$ intercombination lines for
certain cases~\cite{atoms2020215}.  Perturbative analysis by performing
smaller calculations with neglected correlation effects is also useful to estimate
uncertainties~\cite{atoms2010001}.

Furthermore, it is necessary to include uncertainties due to physical effects not
included in the calculation, such as additional classes of excitations, or
quantum electrodynamic corrections.  In recent work authors such as Safronova
\etal~\cite{Safronova} and K\'allay \etal~\cite{Sahoo} have
made progress towards a comprehensive programme for the assessment of
uncertainties that goes beyond the simple assessment of convergence
uncertainties.  The objective is to estimate an uncertainty that is
independent of the actual difference between theory and experiment.  For
uncertainties of this type, the central value is not necessarily the most
probable.  For example, if QED corrections of order $\alpha^3$ have been
omitted, then one can expect further corrections of order $\pm$$c\alpha^3$,
where $c$ is a nonzero coefficient whose value can often be estimated from
other similar calculations, or from general scaling rules with $n$ and $Z$.
It is often possible to establish similar ``reference ions'' where
experimental data exist for comparison with theoretical estimates of the
uncertainties.  The aim is to obtain reasonable estimates of the
uncertainties, not rigorous bounds on the actual difference between theory and
experiment (i.e.\ the error).

\subsection{Molecular electronic ground state properties}~\label{subsec:mstruct}
Without the so-called Born-Oppenheimer (BO) separation~\cite{27BoOp,1954BoHu}
of nuclear and electronic motions the traditional concept of a molecular
structure would basically be lost, as only a murky quantum soup of delocalized
particles would exist.  As a consequence of the BO approximation, electronic
structure theory and nuclear motion theory emerge as the two main subfields
of molecular quantum chemistry.  These two fields are linked by potential
energy surfaces (PESs), plus any beyond-BO corrections that may deemed
appropriate for a given problem~\cite{jt186,jt236,jt566,Matyus14}.  Given that for all
but a few simple problems the converged absolute energy of a molecular system
cannot be obtained, it is important to note that molecular structure
computations are always concerned with relative rather than total energies, and
the same must be the case for the uncertainty estimates.

Much of modern applied molecular quantum chemistry is aimed at mapping out,
locally or globally, PESs of molecular species or reaction complexes
(scattering systems) by means of sophisticated numerical techniques
\cite{CMS,Dream}.  For studies of molecular structures the PES is needed
mostly in the vicinity of a minimum.  The widespread availability of analytic
gradients and higher derivatives in standard electronic structure
codes~\cite{94YaOsGoSc} has substantially increased the utility of quantum
chemistry for the exploration of PESs.  For studies of scattering systems it
is important to have a full-dimensional representation of the surface
throughout the accessible region.  Fundamental work in this area was done by
Murrell and coworkers~\cite{84MuCaFaHu}; see ref.~\cite{braams2009}
for more recent developments.

\begin{figure}[b]
\centering
\includegraphics[width=0.495\textwidth]{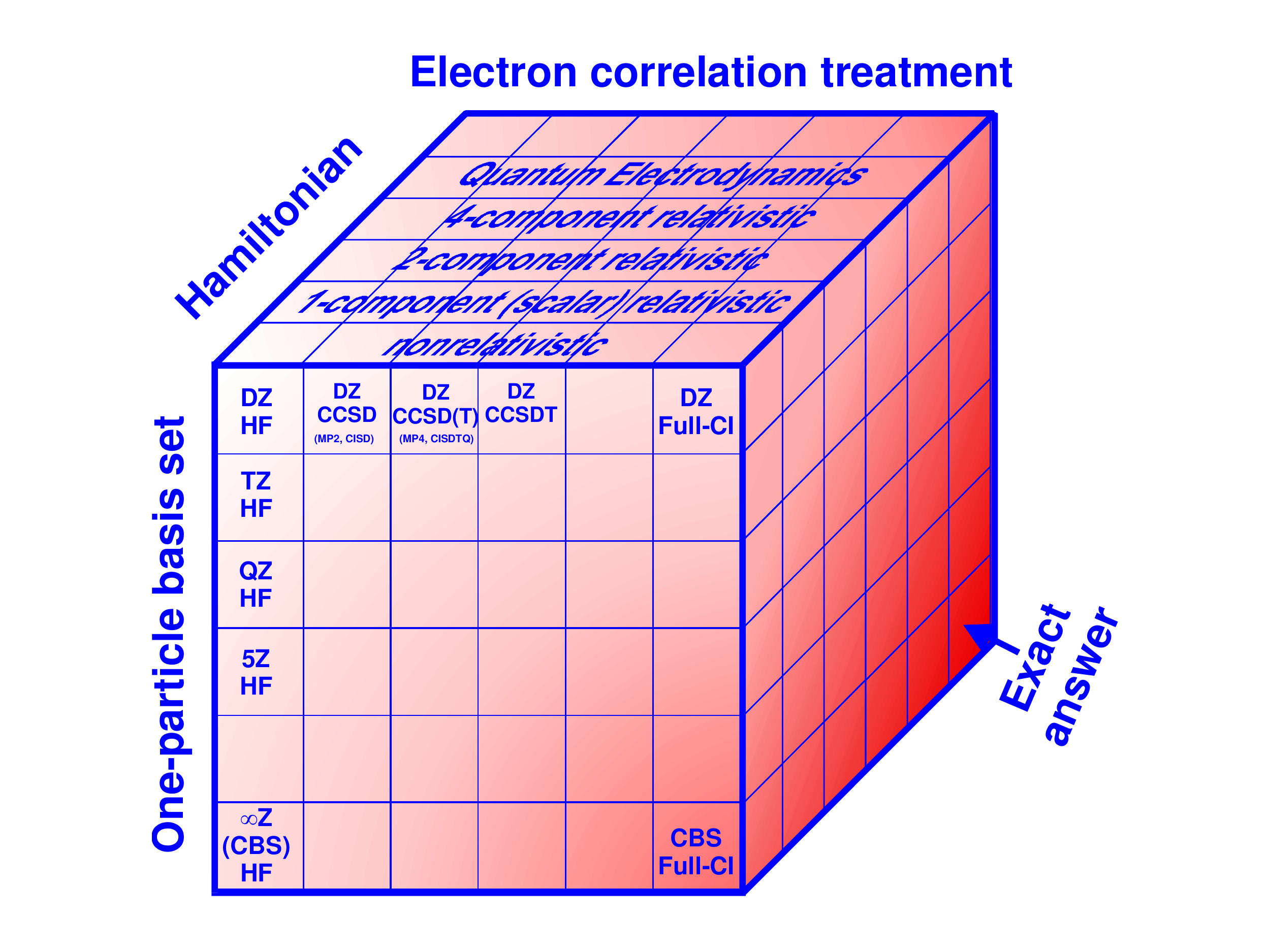}
\caption{The three axes of the computational cube of wavefunction-based
  electronic structure theory represents three important approximations.  The
  figure was first published in~\cite{magiccube}.}
\label{fig:MagicCube}
\end{figure}

For all systems of chemical interest, the exact solution to the
(nonrelativistic, time-independent) electronic Schr\"odinger equation cannot
be obtained; thus, a hierarchy of increasingly accurate wave~function
approximation methods is needed beyond the BO separation of nuclear and
electronic motions~\cite{CMS}.  Basic to the understanding of this hierarchy
and of the uncertainty at any given level is the computational cube depicted in
figure~\ref{fig:MagicCube}.  It demonstrates that there are three fundamental approximations in
polyatomic electronic structure theory:
\begin{itemize}\setlength\itemsep{0em}
\item choice of the electronic Hamiltonian;
\item truncation of the one-particle basis (often referred to as the atomic
  orbitals);
\item the extent of the electron correlation treatment, the $n$-particle basis.
\end{itemize}
The target result corresponding to the three simultaneous limits is approached as
closely as possible by choosing an appropriate Hamiltonian and extending both
the one-particle basis set and the many-electron correlation method
($n$-particle basis) to technical limits.  For lighter elements, perhaps up to
Ar(Z=18), the effects of special relativity will not be consequential (see the
previous section), except in electronic structure studies seeking ultimate
accuracy.

The {\it ab initio} limit can be approached by composite schemes that employ
multiple electronic structure computations at different levels of theory to
arrive at a single energy for a given molecular geometry.  A general composite
scheme that is highly successful is the FPA approach~\cite{98CsAlSc}.  A
fundamental characteristic of this approach is the dual extrapolation to the
one- and $n$-particle limits of electronic structure theory.  The process
leading to these limits can be characterized as follows:
\begin{itemize}\setlength\itemsep{0em}
\item use of a family of basis sets, such as (aug)-cc-pV$X$Z~\cite{Dunning1989}, 
  which systematically approaches completeness through an
  increase in the cardinal number $X$, as a key aspect of FPA is the assumption
  that the higher order correlation increments show diminishing basis set
  dependence;
\item application of lower levels of theory (typically, HF and MP2
  computations) with very extensive basis sets;
\item execution of a sequence of higher order correlation treatments with the
  largest possible basis sets;
\item layout of a two-dimensional extrapolation grid based on the assumed
  additivity of correlation {\it increments}, that is, the differences between
  correlation energies given by successive levels of theory in the adopted
  hierarchy.
\end{itemize}

Within the FPA approach one considers the consequences of several ``small'' physical effects:

\begin{itemize}\setlength\itemsep{0em}
\item core electron correlation;
\item special relativity;
\item adiabatic and nonadiabatic corrections to the BO approximation;
\item quantum electrodynamics (QED).
\end{itemize}

In diatomic molecules containing first-row atoms several effects due to core
correlation have been established~\cite{CsaszarAllen1996}.  Equilibrium bond
lengths experience a contraction of about 0.001\,\AA\ for single bonds and
0.002\,\AA\ or more for multiple bonds.  The {\it direct} effect of core
correlation is a correction function to the valence diatomic potential energy
curve that has negative curvature at all bond lengths around the equilibrium
position.  Core correlation decreases all higher order force constants.

For extremely accurate structural studies the corrections to the BO
approximation cannot be neglected, especially if light atoms are present in
the system.

QED also provides electronic radiative corrections (or Lamb shifts) arising
from the interaction of the electron with the fluctuation of the
electromagnetic field in vacuum.  Studies of atoms, see above, and simple
molecules~\cite{jt265} have indicated that QED effects are generally orders of
magnitude smaller than scalar relativistic corrections.

Molecular properties are also an important result of electronic structure
calculations, not least because they provide input into subsequent scattering
calculations.  FPA-type approaches are now being used to provide uncertainty
estimates for permanent dipole moments~\cite{jt424,jt509}.  There are two
viable methods of calculating dipole moments associated with a given
electronic wave~function.  The most straightforward method, implemented
directly in standard quantum chemistry codes, is to compute the dipole moment
as an expectation value (EV).  An alternative method is to compute the dipole
moment studying the response to the application of a (small) electric field
placed in appropriate directions by finite differences (FD) of the perturbed
energies.  The methods are related by the Hellmann-Feynman theorem~\cite{06Jensen}, 
but in general this theorem only holds when exact wave
functions are used.  In practice, differences between the two methods can be
large~\cite{02Lipins.ai}.  EV dipoles are cheaper to compute, indeed they are
essentially free once a wave~function is available, whereas FD dipoles require
the computation of extra points with a finite field.  However, minor
contributions to the dipoles, e.g., non-BO or relativistic effects, can be
evaluated in the FD approach using energy differences even when their
contribution to the electronic wave~function is unknown.  Furthermore, there
is a general acceptance~\cite{83WeRoRe.OH,92ErMaPe.ai,jt475} that the FD
approach converges more quickly to the true answer for a given (approximate)
wave~function.  We therefore recommend the adoption of this approach to the
uncertainty assessment for dipole moments.  We note that, unlike the situation
with the use of different gauges for photo\-ionization calculations
\cite{75PiKe,09GrBaPi}, thus far comparison of EV and FD approaches have
provided little insight into the uncertainty in a given calculation.

Even less effort has been dedicated to the computation of transition dipole
moments, despite their importance for electronic spectra and as inputs to
scattering calculations.  However, FD methods for evaluating transition
dipoles are available~\cite{98AdZaSt.methods} if not extensively used.
Studies~\cite{jt599,jt623} suggest that while the FD approach for transition
dipoles shows improved convergence behavior compared to the EV approach,
perhaps more so than for the diagonal dipole moments, there are technical
issues with their use that still need to be overcome~\cite{jt623,jt632}.

\begin{figure}[t]
\centering
\includegraphics[width=0.48\textwidth]{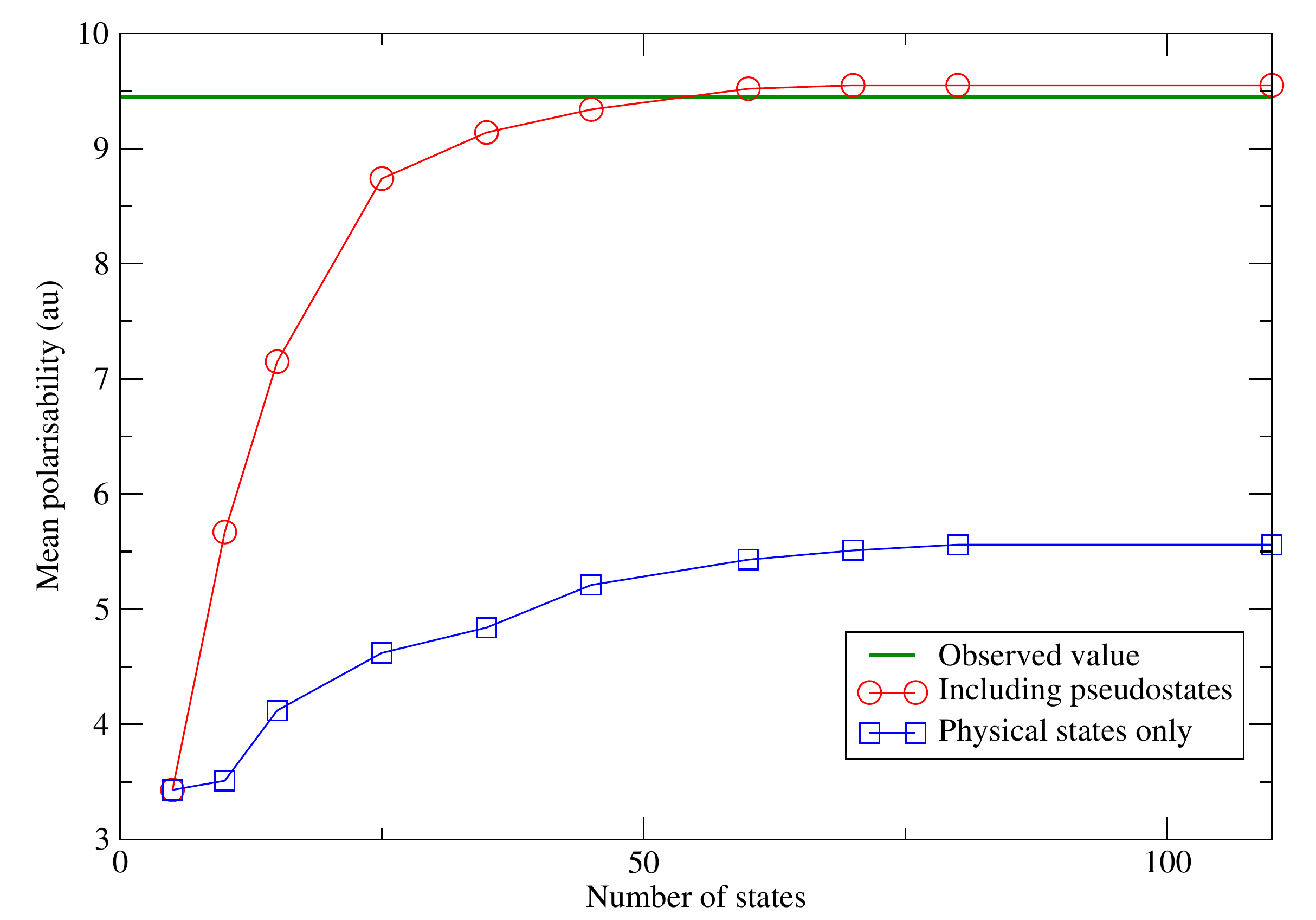}
\caption{\label{fig:pol} Spherically averaged polarizability of water in its
  equilibrium geometry computed using sum-over-states formula~(\ref{eq:sos})~\cite{jt468}; 
 the experimental value is corrected for vibrational effects~\cite{jna97}.}
\end{figure}

Target polarization is an important property for scattering calculations.
However, the target polarizability rarely enters directly into the scattering
model.  Even when it does, it usually enters only in the long range part of the
potential~\cite{91JaGi}.  How well a given scattering model represents the
target polarizability can be used as a proxy for how converged the
polarization potential is as a whole.  If the model gives a poor
representation of the target polarizability, then the representation of the
overall polarization effect is likely to be poor.  The components of the
dipole polarizability tensor can be computed using the formula
\begin{equation}
\alpha_{rs}=2\sum_{n>0}\frac{<0|\mu_r|n\!><n|\mu_s|0\!>}{E_n-E_0},
\label{eq:sos}
\end{equation}
where the $\mu$ are dipole operators, and $r$ and $s$ represent Cartesian
components.  Here $E_n$ and $|n\!>$ represent the electronic energy and
associated electronic wave~function for the $n$-th electronic state of the
system; the state for which the polarizability is being calculated is labeled
$|0\!>$, but it does not have to be the ground state.  For the ground state this
series converges from below, provided enough states are included in the
expansion.  Experience~\cite{jt468,jt585} shows that (a) convergence to the
correct value requires consideration of the continuum; and (b) sums running
over only bound states show apparent convergence to a value that is too low.
These issues are illustrated in figure~\ref{fig:pol}.

Use of Eq.~(\ref{eq:sos}) can therefore demonstrate the adequacy of a chosen
close-coupling expansion.  However, when doing this one also should note that
approximate wave~functions, such as those given by the HF approximation, are
usually more polarizable than accurate wave~functions.  A cancellation of
errors can therefore arise, whereby an inaccurate target representation is
combined with an incomplete sum over states yielding a polarizability in
apparent agreement with experiment or better computations.

\subsection{Molecular electronic excited state properties}
Excited electronic states are of interest in their own right and form an
important component of scattering calculations, where their representation is
important both for electronic excitation studies and as part of close-coupling
expansions.  There are far fewer systematic studies of the convergence of
excited state calculations with respect to the various components discussed
above.  FPA has been used for the study of properties of excited electronic
states only to a limited extent~\cite{DeYonker2014,Valeev2001,Bokareva2009}.
Indeed an issue for many scattering studies, both theoretical and
experimental, is that generally there are far fewer studies of excited states
of different spin symmetry than the ground state, since excitation of such
states is optically forbidden.  However, the lowest excited state is usually
in this class.

Molecular excited electronic states can be classified as valence, roughly
corresponding to rearrangement of the electrons within the valence orbitals,
and Rydberg, corresponding to a loosely bound electron orbiting a parent ion.
Different techniques are required to give good representations of these two
types of states~\cite{jt560}, even though many molecular states are either a
mixture of the two or change their character as a function of bond length.  We
note that Rydberg states form regular series, which are often well represented
by quantum defect theory~\cite{J96}.  Experience shows that uncertainties in
these states can also often be better represented in terms of quantum defects
rather than absolute energies~\cite{jt189,jt260}, although other methods can
be used for calculations including assessment of uncertainties
\cite{10CoWhPe}.

\section{Uncertainty Estimates for Electron Scattering Calculations}~\label{sec:UQcoll}
Before going into uncertainty assessment for collision processes, it is
advisable to recognize that there are a number of energy ranges for which
particular methods have been developed and are believed to be particularly
suitable.  The confidence in a given method is usually based on general
scattering theory, some numerical examples, and -- last but not least --
comparison with experimental benchmark results.  There is, however, never a
guarantee in collision physics, although some variational principles exist
(e.g., for the eigenphase sum).

In general, the energy ranges of interest are:
\begin{itemize}\setlength\itemsep{0em}
\item low energy collisions, with incident projectile energies well below the
  first electronic inelastic threshold;
\item low energy, near-threshold collisions, with projectile energies well
  below the first ionization threshold;
\item intermediate energy collisions, with incident projectile energies from
  about the first ionization threshold to a few times that value;
\item high energy collisions, with projectile energies exceeding several times
  the first ionization threshold;
\item collisions with relativistic energies, in which the kinetic energy of
  the incident projectile should no longer be described by the
  nonrelativistic formula.
\end{itemize}

There is a wealth of literature available on methods for electron scattering
calculations; hence we refer to recent
reviews~\cite{0022-3727-46-33-334004,1402-4896-90-5-054006,Burke2011,jt474}.
However, we emphasize again that there is no unambiguous rule regarding the
reliability of a particular method.  As will be further discussed below, there
are simply too many parameters other than the collision energy that may come
into play.  Nevertheless, it seems useful to provide some general guidelines
based on this one parameter.

For low energy collisions a one-state close-coupling expansion may provide a
good start.  In contrast to potential scattering (which is a further
simplification if the potential is chosen as local, i.e.~only depending on
the position of the scattering projectile), the approach can properly contain
exchange effects.  On the other hand, even the closed channels can have a
major influence by polarizing the target.  This effect is often accounted
for by some real-valued ``optical'' potential (local or nonlocal).  In fact,
the method can be pushed toward higher energies by including an imaginary
``absorption'' potential to account for loss of flux into inelastic channels.

Moving on to the near-threshold regime, the close-coupling expansion
containing a number of $n$~discrete states (to be referred to as ``CC{\it n}''
below) has been the method of choice for many years.  It is often highly
successful in the description of resonances associated with low-lying
inelastic thresholds.  However, the method may have problems in the low energy
regime if significant polarization effects originate from coupling to
higher-lying discrete states and, in particular, the ionization continuum.

For intermediate energy collisions, the above-mentioned effect of coupling to
discrete states omitted in the CC{\it n} expansion, and even more importantly
to the ionization continuum, should be accounted for in some way.  One way to
do this is to extend the CC expansion by including a number of so-called
``pseudo-states'', which are essentially finite-range states that are forced
to fit into a box.  For the general idea, details of the box are not
important; it only matters that the states are square-integrable and provide a
way to discretize the (countable) infinite Rydberg and the continuous
ionization spectra.  This is the basic idea behind the ``convergent
close-coupling'' (CCC)~\cite{Bray2012135} and ``\hbox{R-matrix} with
Pseudo-States'' (RMPS)~\cite{0953-4075-29-1-015,jt341} approaches.  While the
implementations may vary greatly, the critical idea is exactly the same in
both methods.  Hence, if the same states (physical and pseudo) are included in
the expansion, the final results should be the same -- except for numerical
issues that may remain in practice.

Other ways to account for the possibility that two electrons may leave the
target after the collision (but only one of the electron wave~function
fulfills the correct boundary conditions in the CC approach) include
``time-dependent close-coupling'' (TDCC)~\cite{PhysRevA.65.042721} and
``exterior complex scaling'' (ECS)~\cite{Rescigno24121999}.  In the former, a
wavepacket is used for the projectile and the formalism is expressed as an
initial value rather than a boundary value problem.  In the latter, the
coordinate system is changed from a real to a complex radial grid in order to
transform the oscillatory character of the positive-energy continuum wave
function to an exponentially decreasing character that, once again, allows for
proper evaluation of certain integrals.  CCC, RMPS, TDCC, and ECS have been
highly successful in handling ionization processes in particular, although the
extraction of the relevant information is by no means trivial.  For details we
refer to some of the references given, which however should only be considered
as a starting point.  Regarding actual applications to date, ECS has not
really been used for production calculations of atomic data relevant for
plasma modeling, TDCC has been used mostly to check other approaches for
quasi-one and quasi-two electron targets, CCC has been applied over a wide
range of the latter targets, while RMPS has been applied also to targets with
more complex structure, in particular the noble gases beyond helium as well as
other open-shell systems.

Moving on to the high energy regime, perturbative methods based on some form
of the Born series are generally the method of choice.  In this case, the
projectile is either described by a plane or a distorted wave, and then the
transition matrix elements are obtained by relatively straightforward
integrations.  The first-order Distorted-Wave Born Approximation (DWBA)
\cite{taylor1972scattering,joachain1975quantum,madison1973distorted,itikawa1986distorted}
has the advantage over the corresponding plane-wave (PWBA) version
\cite{bote2008calculations} in that it accounts for some higher order terms of
the plane-wave series.  In practice, production calculations of atomic data in
the high energy regime are mostly being performed in the DWBA approach~\cite{Gao06,AlH09,Toth11,Zhang14}.  If
possible, a good check of the applicability of the method involves pushing it
toward the intermediate energy regime and then comparing the predictions to
those from more sophisticated methods.  The present implementation of the CCC
approach in momentum space is particularly useful in this respect, since the
limiting case of the CCC T-matrix elements for high energies is actually the
DWBA or PWBA result.

Regarding the high energy range, full-relativistic implementations of RMPS
\cite{badnell2008dirac}, CCC~\cite{fursa2008fully}, and DWBA
\cite{zuo1991relativistic} exist and are frequently used, especially for heavy
targets and when the description of explicitly spin-dependent effects (beyond
exchange) is desirable.  This may, indeed, be necessary since (in a classical
picture) the kinetic energy of the projectile may be relativistic near the
nucleus even if it is nonrelativistic in the asymptotic regime far away from
the target.  In this paper we will not consider collisions for which the
initial energy is already relativistic.

Finally, we mention the existence of semi-empirical methods, such as the
``Binary Encounter f-scaling'' (BEf)~\cite{PhysRevA.64.032713} and ``Binary
Encounter Bethe'' (BEB)~\cite{PhysRevA.50.3954} approaches to electron impact
excitation and ionization.  While these methods are highly useful in practice, they are
somewhat limited in scope.  For example, BEf can only be used for optically
allowed transitions and also requires experimental or reliable theoretical
data for rescaling.  We do not feel comfortable to suggest a method for an
uncertainty assessment for these approaches.

As mentioned above, there are many issues that contribute to the problem of
uncertainty assessment in scattering calculations.  These include:
\begin{itemize}\setlength\itemsep{0em}
\item Target properties (energy levels, polarizability, dipole and higher
  moments), which are ultimately associated with the quality of the wave
  functions used.
\item Model contributions, including:
  \begin{itemize}\setlength\itemsep{0em}
  \item The need for a consistent treatment of the $N$-electron target
    vs.\ the $(N+1)$-electron collision problem, which is a critical issue in
    obtaining accurate resonance positions;
  \item accounting for the nuclear motion in electron-molecule collisions.
  \end{itemize}
\item Numerical uncertainty.
\end{itemize}

Some of these issues will be elaborated further below.  Not surprisingly,
the major challenge is to propagate the uncertainty associated with the above
lists to give a final uncertainty on the quantities of interest (see below).
We suggest that:
\begin{itemize}\setlength\itemsep{0em}
\item Calculations be performed for a range of target models, thereby
  reflecting the underlying uncertainty in the target properties.
\item Attempts be made to quantify uncertainties associated with the choice of
  the scattering model; this will need to be done on a case-by-case basis (see
  below).
\item Numerical uncertainties be quantified similarly to the FPA procedure
  described above.
\end{itemize}

\subsection{Electron -- atom/ion scattering}~\label{subsec:acoll}
There is a wealth of experimental observables in the field of electron
collisions with atoms, ions, and molecules.  For a fixed incident projectile
energy and direction (even those could, of course, be represented by some
distributions), the most general (and hence least specific) observable is the
grand total cross section, obtained by integrating over all processes,
energies, angles, angular momenta, spins, their components, etc..  Such a
cross section is certainly relevant and can sometimes (but not always) be
measured with high accuracy in transmission cells or via the loss of the
target species in traps.  The grand total cross section is made up of sums or
integrals over unobserved quantities, where the lack of observation is not a
requirement of quantum mechanics, but rather a choice of the experimenter.
This choice may be voluntary or involuntary.  In the former case, one might
only be interested in a rather global set of parameters to model a system,
while the latter case might be forced if the signal rate is simply inadequate
to measure what one would really like to know.

It is clearly unrealistic to discuss all possible cases, including also those
not even specified above, where the initial projectile and target beams might
have been prepared beyond an unpolarized ensemble.  We therefore restrict our
discussion to angle-integrated state-to-state cross sections and in some cases
the rate coefficients that can be derived from them by performing an integral
over the incident projectile energy.

For electron collisions with atoms and ions, the processes of interest for the
present paper, including initially excited states, are:
\begin{itemize}\setlength\itemsep{0em}
\item elastic + momentum transfer;
\item inelastic (excitation);
\item inelastic (ionization);
\item dielectronic recombination.
\end{itemize}
A few illustrative examples about how one might attempt to quantify
uncertainties in theoretical predictions for these processes will be given
in section~\ref{sec:results}.

\subsection{Electron -- molecule scattering}~\label{subsec:mcoll}
Many of the issues involved in uncertainties for electron-molecule scattering
are similar to those for atoms so below we concentrate on those that differ.

Processes of interest, including those starting from initially excited states,
are
\begin{itemize}\setlength\itemsep{0em}
\item elastic and momentum transfer collisions;
\item inelastic, rotational excitation;
\item inelastic, vibrational excitation;
\item dissociative electron attachment or recombination;
\item inelastic, electronic excitation;
\item impact dissociation, which normally goes via electronic excitation;
\item ionization.
\end{itemize}
These processes (listed in approximate order of increasing collision energy)
involve a mixture of electronic excitation (either directly or via impact
dissociation or ionization) and excitation of the (rotational or vibrational)
nuclear motion.  There is no current, general method that solves for all these
processes simultaneously in a unified self-consistent manner.  For example,
most treatments of electronic excitation or ionization are performed at the
fixed nuclei level whereas treatments of dissociative attachment or
recombination use specially adapted nuclear motion techniques employing
resonance (potential energy) curves which are computed in electron collision
calculations.  See refs.~\cite{jt527,jt586} for example.

In practice nuclear motion is often introduced in a somewhat {\it ad hoc}
fashion deemed appropriate for the process of interest.  For example,
resonances greatly enhance vibrational excitation cross sections and these can
be computed in a relatively straightforward fashion using resonance curves,
see ref.~\cite{jt586}.  Conversely, nonresonant vibrational excitation can be
treated by vibrationally averaging T-matrices as a function of geometry
\cite{chang1972theory,84MoFeSa}.

The vibrational averaging of the geometry-fixed scattering (or T-) matrices is
a part of the frame transformation approach
\cite{chang1972theory,atabek1974quantum,greene85}, developed in the 1970s and
1980s to account for non-BO couplings of the incident electron with the
vibrational and rotational motion of the target molecule.  For collisions with
molecular ions, the frame transformation can be combined with multi-channel
quantum defect theory (MQDT)~\cite{seaton66,aymar96} to give an approach that
unifies nonresonant and resonant processes in electron-molecule scattering
\cite{atabek1974quantum,greene1979general,greene85,gao1989energy,gao1990alternative}
including rovibrational and electronic resonances, and that accounts for non-BO
couplings and vibrational excitation of the target by the incident electron.
Full vibrational close-coupling also provides a means of treating resonant and
nonresonant processes simultaneously for electron collisions with neutral
molecules, but it is rarely used~\cite{95SuMoIs}.

Rotational motion and excitation of the target molecule are often treated by
means of a transformation from the body-fixed frame to the laboratory frame by
simple angular momentum recoupling~\cite{cha56}, which can be viewed as a part
of the general frame transformation approach discussed above
\cite{jt439,faure09,kokoouline10a}.  If one uses the rigid-rotor
approximation, the purely rotational frame transformation is analytical for
linear, spherical, or symmetric top molecules and, therefore, is easy to
implement~\cite{jt439,faure09}.  The rotational frame transformation approach
has been demonstrated to work very well when compared to full close-coupling
treatments~\cite{jt439}.  For molecules with permanent dipole moments,
however, rotational excitation requires special treatment because the
long-range interaction of the electron with the target dipole moment means
that a large number of partial waves should be taken into account.  Special
hybrid treatments are used to provide the contribution of the higher partial
waves~\cite{np82,polydcs}.

Along with rotational and vibrational excitation, dissociative electron
attachment and dissociative recombination (DR) are the dominant low energy
processes.  The cross sections for these dissociative processes are very
sensitive to the locations of curve crossings between the dissociative
resonance state(s) and the target curve; the resulting cross sections are
known to be highly sensitive to this aspect of the calculation
\cite{omalley66,bates1991relative,guberman1991generation,mn84,jt260}.  


There is a hierarchy of non-perturbative low-energy electron-molecule
collision models. The simplest one currently in use is the static exchange (SE)
model, which considers electron collisions with a target represented
by a Hartree-Fock wavefunction. In the SE model the electron is allowed
to occupy empty (``virtual'') target orbitals but the target itself
remains frozen. The SE model is well-defined, which makes it useful
for cross-comparison of codes but limited in the amount of physics included.
For example SE calculations can give low-lying shape resonances but usually
they are too high in energy and too broad; Feshbach resonances, which
involve simultaneous target excitation and trapping of the scattering
electron, cannot be represented in this model.
Inclusion of polarization effects using the
static exchange plus polarization (SEP) model is often found to give reliable
parameters for low-lying shape resonances; converging SEP calculations
usually
requires the inclusion of many more virtual orbitals than are required
to converge the simple SE model for the same system \cite{jt533}.
 Conversely Feshbach resonances,
which dominate the DR process, are best represented by models that contain
their parent state as part of a close-coupling expansion.

For collisions with a molecular ion having a closed electronic shell, the
energy surface of the neutral dissociative potential usually crosses the ionic
surface far from the geometry of the equilibrium of the target ion.  In this
case, the actual geometry at which the ionic and dissociative potential
surfaces cross is irrelevant, because the DR cross section is determined by the
probability of electron capture into a state different than the dissociative
state.  During such a process the target ionic core is excited rovibrationally
and the electron is captured into a weakly-bound Rydberg state.  This is
the so-called indirect DR mechanism
\cite{bardsley1967ionization,bates1991relative}, which is dominant for many
closed-shell molecular ions
\cite{kokoouline11a,douguet12a,douguet12b,samantha14}.  The accuracy of the
theoretical DR cross section in the indirect process, via intermediate
molecular Rydberg states and rovibronic resonances, is mainly determined by
the accuracy of representing the non-BO coupling responsible for the incident
electron capture.  Expressed in terms of the electron-molecule scattering
matrix $S_{i'v';iv}$ the DR cross section is $\sigma\sim\vert
S_{i'v';iv}\vert^2/E$, where indices $v'$ and $v$ represent final and initial
vibrational states of the ion during the capturing process, and $i'$ and $i$
describe electronic states.  The matrix element $S_{i'v';iv}$ is obtained by
integrating the geometry-fixed scattering matrix over vibrational states $v'$
and $v$.  For small molecules, the numerical accuracy of vibrational wave
functions is usually relatively good, and the uncertainty of the final cross
section is mainly determined by the quality of the geometry-fixed scattering
matrix.  For larger polyatomic ions, the inaccuracy of wave~functions, which
are usually calculated using the normal-mode approximation
\cite{kokoouline11a,douguet12a,douguet12b,samantha14}, may contribute
significantly to the uncertainty of the final DR cross section.  Therefore,
assuming that the accuracy of the vibrational wave~functions is good, the uncertainty
of the final DR cross section for the indirect mechanism (for most
closed-shell molecular ions) is $\Delta\sigma/\sigma\sim2\vert\Delta
S\vert/\vert S\vert$, where $S$ and $\Delta S$ are the geometry-fixed
scattering matrix and its uncertainty.  The scattering matrix for DR
calculations can be computed using electron scattering codes, such as
R-matrix, complex Kohn, or variational Schwinger methods.  Recent examples,
include Fonseca dos Santos \etal~\cite{samantha14} who obtained their
geometry-fixed scattering matrix using the complex Kohn calculations, and
Little \etal~\cite{jt591} who performed similar calculations based on
\hbox{R-matrix} computations.  Comparisons have shown that these two methodologies
yield very similar results for a given scattering model~\cite{jt288}.  A second
method, used extensively in earlier studies
\cite{kokoouline03a,kokoouline03b,kokoouline11a,douguet12a,douguet12b}, is
based on quantum defects extracted from energies of Rydberg states of the
corresponding neutral molecule.  The \hbox{Rydberg}-state energies are usually
obtained {\it ab initio}, but experimental energies have also been used
\cite{jungen08a,jungen08b,jungen09}.

At collision energies near threshold electronic excitation provides an
important new channel.  For this situation CC{\it n}-type models are usually
employed.  Such calculations face all the difficulties described above for
atoms plus complications introduced by loss of symmetry and nuclear motion
encountered in molecules.

The intermediate energy region is beginning to be explored with fully {\it ab
initio} methods but only for rather simple systems~\cite{jt354,jt434,12PiAbLu,12CoPi}.  
Conversely, extensive studies for the
high energy region were performed using various perturbative approximations such as
the Born or DWBA approximations~\cite{Gao95,Gao06,Kaiser07}.

\section{Uncertainty Assessment for Charge Transfer Collisions}~\label{sec:hcoll}
When the incident electron in a collision with an atomic or a molecular target
is replaced by a positively charged ion a new channel appears: electron
transfer.  Since this channel is the most important one for plasma and related
applications we will concentrate on such charge transfer collisions in this
section.

It goes without saying that an accurate solution of the full Schr\"odinger
equation is not feasible, except maybe for the simplest charge transfer
collision systems involving just two nuclei and one electron.  Accordingly,
and similarly to what has been discussed for electron scattering in
section~\ref{sec:UQcoll}, different approximation methods have been developed,
which are deemed suitable in different energy ranges.

The situations of interest for charge transfer collisions are:
\begin{itemize}\setlength\itemsep{0em}
\item
very low energy collisions, in which the de Broglie wavelength associated with
the projectile motion is comparable with the length scale that is
characteristic for electronic processes;
\item
low energy collisions, in which the projectile de Broglie wavelength is too
small to resolve electronic processes, but the projectile-target interaction
time is still long compared to the characteristic electronic time scale;
\item
intermediate energy collisions, in which the relative projectile-target
speed is comparable with the orbital speeds of the active electrons;
\item
nonrelativistic high energy collisions, in which the previous condition is no
longer fulfilled;
\item relativistic energy collisions.
\end{itemize}
Note that we are using a similar nomenclature as in section~\ref{sec:UQcoll}, although the
actual magnitudes of the collision energies are very different for electron
vs.\ heavy particle projectiles.

An authoritative overview of the entire spectrum of theoretical
charge transfer methods available by the early 1990s was given by Bransden and
McDowell~\cite{Bransden92}.  For more recent, but somewhat more specialized
accounts we refer the reader to~\cite{Belkic08a,Loreau14} and references
therein.  The following paragraphs are meant to provide a (necessarily
incomplete) mini-survey of what is discussed in those works and what else is
of relevance in the context of this article.

The gold standard for the calculation of charge transfer cross sections from
very low up to intermediate projectile energies has long been one or another
variant of the CC expansion.  Accordingly, limitations in basis-set convergence
are the main source of numerical uncertainties.  Most of these CC calculations
are also afflicted by model uncertainties, because it is normally not the full
Schr\"odinger equation that is cast into matrix-vector form.

One gets closest to the ideal of a calculation free of model uncertainty in
the very low energy regime in which a fully quantum mechanical description of
the scattering system is required.  In this region, electron transfer
usually dominates the dynamics and can be understood by considering the real
and avoided crossings of a small number of potential energy curves of the
quasimolecular system of projectile and target.  Accordingly, an expansion in
terms of products of molecular electronic states and nuclear wave~functions is
the standard method of attack.  In its original form this so-called perturbed
stationary state (PSS) approach has inherent defects, because individual terms
in the expansion do not satisfy the boundary conditions of the scattering
problem, thereby introducing spurious origin-dependent couplings in a finite
matrix representation of the Schr\"odinger equation~\cite{Delos81}.  These
defects can be remedied by including electron translation factors (ETFs) or by
using reaction coordinate techniques~\cite{Delos81,Errea94}.  An alternative
method is the hyperspherical close coupling (HSCC) approach, in which a
rescaled Schr\"odinger equation written in terms of hyperspherical coordinates
is solved (see ref.~\cite{Liu05} and references therein).

In modern applications to few-electron systems the molecular states and
couplings are calculated with sophisticated quantum chemistry methods, which
implies that electron correlations are taken into account and the general
approach can be called {\it ab initio}~\cite{Zygelman97}.  It has become
customary, albeit somewhat inaccurate, to refer to these modern versions of
the PSS approach as quantum mechanical molecular-orbital close-coupling
(QMOCC) calculations~\cite{Li15}, and we follow this convention.

Moving up in collision energy to, say, 1 keV/amu and higher, fully
quantum mechanical methods become challenging because they normally involve
partial-wave or other expansions of the scattering amplitude that become very
large in the low energy region.  Very recently, the three-body problem of
proton-hydrogen scattering has been addressed in a fully quantum mechanical
CCC approach that solves the Lippmann-Schwinger integral equations for the
scattering amplitudes~\cite{Abdura15}.  The more traditional approach is to
make use of the smallness of the projectile de Broglie wavelength by adopting
a semi\-classical approximation.  As long as one is interested in total (i.e.~
integrated over projectile scattering angle) cross sections only, the semiclassical
approximation amounts to reducing the full scattering problem to a
time-dependent Schr\"odinger equation (TDSE) for the electronic motion in the
field of classically moving nuclei.  The classical trajectories can be
determined by considering the nonadiabatic coupling of the electronic and the
nuclear motion as is done in the electron nuclear dynamics (END)
method~\cite{Deumens94,Cabrera10}, or by using Coulomb or model scattering
potentials~\cite{Green82a,Green82b,Fritsch84,Hemert85}.  At collision energies
of a few keV/amu and higher, simple straight-line trajectories are just as
good, i.e.~the numerical error introduced by replacing a curved trajectory by
a rectilinear one is negligibly small compared to errors associated with
basis-set convergence issues or other numerical uncertainties.  The same can
be said about the semiclassical approximation itself: at least for total cross
section calculations it is essentially exact in and above the low energy
regime.

In the low energy regime electron transfer still is the strongest electronic
process and molecular state expansions (including ETFs) still are the most
widely used methods~\cite{Fritsch91}.  Within the semiclassical framework they
are often referred to as MOCC methods (without the 'Q').

Once direct target ionization, i.e.~transitions into the continuum, become
important other CC techniques or fully numerical methods for the solution of
the semiclassical TDSE gain importance.  A common feature of the former is
that, similar to what has been discussed for electron scattering, positive
energy pseudo-states are included to discretize the continuum.  Even if one is
not interested in direct target ionization one cannot simply close the
ionization channel in a calculation without running the risk of degrading the
results for target excitation and electron transfer.  It is characteristic of
the intermediate energy regime that all channels are coupled and have to be
taken into account simultaneously.  Examples of suitable intermediate energy
CC methods are the two-center atomic-orbital, AOCC, method~\cite{Fritsch91}
and the two-center basis generator method (TC-BGM)~\cite{tcbgm}, both of which
include bound (atomic) target and bound (atomic) projectile states, endowed
with ETFs, and sets of pseudo-states whose explicit forms vary.

Notwithstanding considerable success in applications to charge transfer
collisions these methods can be criticized for being built on formally
overcomplete basis sets.  Indeed, there are known cases in which too large
basis sets on both centers (perhaps combined with insufficient numerical
accuracy in the calculation of matrix elements) have led to spurious couplings
and unphysical results~\cite{Kuang97}, meaning that the bigger (the basis),
the better (the convergence) is not necessarily true for these two-center
methods.  The insight that completeness of a basis is not necessary, in
principle, for following the evolution of the time-dependent state vector
exactly~\cite{BGM99} does not help in practice, since there is no other
practical criterion available than checking for changes in the results when
more basis states are added.  One-center expansions are not afflicted by the
overcompleteness problem, but are in practice inferior to two-center methods
when it comes to separating electron transfer from ionization to the
continuum.

As indicated above, direct numerical approaches to the solution of the
semiclassical TDSE offer an interesting alternative to CC expansions.  The
basic idea is straightforward: represent the electron wave~function on a grid
(usually in coordinate space) and propagate it in time by application of the
time-evolution operator over a large number of small time steps.  This can be
done in different ways, e.g., by using the split-operator Fast Fourier
transform method.  Whichever technique is used, most time-dependent lattice
(TDL) methods share the following features: (i) the Coulomb potentials of the
nuclei are replaced by soft-core potentials; (ii) absorbers are introduced to
avoid unphysical reflections of the wave~function at the boundaries of the
numerical box; (iii) numerical accuracy mostly depends on the spatial grid
parameters (provided a sufficiently small time step size is used for the
propagation).  One attractive feature of TDL approaches is that they provide a
view on the electron density distribution in the continuum, i.e.~insight into
electron emission characteristics, but they have also been applied
successfully to charge transfer problems~\cite{Minami06,Pindzola15}.

As in the case of electron scattering, perturbative methods and distorted-wave
approaches are the principal methods of choice in the nonrelativistic 
high energy regime.  They can be formulated on the level of the semiclassical
approximation or for the full quantum mechanical problem, and at least for
some of the methods put forward over the years both options can be shown to be
(essentially) equivalent~\cite{Bransden92}.  Numerical uncertainties are
usually well controlled, at least in first-order models, but model
uncertainties can only be estimated by extensive comparisons with other
(preferably nonperturbative) calculations and with experimental data.

Most of the approaches discussed in this section have been generalized to deal
with collisions at relativistic energy.  The principal motivation for studying this
regime is the fundamental interest in relativistic dynamics and phenomena such
as radiative charge transfer and electron-positron pair production.  While the
former process can also be of importance at very low collision energies
\cite{Bransden92,Li15}, the latter is, of course, a truly relativistic effect.
Since relativistic collisions are less relevant for applications, we will not
discuss them further in this article.

We end this brief survey of charge transfer methods with a few general
comments.  First, the majority of methods touched upon in this section deal
with true or effective one-electron problems.  The two-electron problem has
been addressed in a number of perturbative models~\cite{Belkic08b}, and also
in the framework of the semiclassical CC approach~\cite{Fritsch91}.  As
mentioned above, modern (Q)MOCC methods can deal with many-electron systems in
an {\it ab initio} fashion, but they have mostly been applied to one-electron
transitions, i.e.\ single electron transfer~\cite{Li15,Zygelman97}.  For truly
many-electron problems, such as multiple electron transfer to a highly charged
ion, simplifications are unavoidable, which implies that further modeling,
usually on the level of the semiclassical TDSE, is necessary.  An obvious idea
is to replace the many-electron Hamiltonian by a sum of effective one-electron
Hamiltonians, i.e.~to solve the problem on the level of the independent
electron model (IEM).  This has worked quite well in several instances, but it
is not obvious how to carry out a reliable uncertainty assessment of an IEM
calculation.  One way to go about this is to consider several variants of IEM
calculations, e.g., by varying the effective potentials used within reasonable
bounds, and monitor the spread of results obtained.  It will be illustrated in
section~\ref{subsec:hcoll-results} that this is a useful procedure, although it
can give at most qualitative information on the uncertainty of a given IEM
calculation.

Second, echoing a comment made in section~\ref{sec:UQcoll} for electron
scattering, we mention that simpler, sometimes semi-empirical and/or classical
methods for calculating charge transfer cross sections have been widely used
over many years.  Among them are two-state quantum mechanical models (see
ref.~\cite{Bransden92}), variants of the classical over-barrier model
\cite{Niehaus86}, and the classical trajectory Monte Carlo method
\cite{Olson77}.  The latter in particular has been highly successful in many
applications, even in the low energy regime~\cite{Otranto14} in which quantum
effects are deemed important.  An implication of this somewhat surprising
observation is that uncertainty estimates have to rely on extensive
comparisons with more rigorous quantum mechanical methods and with
experimental data.

Third, in many cases the observables of interest are not electron transfer
cross sections, but cross sections associated with post-collisional events
such as radiative de-excitation of excited projectile states or the
fragmentation of the target in ion-molecule collisions.  This requires further
modeling, and hence it introduces further uncertainties and also the problem of
uncertainty propagation.  Again, it seems that the only known and practical
way of dealing with these issues is to perform computations for a range of
models and monitor the spread of results.  An example for this will be given
in section~\ref{subsec:hcoll-results}.

\section{Illustrations}\label{sec:results}
\subsection{Structure}

The best examples of uncertainty estimates in high-precision theory are
provided by few-electron atoms.  Hydrogen is a special case because there the
Schr\"odinger (or Dirac) equation can be solved exactly, and so the
uncertainty comes entirely from higher-order QED terms or nuclear structure
effects not included in the calculation (see section~\ref{subsec:astruct}).  For
the two- and three-electron cases of helium-like and lithium-like atoms
complete calculations in Hylleraas coordinates 
have been performed.  Recent high-precision measurements~\cite{Sanchez_2006,Sanchez_2009} 
and theory~\cite{YanDrake_2008,YanDrake_2008E,Puchalski_2010} 
for the $1s^22s\;^2S - 1s^23s\;^2S$ 
two-photon transition in lithium provide an excellent example of
what can be achieved.  Table \ref{tab:Lithium} lists the various contributions
to the transition frequency, expressed as a double power series in powers of
$\mu/M$ and $\alpha$, where $\mu/M$ is the ratio of the reduced electron mass
to the nuclear mass, and $\alpha$ is the fine structure constant (see table
caption for numerical values).  The sum of the first three entries gives the
total nonrelativistic transition energy, including first- and second-order
finite nuclear mass corrections of order $\mu/M$ and $(\mu/M)^2$.  These
account for the nonrelativistic part of the isotope shift, but at this level
of accuracy, the usual normal and specific isotope shifts of order $\mu/M$ are
not sufficient, and therefore the second-order $(\mu/M)^2$ term must also be
included.  The associated uncertainties shown in Table \ref{tab:Lithium} were
reliably estimated from the rate of convergence of the calculation with the
size of the Hylleraas basis set.  Next comes the leading relativistic
correction of order $\alpha^2$ relative to the nonrelativistic energy from the
Breit interaction, and the relativistic recoil term of order $\alpha^2\mu/M$.
The uncertainties from all these terms can be accurately estimated from the
rate of convergence with the size of the Hylleraas basis set.  At the next
level are the QED corrections of order $\alpha^3$, corresponding to the Lamb
shift in hydrogen.  The theory for these terms is complete in terms of known
expectation values of operators, including both the electron-nucleus and
electron-electron QED contributions (the Araki-Sucher terms).  The dominant
source of uncertainty are the Bethe logarithms for the states of lithium
\cite{YanDrake_QED}.  The terms of order $\alpha^3\mu/M$ are radiative recoil
corrections due to the finite nuclear mass.  These can be calculated to more
than sufficient accuracy in terms of known operators~\cite{YanDrake_QED}.  The
dominant source of uncertainty in the theoretical transition frequency comes
from the higher-order QED corrections of order $\alpha^4$ and $\alpha^5$,
since the basic theory for these terms has not yet been developed.  However,
estimates can be obtained from the corresponding QED shifts in hydrogen with
appropriate scaling with the nuclear charge and electron screening, as shown
in the table, with 10\% and 25\% uncertainties assigned respectively for these
two terms.  Finally, the correction due to the finite nuclear charge radius is
included.

\begin{table}
\caption{Theoretical contributions to the $1s^22s\;^2S - 1s^23s\;^2S$
  transition energy (cm$^{-1}$) of $^7$Li~\cite{YanDrake_2008,YanDrake_2008E,Puchalski_2010}, 
  and comparison with
  experiment~\cite{Sanchez_2006}.  The entries on the left indicate the powers
  of $\mu/M$ and $\alpha$ that give rise to each contribution relative to the
  nonrelativistic energy for infinite nuclear mass, where $\mu/M =
  7.820\,202\,988(6)\times 10^{-5} $ is the ratio of the reduced electron mass
  to the nuclear mass for an atomic mass of 7.016\,003\,4256(45) u, and
  $\alpha = 1/137.035\,999\,139(31)$ is the fine structure constant.  The
  contributions of order $\alpha^4$ and $\alpha^5$ are estimates.  Nucl.\ size
  is the finite nuclear-size correction for an assumed nuclear charge radius
  of 2.390(30) fm.}
\label{tab:Lithium} 
\begin{tabular}{l r@{}l }
\hline
\hline
Contribution      & \multicolumn{2}{c}{Transition Energy (cm$^{-1}$)}\\
\hline
Infinite mass     &\qquad27\,206&.492\,847\,9(5)       \\
$\mu/M$           &    --2&.295\,854\,362(2)     \\
$(\mu/M)^2$       &      0&.000\,165\,9774       \\
$\alpha^2$        &      2&.089\,120(23)         \\
$\alpha^2\mu/M$   &    --0&.000\,003\,457(9)     \\
$\alpha^3$        &    --0&.187\,03(26)          \\
$\alpha^3\mu/M$   &      0&.000\,009\,74(13)     \\
$\alpha^4$ (Est.) &    --0&.005\,7(6)            \\
$\alpha^5$ (Est.) &      0&.000\,52(13)          \\
Nucl.\ size       &    --0&.000\,390(10)         \\
Total             &27\,206&.093\,7(6)            \\
Expt.~\cite{Sanchez_2006}&27\,206&.094\,082(6)  \\
\hline
\hline
\end{tabular}\\
\end{table}

The final theoretical value 27\,206.093\,7(6)cm$^{-1}$ is in good agreement
with the substantially more accurate (2~parts in $10^{10}$) measurement
27\,206.094\,082(6) cm$^{-1}$ (see Table~\ref{tab:Lithium}.)  However, the
important lesson to be learned from the comparison is that, since the
theoretical uncertainties in the lower order terms are well controlled, the
comparison between theory and experiment provides an experimental value for
the higher order QED terms of order $\alpha^4$ and $\alpha^5$ that cannot yet
be calculated directly.  This same principle has been applied with great
effectiveness to determine the nuclear charge radius for a range of halo
nuclei from ${}^6$He to ${}^{11}$Be from the measured isotope shifts
\cite{Drake_LiShift,Drake_RMP}.  Here, the otherwise dominant uncertainties
from the mass-independent QED terms in Table \ref{tab:Lithium} cancel when
taking the difference between isotopes with different $\mu/M$, and so the
residual difference between experiment and theory with much smaller
uncertainties provides an accurate determination of the relative nuclear
charge radii.

The pair of articles by Safronova \etal~\cite{Safronova,Safronova2}
describes high accuracy computations with uncertainty estimates for energies
of low-lying excited states in Ag-like, Cd-like, In-like and Sn-like ions.
The electronic structure for all these states is characterized by a
[Kr]$4d^{10}$ core and 1, 2, 3, or 4 valence electrons, respectively.  A
$5s^2$ component in the structure of an In-like or Sn-like ion can optionally
be viewed as part of the core, leaving one or two valence electrons.  Three
models are used in refs.~\cite{Safronova,Safronova2}: the linearized
coupled-cluster method including all single, double, as well as partial triple
excitations (All-order SDpT), the configuration-interaction plus all-order
model (CI+All-order) and the configuration-interaction plus many-body
perturbation theory model (CI-MBPT).  All three models start from a frozen-core DF
potential.  The All-order SDpT model is the most computationally intensive, and
it is used only for the monovalent systems, i.e.~the Ag-like ions and the
In-like ions with the $5s^2$ electrons included in the core.  The CI+All-order and CI-MBPT
models are used for monovalent and multivalent systems.  Within each model the
principal convergence issue arises from the truncation in the partial wave
expansion, particularly for the $4f$ shell.  Second order perturbation theory
is used to evaluate the contribution of partial waves with $l>6$.  It is found
that this contribution is approximately equal to the contribution from the
$l=6$ term, which is then used as an approximation.  The Breit term and the
QED corrections are evaluated separately.  Finally, a 25\% uncertainty is
assigned to each of the four corrections (i.e.~\ higher order correlations,
higher partial wave contributions, Breit interactions, and QED corrections),
and the results are combined via sum of squares to obtain the total
uncertainty.
 
\begin{table}
\caption{\label{tab:Atomic} Calculations and their uncertainties of the
  excited state energies of Ce$^{9+}$ and Ba$^{7+}$~\cite{Safronova}.  All
  values are in cm$^{-1}$.}

\begin{tabular}{lrrrr}
\hline \hline
Ion                      & Ce$^{9+}$&          &Ba$^{7+}$ & \\
Level                    &5p$_{3/2}$&4f$_{5/2}$&5p$_{3/2}$&4f$_{5/2}$\\
\hline
Experiment               & 33427&  54947& 23592& 137385\\
All-order SDpT           & 33406&  55419& 23564& 137770\\
Diff (Exp -- SDpT)       &    21&  --472&    28&  --385\\
CI + all                 & 33450&  54683& 23605& 137256\\
Diff (Exp -- CI+all)     &  --23&    264&  --13&    129\\
CI+MBPT                  & 33986&  54601& 24020& 137086\\
High-order correlations  & --147 &  2687& --134&   2224\\
Higher partial waves     &    14& --1011&    12&  --858\\
Breit interaction        & --403& --1595& --293& --1197\\
Uncertainties            &   130&    220&      & \\
\hline
\end{tabular}
\end{table}

The above considerations yield an uncertainty estimate based strictly on
theory.  It was applied by Safronova \etal~\cite{Safronova} to the
In-like Ce$^{9+}$, Pr$^{10+}$, and Nd$^{11+}$ ``monovalent'' ions where the
CI+All-order and the CI-MBPT approach are both applicable.  However, Safronova
\etal\ also rely on comparisons with reference ions for which
experimental data are available.  In ref.~\cite{Safronova} the isoelectronic
Ba$^{7+}$ ion was used as a reference ion.  Results from the monovalent
All-order (coupled cluster) SDpT model and the CI+All-order model with three
valence electrons were compared, and it was found that CI+All-order gave the
better agreement with measurement.  Table~\ref{tab:Atomic} (extracted from
Table~V of~\cite{Safronova}) shows the final computed energies for two levels in
In-like Ce$^{9+}$ and Ba$^{7+}$ together with estimated contributions from
high order correlations, higher partial waves, and Breit interaction, and
along with uncertainty estimates for the states in Ce$^{9+}$.  QED corrections
were considered to be negligible for this system.  The uncertainty estimate
for the calculated Ce$^{9+}$ energies was obtained as the sum of two
contributions: the error (relative to experiment) in the calculation for the
Ba$^{7+}$ reference ion and the absolute change between the reference ion and
the actual ion in the sum of the identified small terms.  Finally, these
calculated uncertainties were compared with the actual deviation from the
experimental energies (which are available for Ce$^{9+}$), and it was found
that the estimate of uncertainties of calculated energies is reasonable for
$4f$ states but significantly larger for $5p$ states.

UQ is being routinely used to determine structural parameters of small molecules within
tight uncertainty bounds. This is done, at least in part, to aid the predictions 
of the rotational spectrum of these species and hence their detection 
in the laboratory and space~\cite{phg08,Demaison2010}.

Table~\ref{tab:Abinitio} illustrates an application of the FPA method to
uncertainty assessment of the calculated dissociation energy ($D_0$) of
H$_2$$^{16}$O.  Full details are given by Boyarkine \etal~\cite{jt549}
who provide similar results for water iso\-topologues.  Subsequent measurement of
$D_0$ for H$_2$$^{18}$O yielded results within the uncertainties of the
predicted value~\cite{15MaKoZoBo}.  The uncertainties listed in
Table~\ref{tab:Abinitio} include a contribution of 1 cm$^{-1}$ due to
nonadiabatic effects, even though these effects were assumed to give a
negligible direct contribution to the value of $D_0$ and actual calculation of
nonadiabatic effects did not form part of the study.  An estimate of the
magnitude of contributions due to effects neglected in a given model is a part
of the uncertainty assessment.

\begin{table}
\caption{\label{tab:Abinitio}\emph{Ab initio} contributions to the first
  dissociation energy of H$_2$$^{16}$O.  All values are in cm$^{-1}$.
  Uncertainties are given in the last column.  Signed contributions are
  incremental values.  Contributions (A) to (H) only concern the electronic
  motion with fixed nuclei.  CBS means complete basis set, FCI is full
  configuration-interaction, DBOC means the diagonal Born-Oppenheimer
  correction and ZPE is the vibrational zero-point energy.  Contributions (I)
  to (N) involve solving the nuclear motion problem for water and for the OH
  diatomic and are therefore nuclear mass dependent (MD).  Full details about
  the components considered in the focal point analysis can be found in
  Boyarkine \etal~\cite{jt549}.}

\begin{tabular}{clrc}
\hline \hline
   &                               &Value & Uncertainty \\
\hline
A & CBS CCSD(T) frozen core        & 43956 & 6    \\
B & Core correlation CCSD(T)       &   +81 & 2    \\
C & All-electron CBS CCSD(T) [=A+B]& 44037 & 6    \\
D & Higher order electron correlation       &   --52 & 3    \\
E & CBS FCI [=C+D]                  & 43985 & 7    \\
F & Scalar relativistic correction &   --53 & 3     \\
G & QED (Lamb shift) correction    &    +3 & 1     \\
H & Spin-orbit effect              & --69.4 & 1     \\
I & Angular momenta coupling, OH   & +31.5 & 0     \\
J & Sum spin effects, OH [=H+I]    & --37.9 & 1     \\
K & DBOC, H$_2$O                   & +35.3 & 0.5     \\
L & ZPE H$_2$O                     &4638.1 & 0     \\
M & ZPE OH                         &1850.7 & 0.5     \\
N & Net ZPE, H$_2$O [=L+M]         &2787.4 & 0.5     \\
U & Nonadiabatic contributions    &     0 & 1     \\
V & Total MD, H$_2$O [=I+K+N+U]    & --2721 & 1     \\
  & $D_0$(H$_2$O) Calc. [=E+V]     & 41145 & 8     \\
  & (Obs -- Calc) $D_0$(H$_2$O)    &    +1     \\
\hline
\end{tabular}
\end{table}

One area where uncertainty assessment is beginning to have significant impact
is the computation of dipole-moment surfaces~\cite{jt424,jt509} and hence
rotation-vibration transition intensities~\cite{jt613}.  The methodology used
here is based on adapting the FPA method for computing the dipole moment
surface (DMS) and then performing multiple computations using different PESs
and DMSs to establish stability of the results~\cite{jt522,jt625}.  These
computations are important as it is often difficult to measure absolute
transition intensities with the accuracy demanded for the interpretation of
modern remote sensing experiments.

\subsection{Electron - atom/ion collisions}
\begin{figure}[t]
\centering
\includegraphics[width=0.48\textwidth]{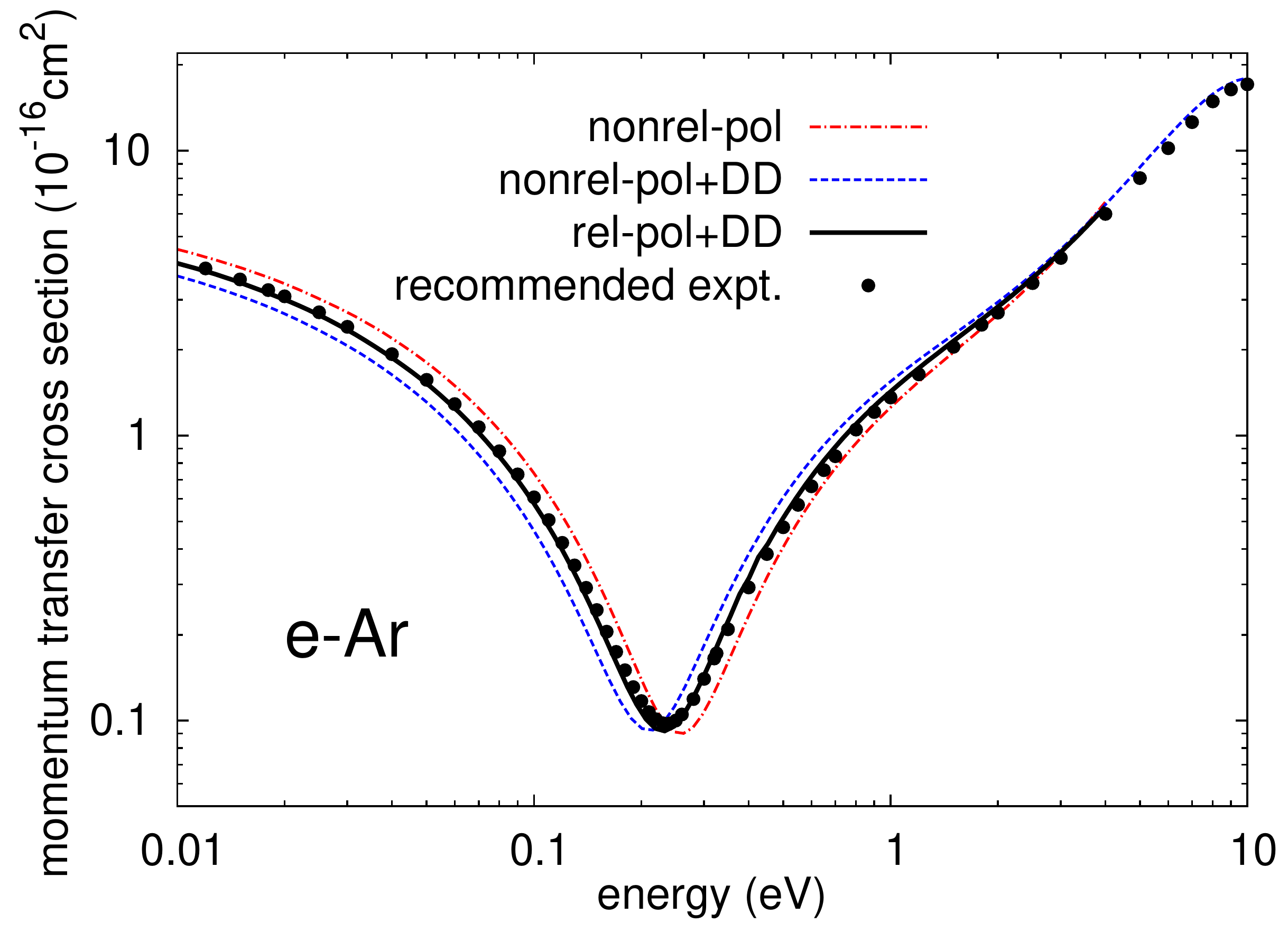}
\caption{Momentum transfer cross section for electron scattering from argon
  atoms in their ground state.  Results from various polarized orbital
  calculations (see text) are compared with a recommended set of experimental
  data~\cite{Landolt} (solid circles).  The curve labeled ``rel-pol+DD'' is a
  fully relativistic model including dynamic distortion.  (Figure taken
  from~\cite{0022-3727-46-33-334004}.)}
\label{fig:argon-momtrans}
\end{figure}

As the first example we consider the momentum transfer cross section for
low energy electron collisions with Ar atoms; an important parameter for many
laboratory plasmas.  Figure~\ref{fig:argon-momtrans} shows a comparison
between experiment and theory.  While the agreement with experiment is nearly
perfect for the presumably best model (fully relativistic including dynamic
distortion (DD) of the target charge distribution by the projectile), the
important issue for the present paper is the fact that i)~a number of
calculations were performed, and ii)~that even a nonrelativistic approach
with a less sophisticated way of accounting for the above effect yields rather
similar results.  Because of this, together with the general confidence in the
polarized-orbital method as an enhanced one-state close-coupling approach that
contains the most important physical effects, one can make reasonable
estimates about the position of the minimum (we suggest $\rm 0.15\,eV \pm
0.05\,eV$) and the value of the cross section away from the resonance (10\% or
better at 0.01$\,$eV and 1.0$\,$eV).

\begin{figure}[t]
\centering
\includegraphics[width=0.485\textwidth]{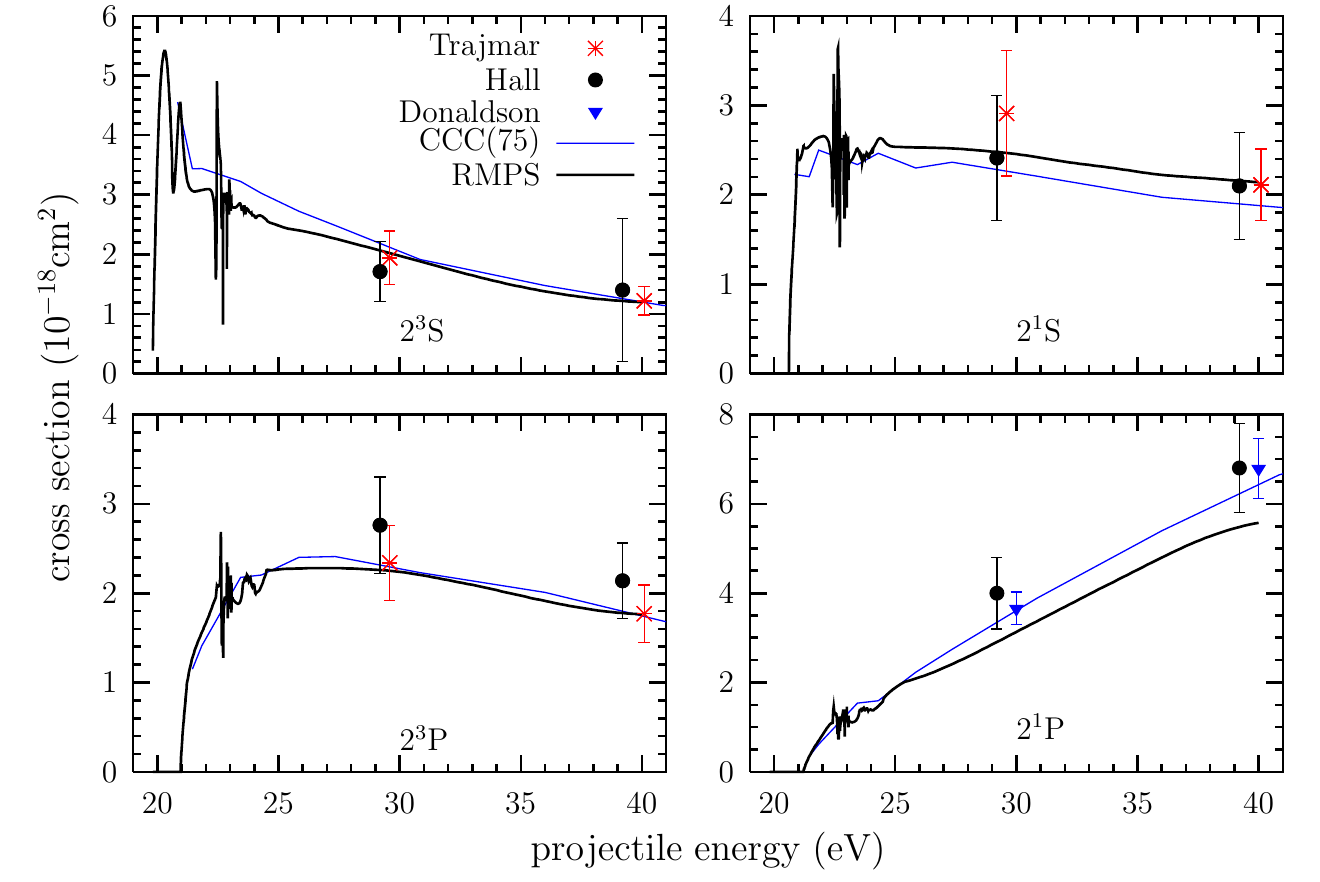}
\caption{Cross section for electron-impact excitation of the $n=2$ states in
  helium from the $\rm (1s^2)^1S$ ground state.  Three sets of experimental
  \hbox{data~\cite{PhysRevA.8.191,Hall1973,0022-3700-5-6-022}} are compared
  with predictions from nonrelativistic CCC~\cite{PhysRevA.52.1279} and
  RMPS~\cite{0953-4075-31-10-005} models.  (Figure adapted
  from~\cite{0953-4075-31-10-005}.)  }
\label{fig:hen2}
\end{figure}

Figure~\ref{fig:hen2} shows predictions from both CCC and RMPS for
electron impact excitation of the \hbox{$n=2$} states of helium.  Once again,
it seems possible to make a reasonable estimate of the uncertainty in the
theoretical predictions.  Even though these computations are nearly 20 years
old, they have indeed withstood the test of time.  This is ultimately not
surprising, since the scattering models contain what we believe is the
essential physics, namely an accurate target description (the relevant energy
levels and oscillator strengths agree with experiment and much more
sophisticated structure-only computations at the 10\% or better level) as well
as channel coupling within the discrete spectrum as well as to the ionization
continuum.  Significant differences occur only in the resonance regime near
the low-lying excitation thresholds, with the principal reason being that the
CCC calculations had only been performed at a few energies.  Looking at the
comparison, one might conclude that the average of the two sets of theoretical
predictions is accurate at least at the 20\% level (most likely better).  This
is something that cannot be said even today for most of the experimental data
points, of which there are only very few anyway.

\begin{figure}[t]
\centering
\includegraphics[width=0.485\textwidth]{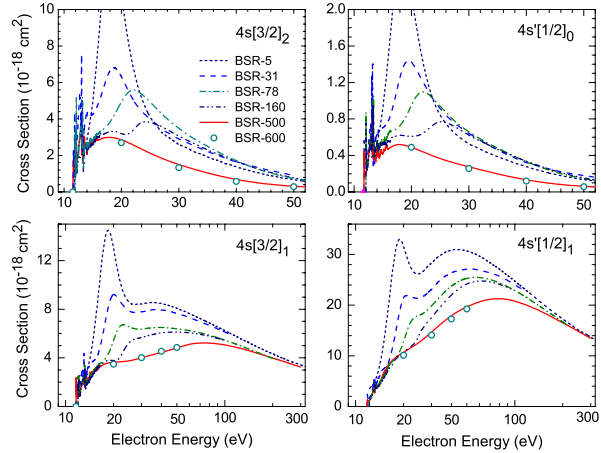}
\caption{\label{fig:Ar-excite} Angle-integrated cross sections for
  electron impact excitation of the $\rm 3p^54s$ states in argon from the
  ground state $\rm (3p^6)^1S_0$.  The results of a number of BSR calculations
  shows the convergence of the predictions from a close-coupling model.
  (Figure taken from~\cite{PhysRevA.89.022706}).}
\end{figure}

Figure~\ref{fig:Ar-excite} is an example of a systematic study regarding the
convergence of the close-coupling expansion~\cite{PhysRevA.89.022706}.  In this particular
implementation, the resulting equations are solved using the so-called
``B-Spline \hbox{R-matrix} with Pseudo-States'' (BSRMPS) approach.  Once again,
however, we emphasize that it is not the implementation of a particular model
that determines the overall uncertainty of its predictions.  If the
close-coupling expansion could literally be driven to an infinite number of
states on an infinitely fine spatial grid, then it should yield the correct
solution of the underlying many-particle Schr\"odinger or Dirac equation.  In
practice, of course, this is not possible.  In the above example, the
structure description for the states of interest, the initial $\rm
(3p^6)^1S_0$ state and the four final $\rm 3p^54s$ states of argon, was
carried out as well as the authors believed was necessary for most of the
uncertainty to come from the finite size of the close-coupling expansion.
This also means that purely numerical errors in solving the equation with a
fixed number of states are believed to be negligible.

Looking at the figure, one can see a very strong effect of adding more and
more states to the close-coupling expansion.  However, going from a 500-state
(BSR-500) to a 600-state (BSR-600) model ultimately indicates some
convergence.  As mentioned several times already, there is no guarantee for
the correctness of the final results.  Nevertheless, the results change in a
systematic way, and it seems as if at the very least the BSR-600 predictions
can be taken as a likely upper limit of the ``true'' solution of the
underlying equations, at least outside of the resonance regime.

\begin{figure}[b]
\includegraphics[width=0.35\textwidth]{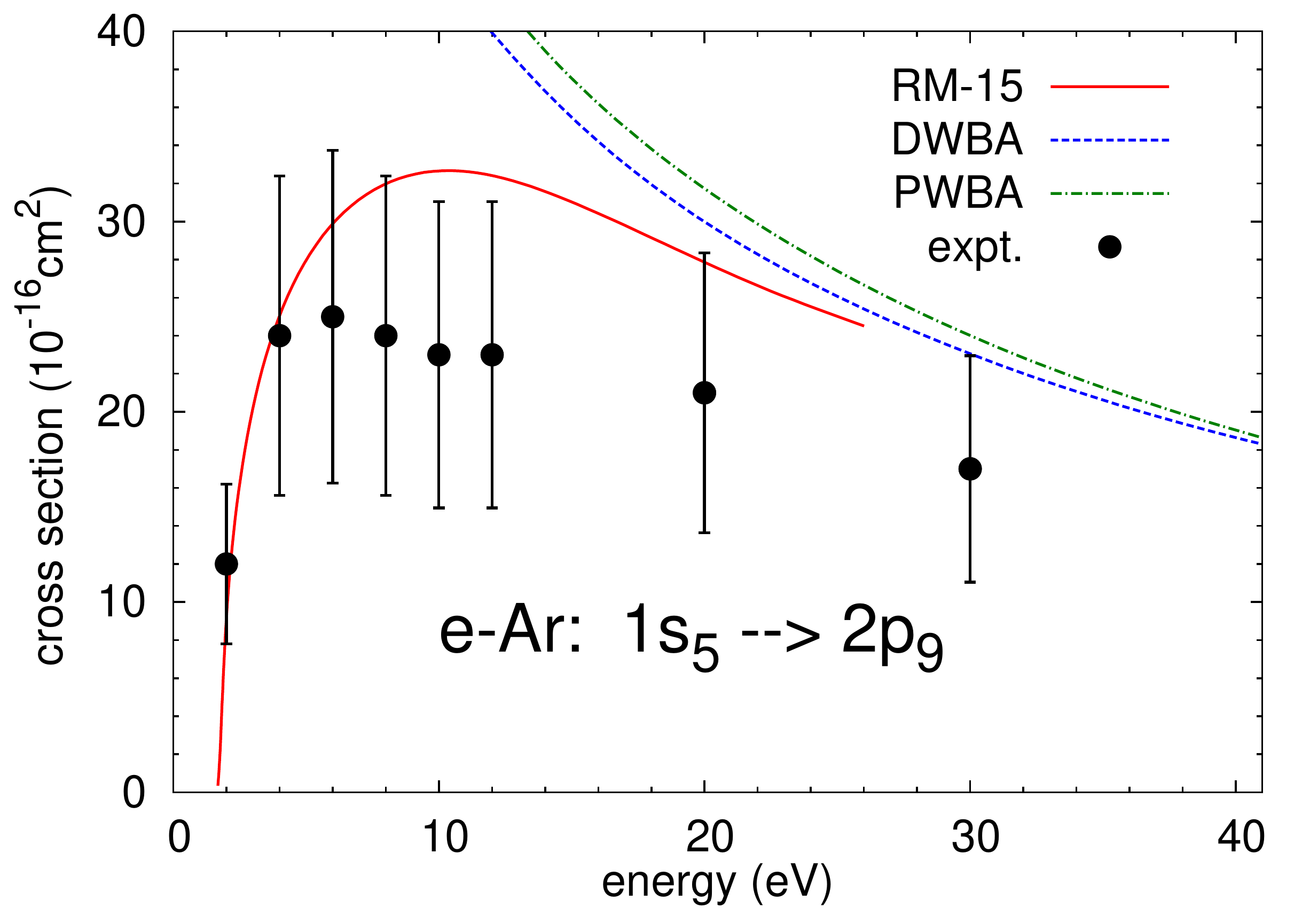}
\includegraphics[width=0.35\textwidth]{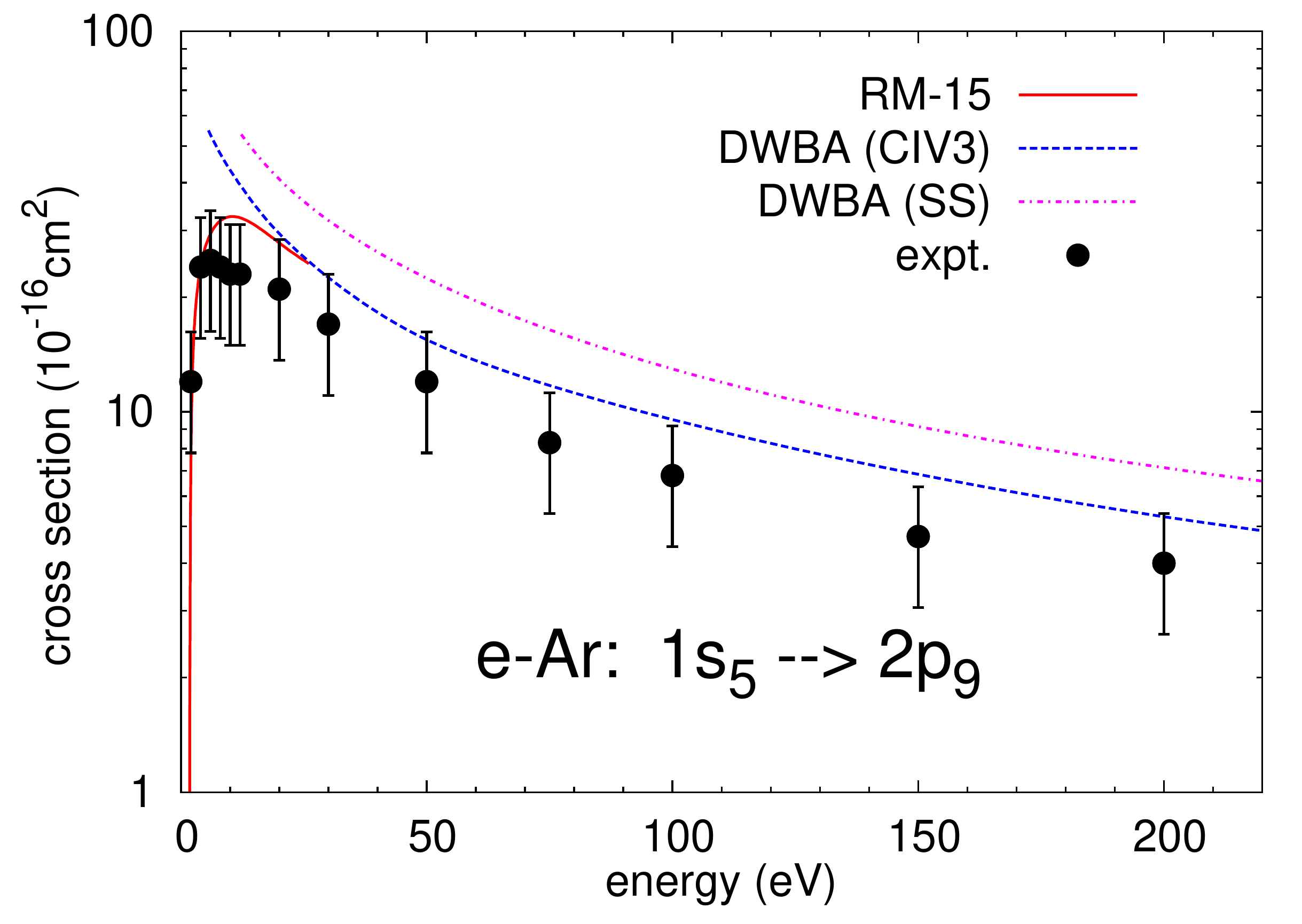}
\caption{
Cross section for electron impact excitation of the 
$\rm   (3p^54s)^3P_2 \to (3p^54p)^3D_3$ ($\rm 1s_5 \to 2p_9$) (in Paschen notation)   
transition in argon.  A number of theoretical predictions are compared with the 
experimental data of Boffard \etal~\cite{PhysRevA.59.2749}. 
(Figure taken from~\cite{0022-3727-46-33-334004}.)  
}
\label{fig:armeta}
\end{figure}

Moving on to the intermediate energy and high energy regimes,
figure~\ref{fig:armeta} shows the angle-integrated cross section for excitation
of the $\rm (3p^54p)^3D_3$ ($\rm 2p_9$) state in argon from the initial
metastable $\rm (3p^54s)^3P_2$ ($\rm 1s_5$) state~\cite{0022-3727-46-33-334004}.  This is a very strong
optically allowed transition with a threshold energy of less than 2~eV.  Hence
it is not the absolute projectile energy that matters in the classification of
the energy regime.  For all practical purposes, this is a high energy
collision, and hence one might assume that perturbative methods should be
appropriate.  Indeed, the top panel shows that results from PWBA, DWBA, and a
\hbox{15-state} $R$-matrix (close-coupling) model quickly converge towards
each other -- provided a very similar target description is being used.  In
fact, the principal reason for the deviation between the various sets of
results in this panel is the lack of unitarization of the DWBA scattering
matrix rather than a fundamental problem with a perturbative approach.  On the
other hand, we see a significant (about 30\% in this case) dependence of the
DWBA predictions when the relevant one-electron orbitals (4s and 4p) were
generated with different atomic structure codes (CIV3)~\cite{CIV3} or
{SUPERSTRUCTURE}~\cite{SS}, respectively) and slightly different optimization
criteria.  This is an instructive example where the reliability of a collision
calculation is effectively determined by the quality of the structure
description rather than the collision model itself.  While the results
obtained with the CIV3 orbitals appear to provide better agreement with
experiment in this particular case, this is by no means the rule.
Furthermore, the uncertainties associated with the absolute experimental
normalization are often substantial.  Clearly, the availability of a reliable
oscillator strength for this transition can be used to rescale the predictions~\cite{PhysRevA.64.032713}
and hence reduce the likely uncertainty of the predictions.

\begin{figure}[b]
\includegraphics[width=0.490\textwidth]{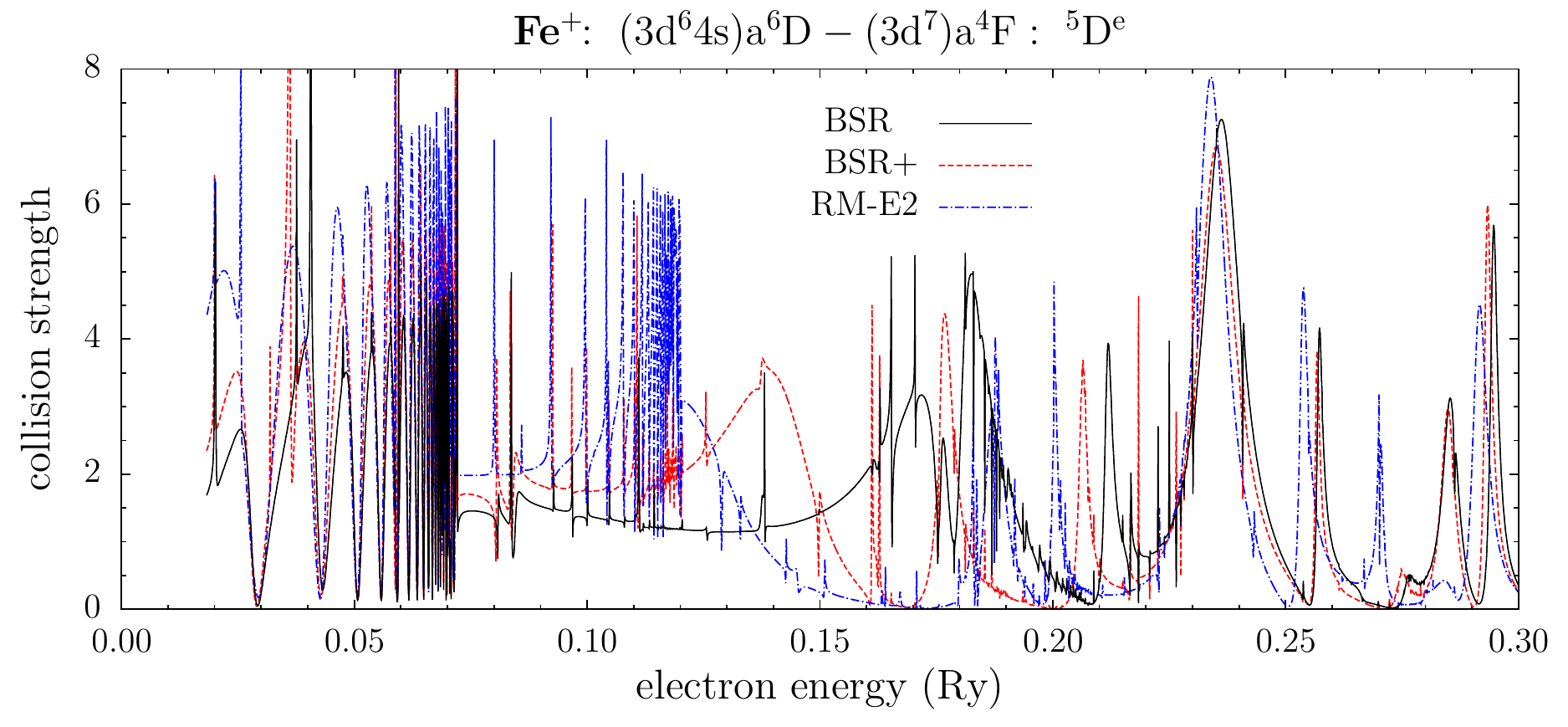}
\caption{
Predictions from two calculations~\cite{ISI:000231564200009,0953-4075-35-16-308} for the $^5D^e$ partial-wave collision strength
of the electron-induced transition $(3d^6 4s)^6D - (3d^7)^4F$ in Fe$^+$.
}
\label{fig:FeII-resonances}
\end{figure}

Next we present a few examples for electron-ion collisions.  The
first one is for electron scattering from Fe$^+$, which is a very complex
target.  Due to this complexity and the ionic character of the target, there
is a wealth of resonance structure as function of the incident projectile
energy.  An example is shown in figure~\ref{fig:FeII-resonances} for just one
partial-wave symmetry.

\begin{figure}[t]
\includegraphics[width=0.46\textwidth]{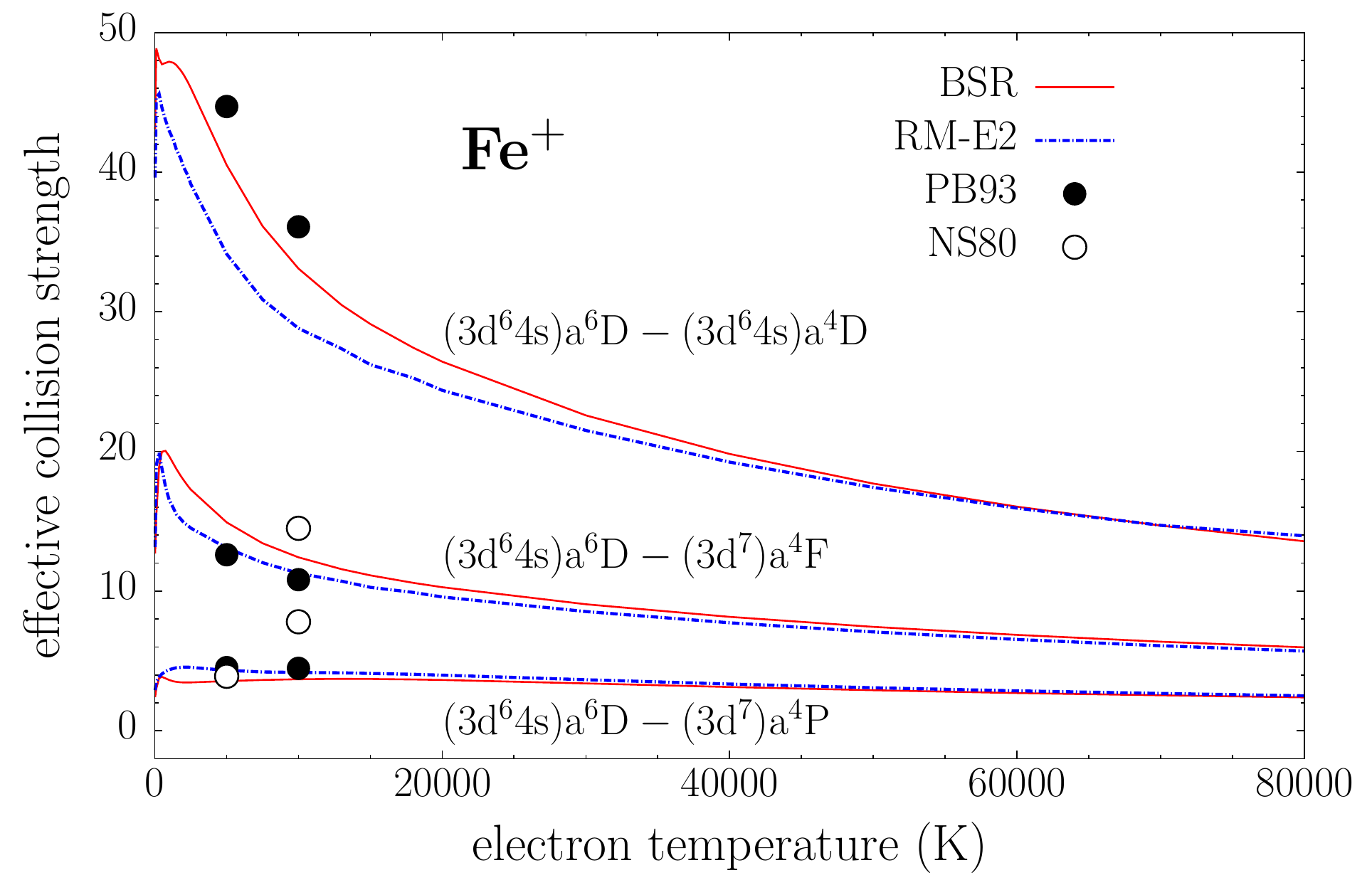}
\caption{Predictions from a number of calculations for effective collision strengths 
as a function of temperature for electron collisions with Fe$^+$.
BSR results~\cite{ISI:000231564200009} are compared with those of Nussbaumer and Storey (NS80)~\protect{\cite{NS80}},
Pradhan and Berrington (PB93)~\protect{\cite{PB93}}, and
Ramsbottom \etal~\protect{\cite{0953-4075-35-16-308}}.
}
\label{fig:FeII-effcol}
\end{figure}

Comparing results from individual calculations makes little sense in this
case.  Instead, one should concentrate on an observable that is more stable
regarding small changes in the individual predictions, but which is still
meaningful in modeling applications.  Such an observable is the effective
collision strength, where an integral over the incident energies weighted over
a Maxwellian (or possibly other) speed distribution is performed for a range
of temperatures.  An example is shown in figure~\ref{fig:FeII-effcol}.
Comparing the results from different models should give some indication about
the uncertainty of the predictions, especially if some additional criteria
regarding the likely quality of the target description and the collision model
are used to give increasing weight to a particular set of results.  In the
example shown, however, the latter may not even be necessary if an uncertainty
of about 20\% is deemed sufficient.

\begin{figure}[t]
\includegraphics[width=0.46\textwidth]{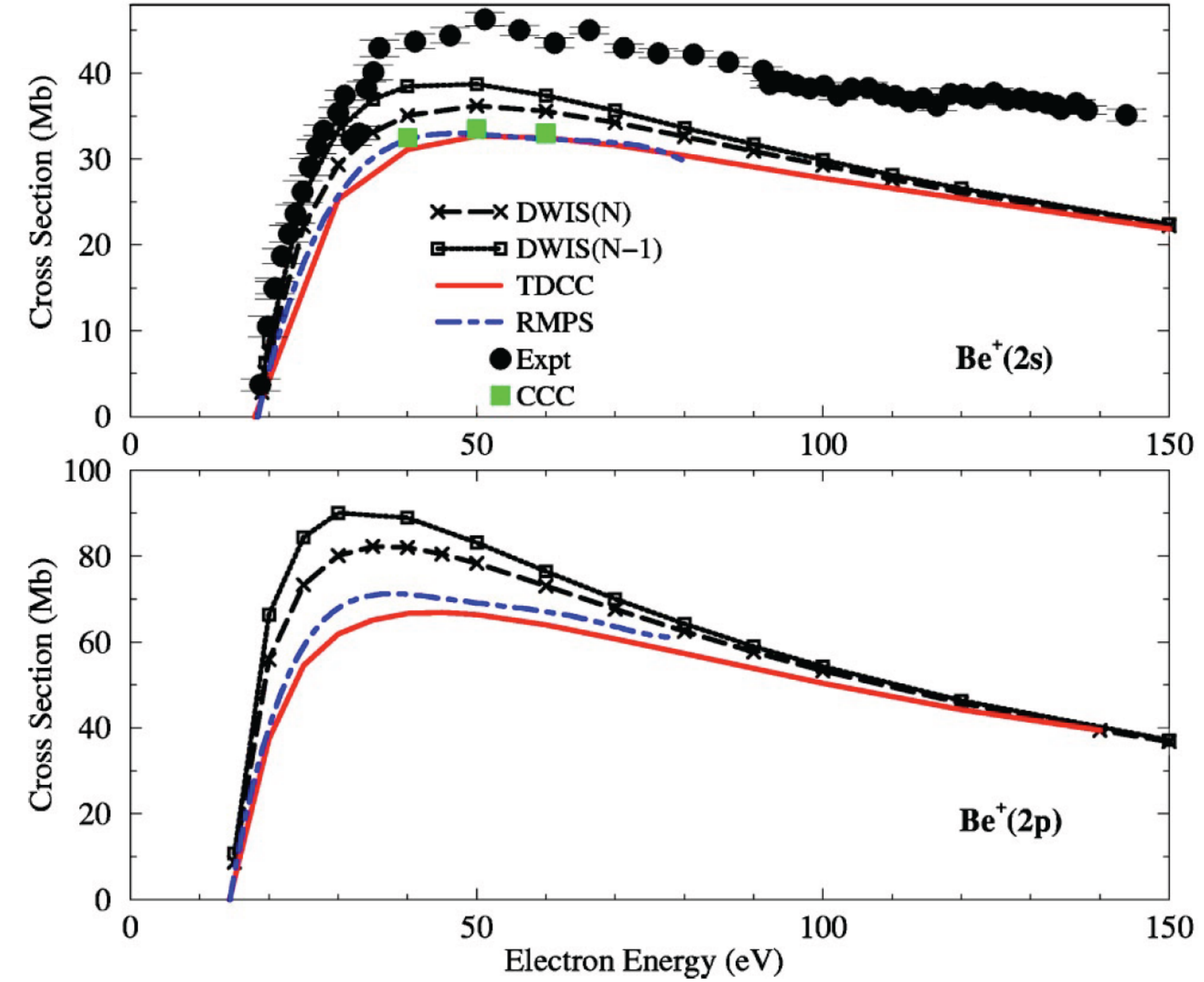}
\caption{Electron impact ionization cross sections for Be$^+$ from
the $(2s)$ ground state and the $(2p)$ excited states. The predictions
from several distorted-wave and close-coupling models are compared with the
experimental data of Falk and Dunn~\cite{PhysRevA.27.754}.
(Figure adapted from~\cite{PhysRevA.68.032712}.)
}
\label{fig:Be+ion}
\end{figure}

For highly charged ions, as electron correlation effects are less important,
perturbative methods such as PWBA, DWBA produce comparable results to
nonperturbative methods as long as the structure description is reliable. 
As seen from figure~\ref{fig:Be+ion}, even for a singly-ionized system such as 
Be$^+$, there is much better agreement between predictions from a number of different
distorted-wave models and highly sophisticated close-coupling theories than between
any of these predictions with the only available set of experimental data~\cite{PhysRevA.27.754}.
Given the importance of beryllium and its ions for fusion devices, additional calculations were recently
performed, all of which essentially confirmed the results published in~\cite{PhysRevA.68.032712}.
The principal reason for this good agreement is the generally fast convergence of
pseudo-state models with the number of pseudo-states included in the close-coupling
expansion~\cite{Mitnik1999}.
It is hence very likely that the theoretical results are more reliable than the experimental
data in this case.

We note that cross sections, which are dominated by resonant processes, can
be very sensitive to the details of the calculation.
This has been explored,
for example, for the e-C$^+$ collision system~\cite{CII}, where a single
low-lying resonance dominates the low-energy behavior.

The process of dielectronic recombination is strongly affected by resonances
and, moreover, an almost unbounded number of states and transitions can be
involved.  This makes it very difficult to calculate cross sections and even
more difficult to estimate uncertainties in the calculated
results. Experimental benchmarks are of the highest importance as may be seen
in recent work for intermediate charge states of tungsten
~\cite{Spruck2014,Badnell2016}.
However, in the recent article~\cite{Badnell2015} 
there is a discussion of uncertainties in calculated rates
of dielectronic recombination for S$^{2+}$ recombining to S$^+$ associated with
uncertainties in the autoionizing level positions.  
The uncertainty was assessed
by performing two additional calculations in which a critical autoionizing
resonance position was shifted to just above threshold, thereby maximizing the
resulting low-temperature rate coefficient, or by shifting it to an
intermediate position.  The authors conclude that ``An observational program,
combined with spectral modeling and a parallel effort in atomic theory, could
make real progress in deriving DR rates for third and fourth row elements with
well-defined uncertainties''.

\subsection{Electron-molecule collisions}
Up until now, uncertainty assessment has been rare in electron-molecule
collision calculations.  An exception is the recent study of electron
collisions with H$_2^+$ by Zammit \etal~\cite{14ZaFuBr}.  This system
has the advantage that it is possible to use (near) exact wave~functions for
the one-electron target.  Zammit \etal\ use a CCC
technique and the adiabatic nuclei approximation to compute
vibrationally resolved dissociative excitation and ionization cross sections
for the system; they obtain results accurate to better than 10\% and 5\%,
respectively.  These uncertainty estimates were derived from considering
(a) the behavior as function of the size of the CC expansion and (b) a smaller
contribution due to their approximate treatment of nuclear motion.  It would
seem that the use of extended close-coupling expansions is the most promising
approach for obtaining uncertainty quantified results for electron-molecule
collisions.

\begin{figure}
\includegraphics[width=0.45\textwidth]{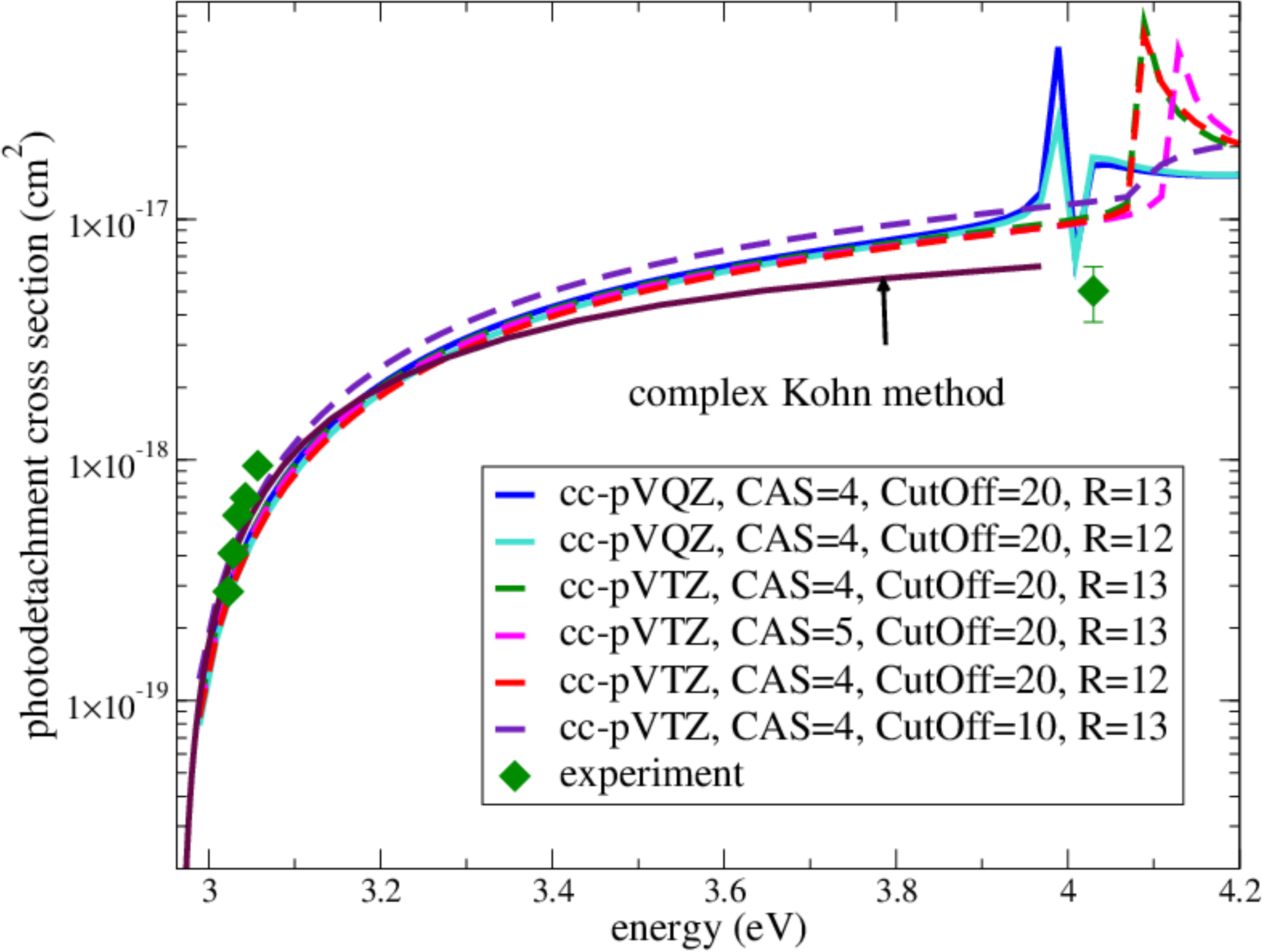}
\caption{
Theoretical (curves)~\cite{khamesyan16} and experimental  (symbols)~\cite{best11} 
cross sections for C$_2$H$^-$ photo\-detachment.  One of the shown theoretical curves 
is obtained using the complex Kohn method~\cite{douguet14}, all other curves are 
from the UK R-matrix~\cite{jt474,jt416} calculations.  The \hbox{R-matrix} results are 
obtained for different values of key parameters controlling accuracy of electron scattering 
calculations at a fixed molecular geometry.  Uncertainty of \hbox{R-matrix} and complex Kohn 
results is about 20\% for the cross section far from the electronic resonance at $4$ eV.}
\label{fig:C2Hm}
\end{figure}

The first example of the uncertainty assessment in electron-molecule
collisions is the calculation of cross section for photo\-detachment of the
C$_2$H$^-$ anion.  The analytical model used in the theoretical treatment of
the process is described in ref.~\cite{douguet14}.  It should be stressed here
that the model does not account for possible rovibrational resonances that
could be present in the photo\-detachment spectrum.  Such a model would
correspond to a low-resolution experiment, similar to the one of
ref.~\cite{best11}, where rovibrational structure is unresolved.  Figure
\ref{fig:C2Hm} shows results of the theoretical calculations using the UK
R-matrix code~\cite{jt474,jt416,jt518} with different values of several
parameters, which control the accuracy of the electron scattering matrix
obtained in the \hbox{R-matrix} code.  The figure also shows the results of the
complex Kohn method (see details in ref.~\cite{douguet14}) and the available
experimental data~\cite{best11}.  Although a systematic uncertainty analysis
of the calculation has not been performed, the \hbox{R-matrix} results shown in the
figure suggest that the accuracy of the computation model is of the order of 15\%
for the cross section far from the electronic resonance at $4$~eV.  The
uncertainty in the position of the resonance is about $\pm0.15$~eV.  The
present uncertainty analysis does not address the uncertainty of the
analytical model itself that neglects the mentioned rovibrational resonances.
However, in an ideal theoretical study of this or a similar process, the
uncertainty of the analytical model should also be discussed.

\begin{figure}
\includegraphics[width=0.45\textwidth]{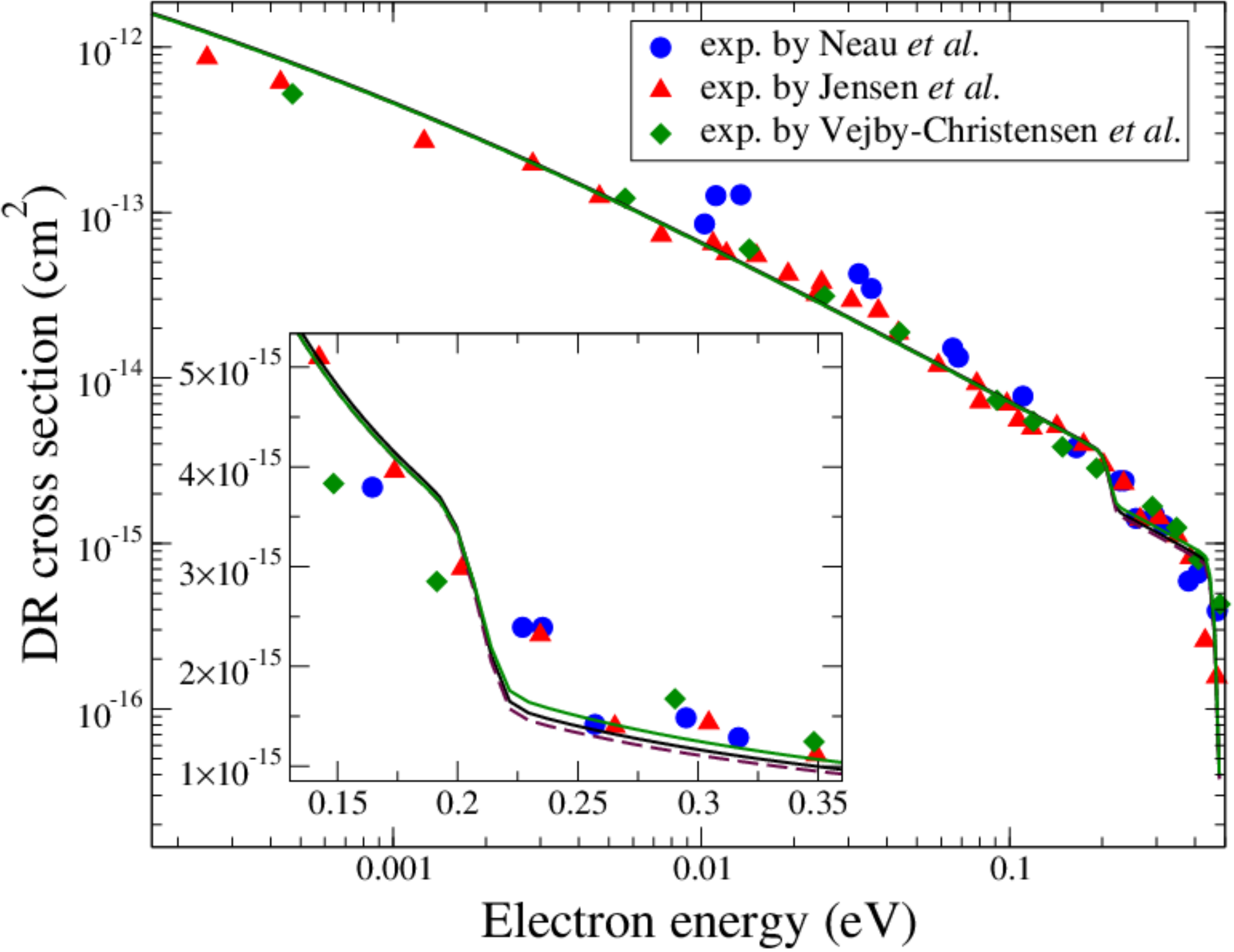}
\caption{Theoretical (curves) and experimental (symbols)~\cite{vejby97,neau00,jensen00} 
  cross section of dissociative recombination
  of the H$_3$O$^+$ ion with electrons.  There are three theoretical curves on
  the figure that are almost indistinguishable in the main graph.  The inset
  zooms a part of the presented data such that three curves are
  distinguishable.  Uncertainty of calculation within the employed analytical
  model (discussed in detail in ref.~\cite{douguet12a}) is about 8\%.
  Uncertainty of the analytical model itself is much larger because the model
  neglects some important physics, such as auto\-ionization and rovibrational
  resonances present in the DR spectrum.  }
\label{fig:H3O}
\end{figure}

The second example is the cross section for dissociative recombination of the
H$_3$O$^+$ ion with low energy electrons.  The analytical and computational
model for the process is described in ref.~\cite{douguet12a}.  Although the
DR process also involves electron scattering, the employed model is very
different from the one discussed above: the scattering matrix for collisions
between an electron and a molecular ion in this approach is obtained from {\it
  ab initio} calculations of excited Rydberg states of the neutral H$_3$O
molecule.  Energies of the lowest electronic state of H$_3$O$^+$ and several
excited electronic states $E_n$ of H$_3$O are obtained using the Columbus code
\cite{columbus08} (see details in~\cite{douguet12a}).  As a second step,
effective quantum numbers $\nu_n({\cal Q})$ of the excited states are computed
from energy difference $\Delta E_n({\cal Q})$ between H$_3$O$^+$ and H$_3$O
energies, where $\cal Q$ refers to a particular molecular geometry.  Functions
$\nu_n({\cal Q})$ are fit with a simple linear (or quadratic) function along
$\cal Q$ and coefficients of the linear (or quadratic) fit are used to obtain
the electron-molecule scattering matrix electron energies and, therefore,
determine the final DR cross section.  The electron-molecule scattering matrix
for positive (relative to the ionization threshold) electronic energies is
therefore computed from negative energies of Rydberg states in the spirit of
quantum defect theory.  In principle, Rydberg states with different principal
quantum numbers $n$ could be used to perform the fit, to construct the
scattering matrix, and to calculate the DR cross section.  In theory, quantum
defects $\mu_n=n-\nu_n$ are slightly different for different $n$, i.e.\ they
are energy-dependent.  Additional uncertainty of $\mu_n$ comes from the
accuracy of the {\it ab initio} calculation.  These are the two major sources
of uncertainty in the calculation of the DR cross section within the
discussed analytical model.  The effect of these uncertainties on the final DR
cross section is demonstrated in figure~\ref{fig:H3O}, where the cross section
is calculated for three different sets of parameters obtained from three
different manifolds of Rydberg states of H$_3$O.  The difference in the
results on the figure is attributed to accuracy of {\it ab initio} energies of
excited electronic states of H$_3$O and to the energy dependence of the quantum
defects.  As in the first example, there is an additional source of
uncertainty due to the employed analytical model, which neglects several
possible processes during a DR event, such as the possibility of
auto\-ionization once the electron is captured by the ion or the influence of
rovibrational resonances.  This uncertainty is not addressed here.

\subsection{Charge transfer collisions}\label{subsec:hcoll-results}
We begin the discussion of examples for uncertainty estimates in
charge transfer collisions with the one-electron C$^{6+}-{\rm H}(1s$) system.
This and similar fully stripped ion - neutral hydrogen atom collision systems
have been the subject of a large number of theoretical investigations over
many years.  In part, this is due to their relevance for applications such as
charge exchange recombination spectroscopy, which is an important tool for the
diagnostics of fusion plasmas (see, for example, refs.~\cite{Igenbergs12,Jorge14} and
references therein).  Another reason for the great interest in these systems
is their benchmark character.  Being true one-electron problems, model
uncertainties can be kept to a minimum and convergence properties of different
approaches, i.e.~numerical uncertainties can be studied.

\begin{figure}[b]
\includegraphics[width=0.9\linewidth]{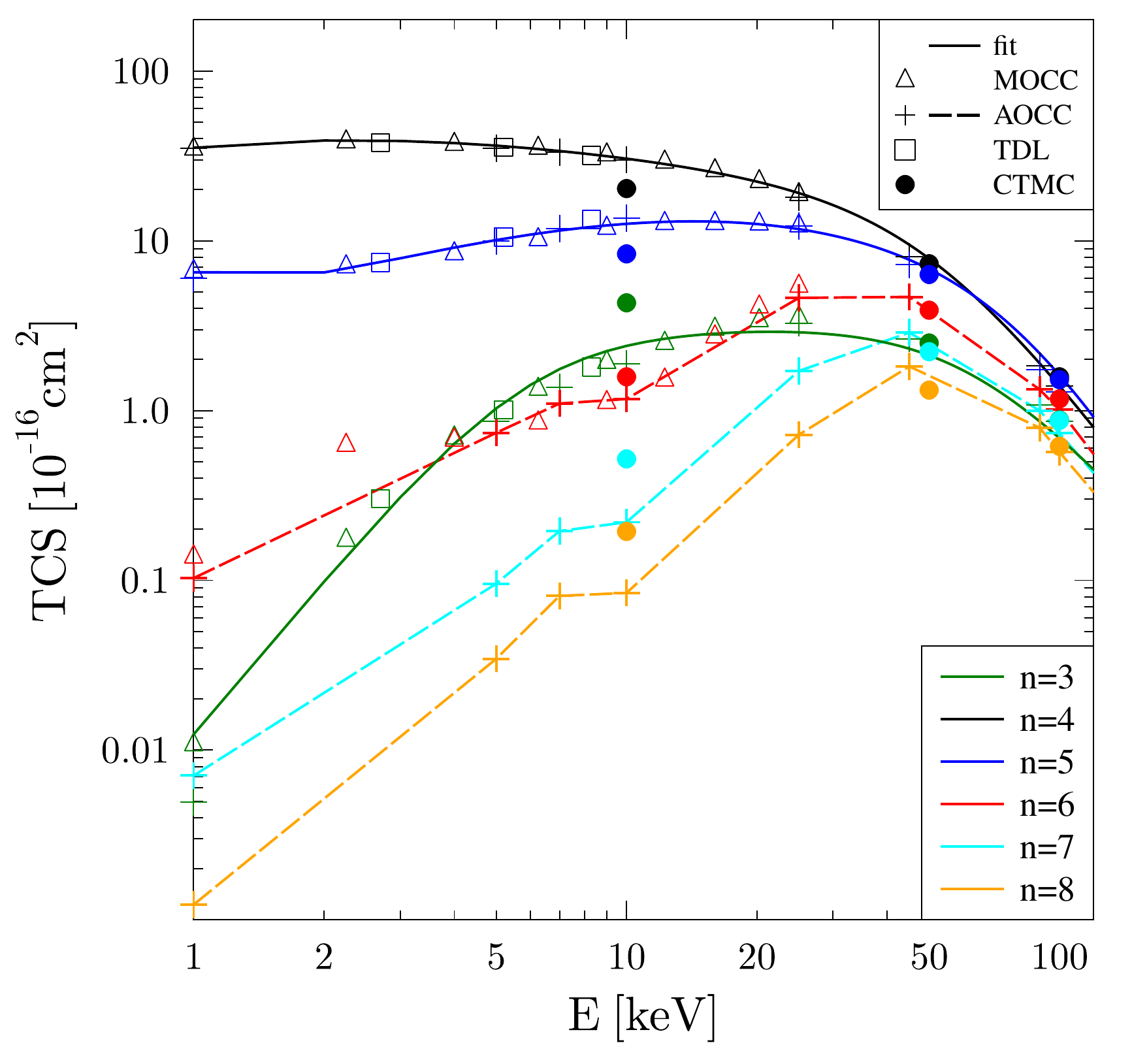}
\caption{\label{fig:cx1} Cross sections for $n$-shell selective charge
  transfer in C$^{6+}$-H($1s$) collisions as functions of impact energy.  Full
  lines: fits according to~\cite{Suno06}, MOCC:~\cite{Harel98},
  AOCC:~\cite{Igenbergs12}, TDL:~\cite{Pindzola15}, CTMC:~\cite{Cariatore15}.}
\end{figure}

Figure~\ref{fig:cx1} displays cross sections for charge transfer from hydrogen
into individual $n$-shells of the hydrogenlike C$^{5+}$ ion in the low to
intermediate energy regimes, in which the semiclassical approximation with
straight-line trajectories is essentially exact.  Based on a large set of
cross section calculations carried out in the 1980s and 1990s, Suno and Kato
constructed recommended data sets that can be fit by simple analytical
functions~\cite{Suno06}.  The recommended data are shown in
figure~\ref{fig:cx1} as solid lines.  In addition, one recent representative is
included for each of the following theoretical methods: MOCC~\cite{Harel98},
AOCC~\cite{Igenbergs12}, TDL~\cite{Pindzola15}, and CTMC~\cite{Cariatore15}.
Experimental data on the $n$-shell resolved level are not available for this
collision system.

It can be observed that predictions from the three semiclassical methods, MOCC, AOCC and TDL,
are in excellent agreement with each other and with the recommended data for
the dominant $n=4$ and the subdominant $n=5$ channels.  From this comparison
one can conclude that the cross section predictions are accurate to within a
few percent.  For the less important $n=3$ and $n=6$ channels the discrepancies are
slightly larger. Except for the $n=6$ MOCC data point at 2~keV, however, the overall
agreement is still very good.

\begin{figure}[t]
\includegraphics[width=0.72\linewidth]{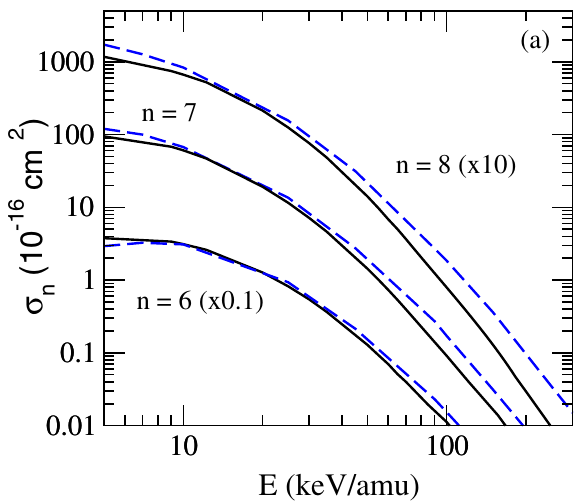} 

\medskip

\includegraphics[width=0.90\linewidth]{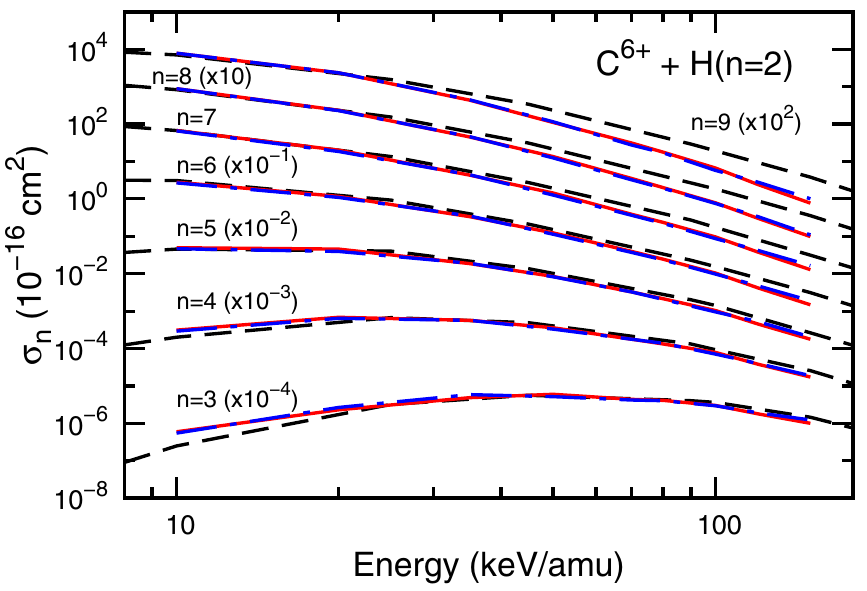}
\caption{\label{fig:cx2} Cross sections for $n$-shell selective charge
  transfer in C$^{6+}-{\rm H}(n\!=\!2$) collisions as functions of impact energy.
  Top panel: full lines: hydrogenic CTMC, dashed lines: AOCC. (Figure taken
  from~\cite{Jorge14}.)  Bottom panel: full lines: hydrogenic-Z-CTMC,
  dash-dotted lines: microcanonical CTMC (both sets of curves are essentially
  on top of each other), dashed lines: AOCC. (Figure taken
  from~\cite{Cariatore15}.)  The AOCC cross sections in both panels are
  from~\cite{Igenbergs12}.}
\end{figure}

The CTMC results included in figure~\ref{fig:cx1} agree very well with the
semiclassical calculations at 50 and 100 keV impact energy.  At lower energies
the discrepancies are larger and different variants of CTMC models give
different results~\cite{Jorge14,Cariatore15}.  This suggests that an
uncertainty assessment solely based on CTMC calculations would be difficult in
this regime.  However, at higher energies and for high $n$ quantum numbers 
the predictions of the different CTMC variants are all in good agreement with
each other and can be viewed as the best results currently available.

Figure~\ref{fig:cx2} shows $n$-shell specific charge
transfer cross sections for C$^{6+}-{\rm H}(n\!=\!2$) collisions.  Due to the weaker
binding of the electron in an excited initial state, higher projectile
$n$-shells are favored, and overall the cross sections are larger when compared
to the ground-state case.  The CTMC calculations of Jorge \etal\ (top
panel) and Cariatore \etal\ (bottom panel) appear to be in good
agreement with each other, while the AOCC cross sections from
ref.~\cite{Igenbergs12} tend to overestimate the CTMC data as the impact
energy and the principal quantum number of the final states increase.  Even
though the two-center basis used includes a total of 340 states, which is a
very large number for a semiclassical CC calculation, the overestimation is
deemed to be a finite-basis effect, i.e.~a convergence problem of the AOCC
method in this energy range~\cite{Jorge14,Cariatore15}.

\begin{figure}[b]
\includegraphics[width=0.9\linewidth]{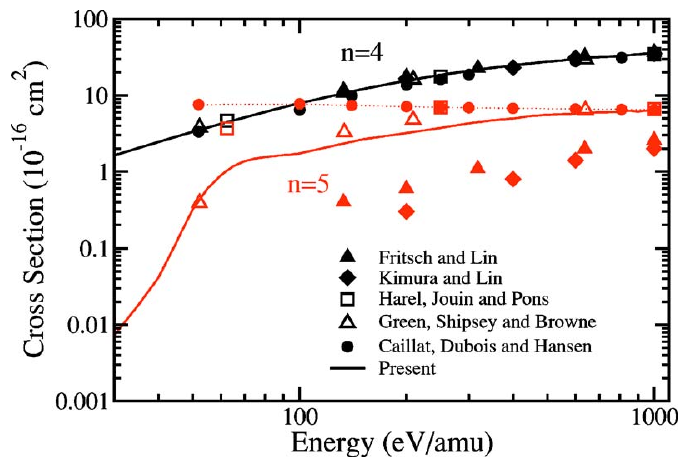}
\caption{\label{fig:cx3} Cross sections for $n$-shell selective charge
  transfer in C$^{6+}$-H($1s$) collisions as functions of impact energy.  Full
  lines (``Present''): HSCC~\cite{Liu05}; Fritsch and Lin: AOCC with curved
  trajectories~\cite{Fritsch84}; Kimura and Lin: (modified) AOCC with
  straight-line trajectories~\cite{Kimura85}; Harel \etal: MOCC with
  straight-line trajectories~\cite{Harel98}; Green \etal: MOCC with
  curved trajectories~\cite{Green82a, Green82b}; Caillat \etal: AOCC
  with straight-line trajectories~\cite{Caillat00}.  (Figure taken from~\cite{Liu05}.)}
\end{figure}

Figure~\ref{fig:cx3} explores the very low energy regime in which
straight-line trajectories, and perhaps the semiclassical approximation
itself, become questionable.  For the dominant $n=4$ channel in the
C$^{6+}$-H($1s$) collision system this does not appear to be an issue and
semiclassical CC calculations based on curved and straight-line trajectories
agree with each other and with a fully quantum mechanical HSCC calculation on
the few-percent level.  However, for $n=5$ qualitative discrepancies are
observed.  They have been traced to both basis-size limitations of the earlier
CC calculations and to trajectory effects~\cite{Caillat00}.  Regarding the
latter it was found that (i) straight-line trajectories lead to a strong
overestimation of the cross section when the calculation is reasonably well
converged with respect to basis size, and (ii) choosing the optimal curved
trajectory involves a certain level of arbitrariness~\cite{Caillat00,Liu05}.
The HSCC method is free of this ambiguity and probably gives the best answer.
Interestingly, it is in close agreement with an early MOCC calculation, which
used a relatively small number of states and curved trajectories based on an
average molecular potential~\cite{Green82a,Green82b}.  By contrast, a larger
MOCC calculation based on straight-line trajectories yields significantly
higher cross sections~\cite{Harel98}.  Given the absence of other fully
quantum mechanical calculations and experimental data a quantitative
uncertainty assessment appears to be difficult in this region.

\begin{figure}[t]
\includegraphics[width=0.9\linewidth]{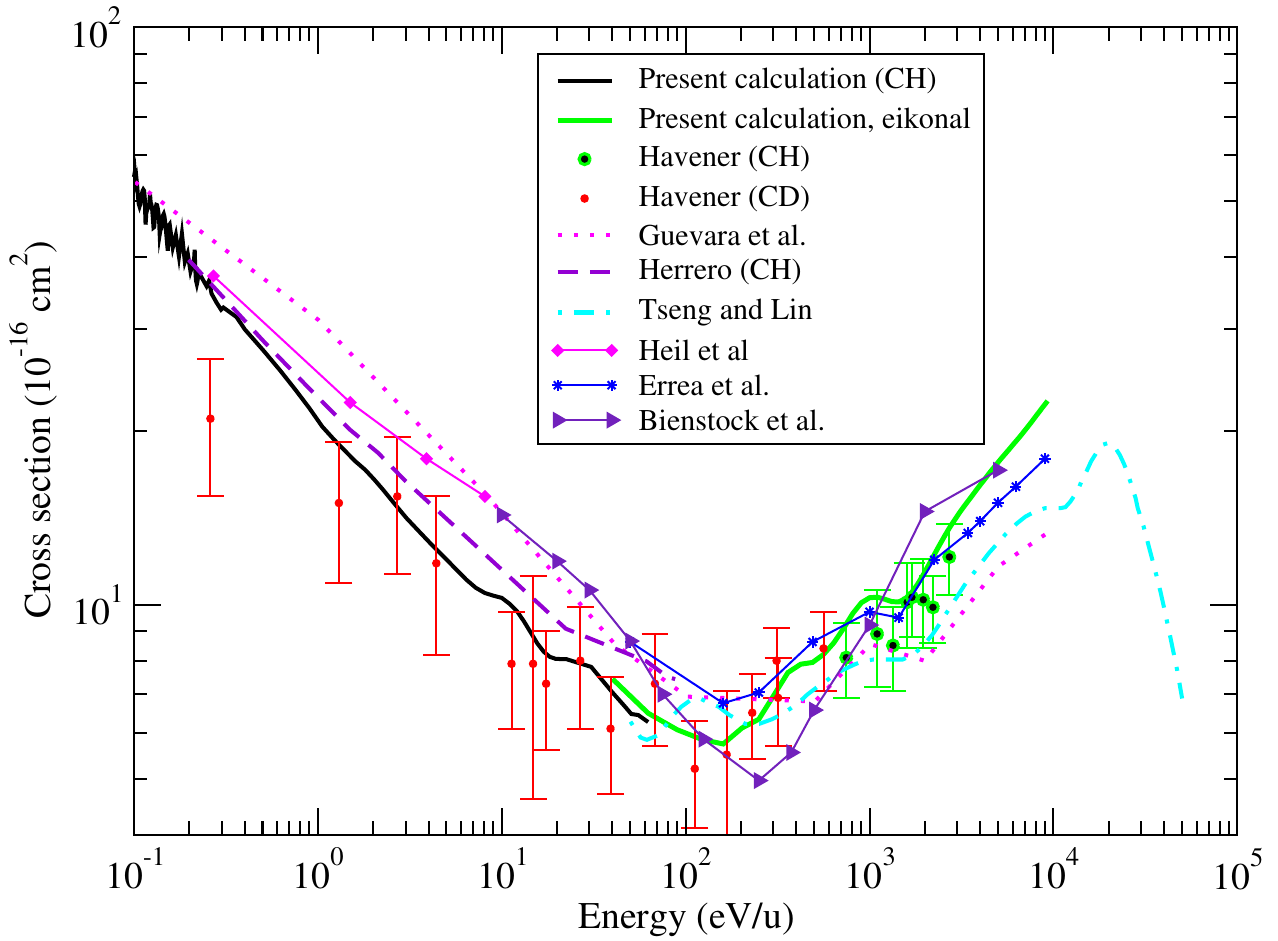}
\caption{\label{fig:cx4} Total cross section for charge transfer in
C$^{3+}$-H($1s$) collisions as function of impact energy.
``Present calculation (CH)'': QMOCC~\cite{Errea15};
``Present calculation, eikonal'': MOCC~\cite{Errea15};
Guevara \etal: END~\cite{Guevara11};
Herrero (CH): QMOCC~\cite{Herrero95};
Tseng and Lin: AOCC~\cite{Tseng99};
Heil \etal: QMOCC~\cite{Heil81};
Errea \etal: MOCC~\cite{Errea91};
Bienstock \etal: QMOCC~\cite{Bienstock82}.
Experimental data for atomic hydrogen [Havener (CH)]
and deuterium [Havener (CD)]:~\cite{Havener95}.
(Figure taken from~\cite{Errea15}.)}
\end{figure}

We now turn to single-electron transfer in the few-electron C$^{3+}-{\rm H}(1s$)
collision system.  Figure~\ref{fig:cx4} displays the total charge transfer
cross section over a wide range of impact energies, spanning the very low energy, 
low energy, and intermediate energy regimes.  A systematic trend can be
observed at energies below 100 eV, which is where the cross section reaches a
minimum.  Of the four QMOCC calculations included the one that uses the most
accurate correlated wave~functions and the largest basis set (a total of ten
many-electron molecular functions) gives the smallest cross section values and
the best agreement with the experimental data.  Except for the data point at
the lowest energy of 0.3 eV/amu the calculated cross section curve lies within
the experimental uncertainty.  In the low energy region, in which the cross
section increases with increasing impact energy, the convergence of molecular
expansions slows down and above 2~keV/amu AOCC expansions are deemed to be
more accurate~\cite{Errea15}.  This is supported by the close agreement of the
calculations of refs.~\cite{Tseng99,Guevara11} with each other and, where
available, with experimental data as well.  Overall, figure~\ref{fig:cx4}
illustrates nicely that an uncertainty estimate for a complicated collision
problem should use the input from several calculations based on different
theoretical methods.

\begin{figure}[t]
\includegraphics[width=0.9\linewidth]{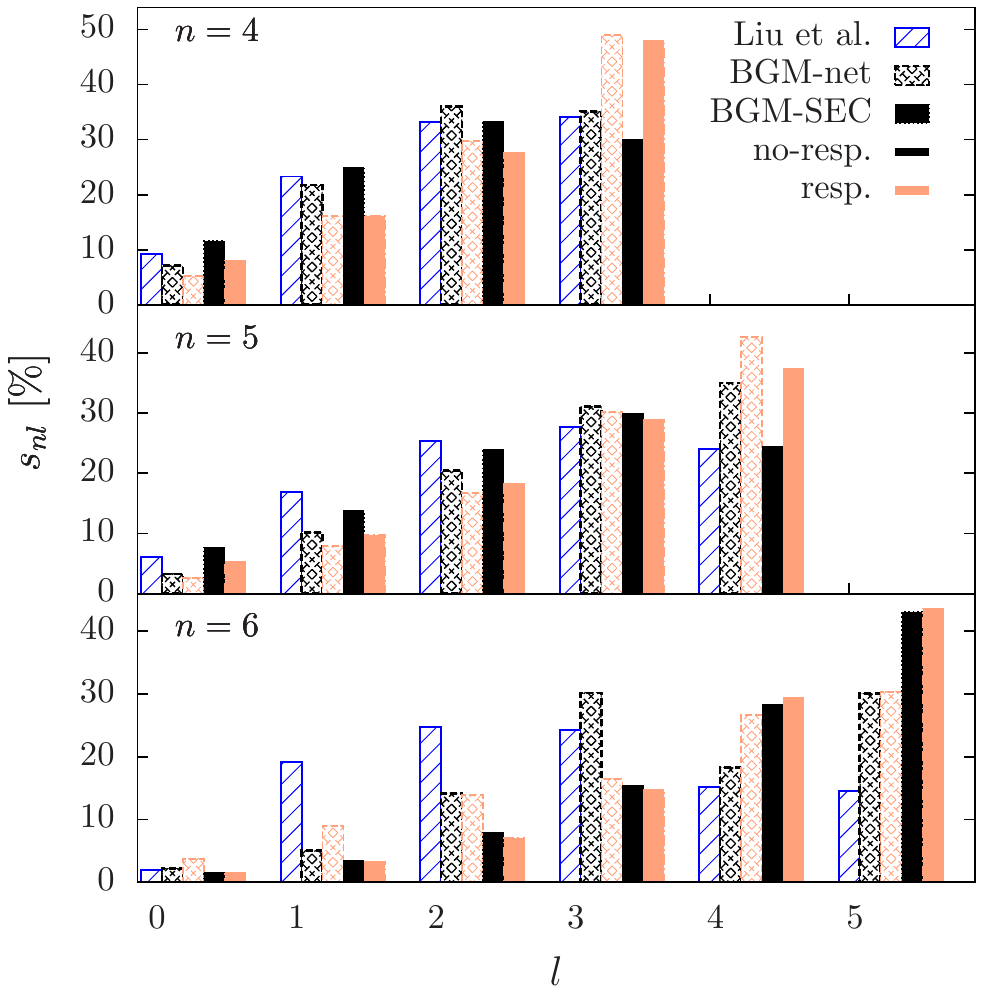}
\caption{\label{fig:cx5} Relative cross sections for $nl$-subshell selective
  charge transfer in Ne$^{10+}$-Ne collisions at 4.54 keV/amu.  Liu {\it et
    al.}: AOCC~\cite{Liu14}, all other results are from TC-BGM
  calculations~\cite{Leung15}. ``BGM-net'' and ``BGM-SEC'' refer to different
  statistical analyses of single transfer, while ``resp'' and ``no-resp''
  calculations do or do not include a time-dependent response potential in the
  single-particle Hamiltonian.  (Figure taken from~\cite{Leung15}.)}
\end{figure}

In figure~\ref{fig:cx5} we show an example for single-electron transfer from a
many-electron target.  The system is 4.54 keV/amu Ne$^{10+}$-Ne and plotted
are relative cross sections for transfer into specific $nl$-subshells.
Results from one AOCC and various TC-BGM calculations are included.  All of
them are on the level of the IEM, but they use different variants of effective
target atom potentials and statistical methods for the calculation of the
cross sections.  It can be argued that what is referred to as BGM-SEC in the
figure represents the most consistent calculation of single transfer within
the IEM~\cite{Leung15}.  However, this is no guarantee for success, since the
IEM represents a model whose accuracy is difficult to determine.  In lieu of
correlated cross section calculations, comparing different IEM variants is the
best one can do to assess the uncertainty of the theoretical results.  A
somewhat conservative estimate would then consist in taking the spread of the
shown relative cross sections as their uncertainty.

\begin{figure}[t]
\includegraphics[width=0.9\linewidth]{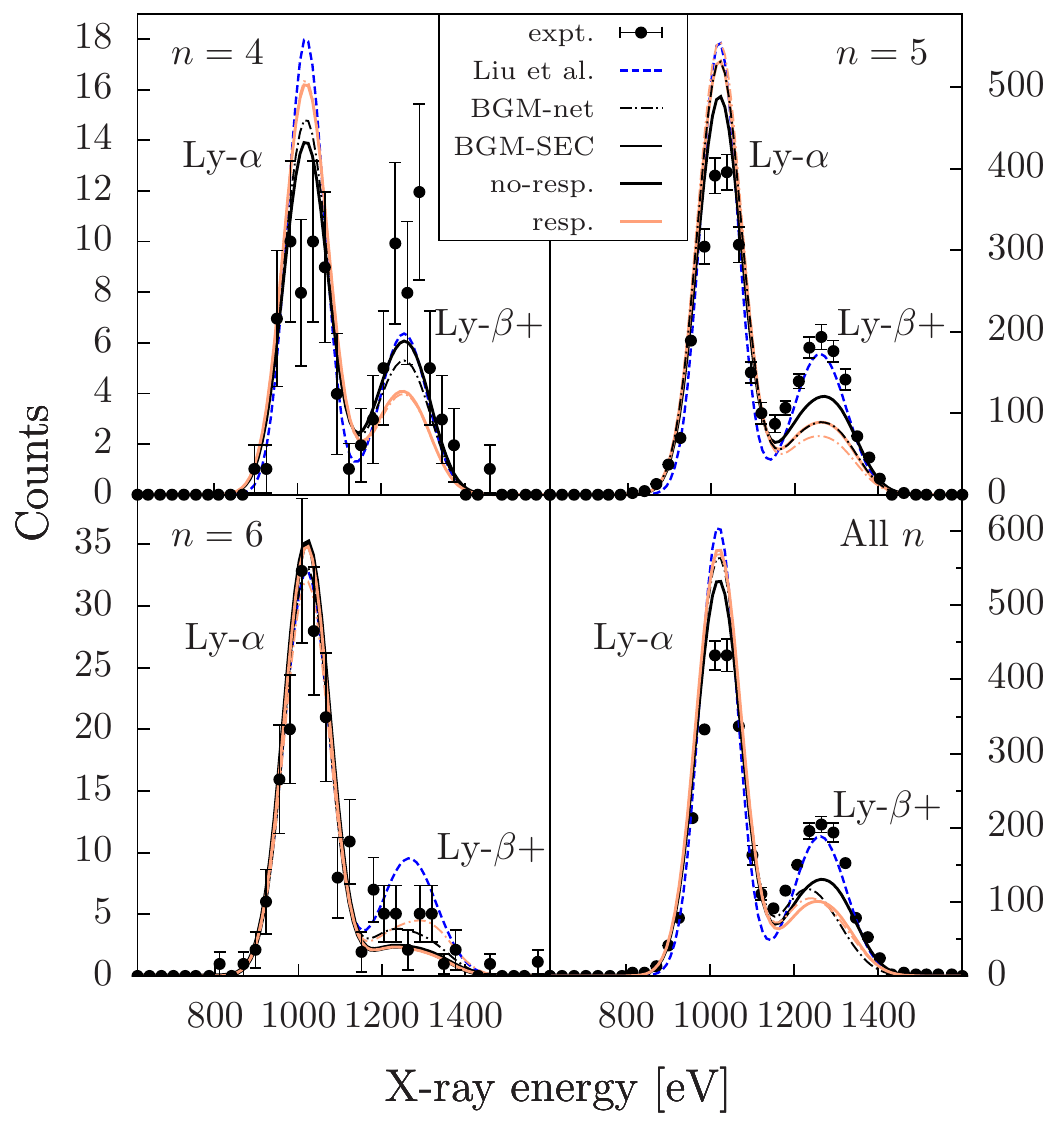}
\caption{\label{fig:cx6} X-ray spectra from single electron transfer to the
  $n$-th shell in Ne$^{10+}$-Ne collisions at 4.54 keV/amu.  Experimental
  data:~\cite{Ali10}; Liu \etal: AOCC~\cite{Liu14}, all other results
  are from TC-BGM calculations (cf.\ figure~\ref{fig:cx5})~\cite{Leung15}.}
\end{figure}

An important motivation for studying these $nl$-distributions is that they
form the starting point for the calculation of x-ray spectra, which in turn
are of interest, e.g., for the understanding of cometary x-ray emission.  In
figure~\ref{fig:cx6} we show the modeled $n$-state selective spectra for the
hydrogenlike Ne$^{9+}$ ion produced in the charge transfer collision.  Results
from all calculations shown in figure~\ref{fig:cx5} as well as experimental data
from ref.~\cite{Ali10} are included.  Overall, the spread in the theoretical
x-ray spectra appears to be similar or even smaller than the spread in the
relative cross sections, implying that error propagation is not problematic
and the x-ray spectra are rather forgiving quantities.  However, this is
partly due to a mutual normalization process used in all of the calculations,
and, perhaps to a lesser extent, to spectral-line convolutions to Gaussian
profiles that were applied as well~\cite{Leung15}.  The ``widths'' resulting
from the various curves shown in figure~\ref{fig:cx6} can again be taken as
theoretical error bars.

\begin{figure}[t]
\includegraphics[width=0.9\linewidth]{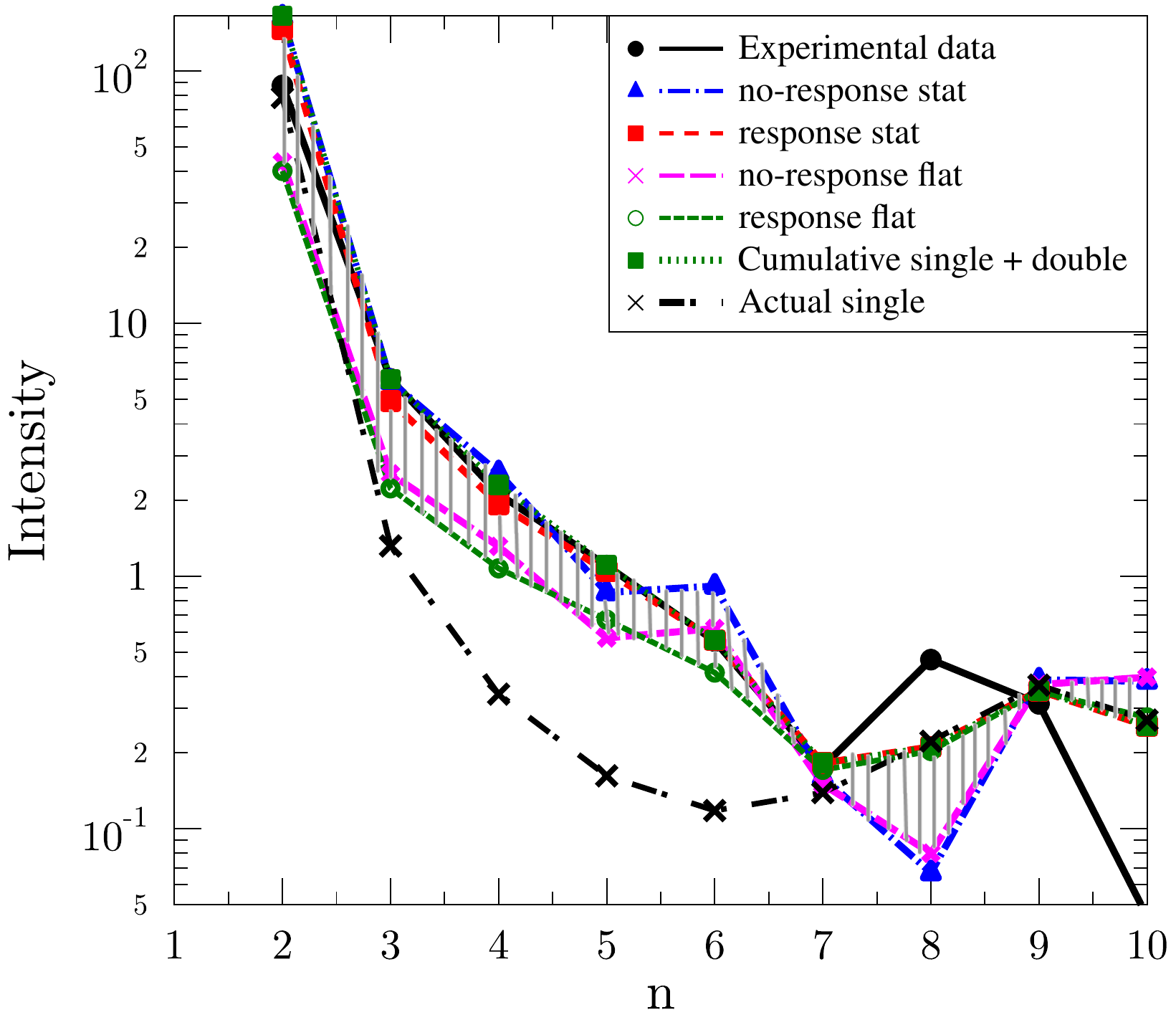}
\caption{\label{fig:cx7} X-ray intensities after charge transfer in
  Ar$^{17+}$-Ar collisions at 7 keV/amu.  Experimental
  data:~\cite{Trassinelli12}.  All theoretical results are from TC-BGM
  calculations (see text)~\cite{Arash13}.}
\end{figure}

The last example of this section concerns a true many-electron problem: x-ray
production after charge transfer in 7 keV/amu Ar$^{17+}$-Ar collisions.  In
this case the available measurements~\cite{Trassinelli12} cannot discern
single vs.\ multiple transfer.  Given that the latter should be a strong
channel for such a highly charged projectile ion, it has to be taken into
account in the theoretical modeling of the measured x-ray spectra.  In
ref.~\cite{Arash13} this was done on the level of the IEM, and again,
different variants of effective potentials (response and no-response models)
and statistical analyses were used in order to get a qualitative idea of the
theoretical error bars.  Further modeling is required in this problem to
obtain x-ray emission spectra, because Auger processes compete with radiative
de-excitation when multiple electron transfer occurs.  Once again, different
model variants were used for this (assuming flat vs.\ statistical $l$-subshell
distributions after Auger electron emission) and the calculations were
assessed by varying the models within reasonable bounds.  

Figure~\ref{fig:cx7}
compares results obtained in this way with the experimental x-ray intensities
for $1snp \rightarrow 1s^2$ transitions in the post-collision Ar$^{16+}$ ion.
One can probably take the shaded area as the theoretical error bar that
results from the comparison of the different calculations.  Except for $n=8$
and $n=10$ the experimental data are within these error bars.  There is one
calculation, however, that is clearly off.  It represents an x-ray spectrum
obtained from neglecting all multiple-transfer contributions in the
calculation.  This suggests that, despite considerable model uncertainties, some
definite conclusions can be drawn from such calculations and comparisons; in
this case, regarding the important role of multi-electron transfer events
\cite{Arash13}.

\section{Summary, Conclusions, and Outlook}\label{sec:summary}

We have reviewed approaches to uncertainty estimates for atomic and molecular
data of the kind that occur in plasma modeling. Model uncertainty is
introduced through the treatment of small terms in the Hamiltonian and (more
importantly in general) through the reduction of a many-body Schr\"odinger
equation to a tractable model. Numerical uncertainty is due to the
representation of the model on a finite mesh or basis. Uncertainties propagate
from structure calculations to predictions of scattering cross sections. We
have summarized the main tools for uncertainty assessment of calculations for
atomic and (small molecule) molecular electronic structure, electron
scattering, and charge transfer in heavy particle collisions.
%
Important tools
include the method of focal point analysis in connection with electronic
structure models and, of course, standard methods of convergence analysis for
the numerical uncertainties.

We discussed some examples of computational work on scattering calculations in
which the authors attempted to provide a reasonable uncertainty
estimate. These examples show that the field is not entirely unexplored. 
On the other hand, for every example of a scattering calculation that is
accompanied by a 
thoroughly performed
uncertainty estimate there are many more where
the authors provide their best calculations without discussion of the
uncertainties. For the case of electron impact collisions we discussed
examples of atomic excitation and ionization where a reasonable uncertainty
estimate is obtained by careful study of convergence in the structure
calculation and in the R-matrix formalism.

Calculations of dielectronic recombination in electron-atom (ion) collisions
are much more complicated than those of excitation and ionization, because
very many states and transitions can be involved. Uncertainty estimates for
calculated dielectronic recombination rate coefficients 
involve not just the convergence of the initial and final state
structure, but also convergence with respect to the number of intermediate
states and the number of transitions that are taken into account. As a
consequence, for the dielectronic recombination process in 
electron-atom collisions the provision of
uncertainties for calculated data (without relying on experimental
benchmarks) is still wide open.

The ultimate aim for a constructive theory of uncertainty quantification for
atomic and molecular data is to develop numerical procedures for structure and
scattering calculations by which reasonable uncertainty estimates are obtained
along with the primary calculated quantities of interest. The final quantities
of interest are cross sections (differential cross sections in general) as a
function of collision energy and perhaps other collision parameters. The
associated uncertainties are structured quantities, not just point values, and
they must be presented in a way that makes it possible to propagate them
through further atomic data processing (e.g., to obtain effective rate
coefficients) and through a plasma simulation. This raises many new issues
that have not been addressed in this review, but that we note here as an
outlook towards future work.

From an operational point of view, in order to propagate uncertainties in
atomic data through a plasma simulation (in a non-intrusive manner, i.e.,
without major changes to the plasma simulation code) one needs to be able to
sample the relevant cross sections with proper account of relevant
correlations in uncertainties. For example, uncertainties in cross sections
for the same process at different collision energies are correlated in some
way that depends on the energies involved. Depending on the application such
functional uncertainties (as opposed to pointwise uncertainties) may be
represented by a Gaussian Process, a polynomial chaos expansion, or a Monte
Carlo sample. The propagation of such uncertainties through a dynamical
calculation is an issue of major interest 
in the field of Uncertainty Quantification as
it is represented, for example, in the National Research Council report
already cited in the Introduction~\cite{NRC-UQ}.
For a perspective from Statistics on
Uncertainty Quantification see also,~\cite{OHagan2013} and for a monograph-length
treatment of Uncertainty Quantification with special attention to
computational fluid dynamics see~\cite{LeMaitre2010}.

In this review we have discussed the state of the art in uncertainty
assessment for calculated atomic and molecular data for plasma
applications. We conclude with what we regard as major issues for future
work. 
First, very broadly, we recommend that 
atomic and molecular physicists develop methods and codes
for scattering calculations in which uncertainty assessment is integrated with
the calculation of the primary quantities of interest. 
Second, more narrowly,
develop uncertainty assessment in a more systematic way for processes
involving resonances: near-threshold processes and dielectronic
recombination. 
Third, develop representations of correlated uncertainties
in atomic and molecular data that are suitable for studies in which those
uncertainties have to be propagated through plasma simulations. 
The latter item
will benefit from a joint effort by atomic physicists and plasma modellers.

\section*{Acknowledgments}
The authors are grateful to the COST Action MOLIM (CM1405) for support.  This
work was also supported, in part, by the United States National Science Foundation
under grants No.~PHY-1403245 (KB), No.~PHY-1520970 (KB), No.~PHY-1506391 (VK),
and the XSEDE allocation
No.~PHY-090031 (KB).  The work of AGC received support from the Scientific
Research Fund of Hungary (Grant No.~OTKA NK83583), and TK~acknowledges 
support of the Natural Sciences and Engineering Research Council of Canada,
Grant No.~RGPIN-2014-03611.  The authors acknowledge helpful correspondence
received from Mariana Safronova and Bijaya Sahoo concerning uncertainty
calculations for many-electron atoms.  Finally, the authors acknowledge helpful
discussions with all the participants in the Joint ITAMP Workshop with IAEA on
Uncertainty Assessment for Atomic and Molecular Data.


\begin{thebibliography}{239}%
\makeatletter
\providecommand \@ifxundefined [1]{%
 \@ifx{#1\undefined}
}%
\providecommand \@ifnum [1]{%
 \ifnum #1\expandafter \@firstoftwo
 \else \expandafter \@secondoftwo
 \fi
}%
\providecommand \@ifx [1]{%
 \ifx #1\expandafter \@firstoftwo
 \else \expandafter \@secondoftwo
 \fi
}%
\providecommand \natexlab [1]{#1}%
\providecommand \enquote  [1]{``#1''}%
\providecommand \bibnamefont  [1]{#1}%
\providecommand \bibfnamefont [1]{#1}%
\providecommand \citenamefont [1]{#1}%
\providecommand \href@noop [0]{\@secondoftwo}%
\providecommand \href [0]{\begingroup \@sanitize@url \@href}%
\providecommand \@href[1]{\@@startlink{#1}\@@href}%
\providecommand \@@href[1]{\endgroup#1\@@endlink}%
\providecommand \@sanitize@url [0]{\catcode `\\12\catcode `\$12\catcode
  `\&12\catcode `\#12\catcode `\^12\catcode `\_12\catcode `\%12\relax}%
\providecommand \@@startlink[1]{}%
\providecommand \@@endlink[0]{}%
\providecommand \url  [0]{\begingroup\@sanitize@url \@url }%
\providecommand \@url [1]{\endgroup\@href {#1}{\urlprefix }}%
\providecommand \urlprefix  [0]{URL }%
\providecommand \Eprint [0]{\href }%
\providecommand \doibase [0]{http://dx.doi.org/}%
\providecommand \selectlanguage [0]{\@gobble}%
\providecommand \bibinfo  [0]{\@secondoftwo}%
\providecommand \bibfield  [0]{\@secondoftwo}%
\providecommand \translation [1]{[#1]}%
\providecommand \BibitemOpen [0]{}%
\providecommand \bibitemStop [0]{}%
\providecommand \bibitemNoStop [0]{.\EOS\space}%
\providecommand \EOS [0]{\spacefactor3000\relax}%
\providecommand \BibitemShut  [1]{\csname bibitem#1\endcsname}%
\let\auto@bib@innerbib\@empty
\bibitem [{\citenamefont {{The Editors}}(2011)}]{PRA}%
  \BibitemOpen
  \bibfield  {author} {\bibinfo {author} {\bibnamefont {{The Editors}}},\
  }\href {\doibase 10.1103/PhysRevA.83.040001} {\bibfield  {journal} {\bibinfo
  {journal} {Phys. Rev. A}\ }\textbf {\bibinfo {volume} {83}},\ \bibinfo
  {pages} {040001} (\bibinfo {year} {2011})}\BibitemShut {NoStop}%
\bibitem [{\citenamefont {{National Research Council Committee on Mathematical
  Foundations of Verification, Validation, and Uncertainty
  Quantification}}(2012)}]{NRC-UQ}%
  \BibitemOpen
  \bibfield  {author} {\bibinfo {author} {\bibnamefont {{National Research
  Council Committee on Mathematical Foundations of Verification, Validation,
  and Uncertainty Quantification}}},\ }\href
  {http://www.nap.edu/openbook.php?record_id=13395} {\emph {\bibinfo {title}
  {{Assessing the reliability of complex models: mathematical and statistical
  foundations of verification, validation and uncertainty quantification}}}}\
  (\bibinfo  {publisher} {National Academies Press},\ \bibinfo {year}
  {2012})\BibitemShut {NoStop}%
\bibitem [{SIA()}]{SIAG-UQ}%
  \BibitemOpen
  \href@noop {} {}\bibinfo {note}
  {{http://www.siam.org/activity/uq/}}\BibitemShut {NoStop}%
\bibitem [{\citenamefont {Dirac}(1929)}]{dirac}%
  \BibitemOpen
  \bibfield  {author} {\bibinfo {author} {\bibfnamefont {P.~A.~M.}\
  \bibnamefont {Dirac}},\ }\href@noop {} {\bibfield  {journal} {\bibinfo
  {journal} {Proc. R. Soc. A}\ }\textbf {\bibinfo {volume} {123}},\ \bibinfo
  {pages} {714} (\bibinfo {year} {1929})}\BibitemShut {NoStop}%
\bibitem [{IAE(2014)}]{IAEA-ITAMP}%
  \BibitemOpen
  \href@noop {} {} (\bibinfo {year} {2014}),\ \bibinfo {note}
  {{https://www-amdis.iaea.org/meetings/ITAMP/}}\BibitemShut {NoStop}%
\bibitem [{Sto(2015)}]{StonyBrook-UQ}%
  \BibitemOpen
  \href@noop {} {} (\bibinfo {year} {2015}),\ \bibinfo {note}
  {{http://www.iacs.stonybrook.edu/uq/pages/workshop}}\BibitemShut {NoStop}%
\bibitem [{\citenamefont {Reiter}\ and\ \citenamefont
  {Janev}(2010)}]{Reiter-HydKin}%
  \BibitemOpen
  \bibfield  {author} {\bibinfo {author} {\bibfnamefont {D.}~\bibnamefont
  {Reiter}}\ and\ \bibinfo {author} {\bibfnamefont {R.~K.}\ \bibnamefont
  {Janev}},\ }\href {\doibase 0.1002/ctpp.201000090} {\bibfield  {journal}
  {\bibinfo  {journal} {Contrib. Plasma Phys.}\ }\textbf {\bibinfo {volume}
  {50}},\ \bibinfo {pages} {986} (\bibinfo {year} {2010})}\BibitemShut
  {NoStop}%
\bibitem [{\citenamefont {{Cs\'asz\'ar}}\ \emph {et~al.}(1998)\citenamefont
  {{Cs\'asz\'ar}}, \citenamefont {Allen},\ and\ \citenamefont {{Schaefer
  III}}}]{98CsAlSc}%
  \BibitemOpen
  \bibfield  {author} {\bibinfo {author} {\bibfnamefont {A.~G.}\ \bibnamefont
  {{Cs\'asz\'ar}}}, \bibinfo {author} {\bibfnamefont {W.~D.}\ \bibnamefont
  {Allen}}, \ and\ \bibinfo {author} {\bibfnamefont {H.~F.}\ \bibnamefont
  {{Schaefer III}}},\ }\href@noop {} {\bibfield  {journal} {\bibinfo  {journal}
  {J. Chem. Phys.}\ }\textbf {\bibinfo {volume} {108}},\ \bibinfo {pages}
  {9751} (\bibinfo {year} {1998})}\BibitemShut {NoStop}%
\bibitem [{\citenamefont {Boyarkin}\ \emph {et~al.}(2013)\citenamefont
  {Boyarkin}, \citenamefont {Koshelev}, \citenamefont {Aseev}, \citenamefont
  {Maksyutenko}, \citenamefont {Rizzo}, \citenamefont {Zobov}, \citenamefont
  {Lodi}, \citenamefont {Tennyson},\ and\ \citenamefont {Polyansky}}]{jt549}%
  \BibitemOpen
  \bibfield  {author} {\bibinfo {author} {\bibfnamefont {O.~V.}\ \bibnamefont
  {Boyarkin}}, \bibinfo {author} {\bibfnamefont {M.~A.}\ \bibnamefont
  {Koshelev}}, \bibinfo {author} {\bibfnamefont {O.}~\bibnamefont {Aseev}},
  \bibinfo {author} {\bibfnamefont {P.}~\bibnamefont {Maksyutenko}}, \bibinfo
  {author} {\bibfnamefont {T.~R.}\ \bibnamefont {Rizzo}}, \bibinfo {author}
  {\bibfnamefont {N.~F.}\ \bibnamefont {Zobov}}, \bibinfo {author}
  {\bibfnamefont {L.}~\bibnamefont {Lodi}}, \bibinfo {author} {\bibfnamefont
  {J.}~\bibnamefont {Tennyson}}, \ and\ \bibinfo {author} {\bibfnamefont
  {O.~L.}\ \bibnamefont {Polyansky}},\ }\href@noop {} {\bibfield  {journal}
  {\bibinfo  {journal} {Chem. Phys. Lett.}\ }\textbf {\bibinfo {volume}
  {568-â569}},\ \bibinfo {pages} {14} (\bibinfo {year} {2013})}\BibitemShut
  {NoStop}%
\bibitem [{\citenamefont {berto Capote}\ and\ \citenamefont
  {Smith}(2008)}]{CAPOTE20082768}%
  \BibitemOpen
  \bibfield  {author} {\bibinfo {author} {\bibnamefont {berto Capote}}\ and\
  \bibinfo {author} {\bibfnamefont {D.~L.}\ \bibnamefont {Smith}},\ }\href@noop
  {} {\bibfield  {journal} {\bibinfo  {journal} {Nuclear Data Sheets}\ }\textbf
  {\bibinfo {volume} {109}},\ \bibinfo {pages} {2768} (\bibinfo {year}
  {2008})}\BibitemShut {NoStop}%
\bibitem [{\citenamefont {Bethe}\ and\ \citenamefont
  {Salpeter}(1957)}]{DrakeBandS}%
  \BibitemOpen
  \bibfield  {author} {\bibinfo {author} {\bibfnamefont {H.~A.}\ \bibnamefont
  {Bethe}}\ and\ \bibinfo {author} {\bibfnamefont {E.~E.}\ \bibnamefont
  {Salpeter}},\ }\href@noop {} {\emph {\bibinfo {title} {Quantum Mechanics Of
  One- And Two-Electron Atoms}}}\ (\bibinfo  {publisher} {Springer-Verlag,
  Berlin},\ \bibinfo {year} {1957})\ pp.\ \bibinfo {pages} {4 --
  46}\BibitemShut {NoStop}%
\bibitem [{\citenamefont {Eides}\ \emph {et~al.}(2001)\citenamefont {Eides},
  \citenamefont {Grotch},\ and\ \citenamefont {Shelyuto}}]{Eides_2001}%
  \BibitemOpen
  \bibfield  {author} {\bibinfo {author} {\bibfnamefont {M.~I.}\ \bibnamefont
  {Eides}}, \bibinfo {author} {\bibfnamefont {H.}~\bibnamefont {Grotch}}, \
  and\ \bibinfo {author} {\bibfnamefont {V.~A.}\ \bibnamefont {Shelyuto}},\
  }\href@noop {} {\bibfield  {journal} {\bibinfo  {journal} {Phys. Rep.}\
  }\textbf {\bibinfo {volume} {342}},\ \bibinfo {pages} {63} (\bibinfo {year}
  {2001})}\BibitemShut {NoStop}%
\bibitem [{\citenamefont {Jentschura}\ \emph {et~al.}(2005)\citenamefont
  {Jentschura}, \citenamefont {Kotochigova}, \citenamefont {Le~Bigot},
  \citenamefont {Mohr},\ and\ \citenamefont {Taylor}}]{PhysRevLett.95.163003}%
  \BibitemOpen
  \bibfield  {author} {\bibinfo {author} {\bibfnamefont {U.~D.}\ \bibnamefont
  {Jentschura}}, \bibinfo {author} {\bibfnamefont {S.}~\bibnamefont
  {Kotochigova}}, \bibinfo {author} {\bibfnamefont {E.-O.}\ \bibnamefont
  {Le~Bigot}}, \bibinfo {author} {\bibfnamefont {P.~J.}\ \bibnamefont {Mohr}},
  \ and\ \bibinfo {author} {\bibfnamefont {B.~N.}\ \bibnamefont {Taylor}},\
  }\href {\doibase 10.1103/PhysRevLett.95.163003} {\bibfield  {journal}
  {\bibinfo  {journal} {Phys. Rev. Lett.}\ }\textbf {\bibinfo {volume} {95}},\
  \bibinfo {pages} {163003} (\bibinfo {year} {2005})}\BibitemShut {NoStop}%
\bibitem [{\citenamefont {Pachucki}\ and\ \citenamefont
  {Jentschura}(2003)}]{Pachucki_2003}%
  \BibitemOpen
  \bibfield  {author} {\bibinfo {author} {\bibfnamefont {K.}~\bibnamefont
  {Pachucki}}\ and\ \bibinfo {author} {\bibfnamefont {U.~D.}\ \bibnamefont
  {Jentschura}},\ }\href@noop {} {\bibfield  {journal} {\bibinfo  {journal}
  {Phys. Rev. Lett.}\ }\textbf {\bibinfo {volume} {91}},\ \bibinfo {pages}
  {113005} (\bibinfo {year} {2003})}\BibitemShut {NoStop}%
\bibitem [{\citenamefont {Yerokhin}\ and\ \citenamefont
  {Shabaev}(2015)}]{YerokhinShabaev_2015}%
  \BibitemOpen
  \bibfield  {author} {\bibinfo {author} {\bibfnamefont {V.~A.}\ \bibnamefont
  {Yerokhin}}\ and\ \bibinfo {author} {\bibfnamefont {V.~M.}\ \bibnamefont
  {Shabaev}},\ }\href@noop {} {\bibfield  {journal} {\bibinfo  {journal} {J.
  Phys. Chem. Ref. Data}\ }\textbf {\bibinfo {volume} {44}},\ \bibinfo {pages}
  {033103} (\bibinfo {year} {2015})}\BibitemShut {NoStop}%
\bibitem [{\citenamefont {Gumberidze}\ \emph {et~al.}(2011)\citenamefont
  {Gumberidze}, \citenamefont {St{\"o}hlker}, \citenamefont {Bana{\'s}},
  \citenamefont {Beyer}, \citenamefont {Brandau}, \citenamefont {Br{\"a}uning},
  \citenamefont {Geyer}, \citenamefont {Hagmann}, \citenamefont {Hess},
  \citenamefont {Indelicato} \emph {et~al.}}]{Gumberidze}%
  \BibitemOpen
  \bibfield  {author} {\bibinfo {author} {\bibfnamefont {A.}~\bibnamefont
  {Gumberidze}}, \bibinfo {author} {\bibfnamefont {T.}~\bibnamefont
  {St{\"o}hlker}}, \bibinfo {author} {\bibfnamefont {D.}~\bibnamefont
  {Bana{\'s}}}, \bibinfo {author} {\bibfnamefont {H.}~\bibnamefont {Beyer}},
  \bibinfo {author} {\bibfnamefont {C.}~\bibnamefont {Brandau}}, \bibinfo
  {author} {\bibfnamefont {H.}~\bibnamefont {Br{\"a}uning}}, \bibinfo {author}
  {\bibfnamefont {S.}~\bibnamefont {Geyer}}, \bibinfo {author} {\bibfnamefont
  {S.}~\bibnamefont {Hagmann}}, \bibinfo {author} {\bibfnamefont
  {S.}~\bibnamefont {Hess}}, \bibinfo {author} {\bibfnamefont {P.}~\bibnamefont
  {Indelicato}},  \emph {et~al.},\ }\href@noop {} {\bibfield  {journal}
  {\bibinfo  {journal} {Hyperfine Interact.}\ }\textbf {\bibinfo {volume}
  {199}},\ \bibinfo {pages} {59} (\bibinfo {year} {2011})}\BibitemShut
  {NoStop}%
\bibitem [{\citenamefont {Klahn}\ and\ \citenamefont {Bingel}(1977)}]{Klahn}%
  \BibitemOpen
  \bibfield  {author} {\bibinfo {author} {\bibfnamefont {B.}~\bibnamefont
  {Klahn}}\ and\ \bibinfo {author} {\bibfnamefont {W.}~\bibnamefont {Bingel}},\
  }\href@noop {} {\bibfield  {journal} {\bibinfo  {journal} {Theor. Chim.
  Acta}\ }\textbf {\bibinfo {volume} {44}},\ \bibinfo {pages} {27} (\bibinfo
  {year} {1977})}\BibitemShut {NoStop}%
\bibitem [{\citenamefont {Klahn}\ and\ \citenamefont {Bingel}(1978)}]{Klahn2}%
  \BibitemOpen
  \bibfield  {author} {\bibinfo {author} {\bibfnamefont {B.}~\bibnamefont
  {Klahn}}\ and\ \bibinfo {author} {\bibfnamefont {W.}~\bibnamefont {Bingel}},\
  }\href@noop {} {\bibfield  {journal} {\bibinfo  {journal} {Int. J. Quantum
  Chem.}\ }\textbf {\bibinfo {volume} {11}},\ \bibinfo {pages} {943} (\bibinfo
  {year} {1978})}\BibitemShut {NoStop}%
\bibitem [{\citenamefont {Schwartz}(2006)}]{Schwartz}%
  \BibitemOpen
  \bibfield  {author} {\bibinfo {author} {\bibfnamefont {C.}~\bibnamefont
  {Schwartz}},\ }\href@noop {} {\bibfield  {journal} {\bibinfo  {journal} {Int.
  J. Mod. Phys. E--Nucl. Phys.}\ }\textbf {\bibinfo {volume} {15}},\ \bibinfo
  {pages} {877} (\bibinfo {year} {2006})}\BibitemShut {NoStop}%
\bibitem [{\citenamefont {Nakashima}\ and\ \citenamefont
  {Nakatsuji}(2007)}]{Nakashima}%
  \BibitemOpen
  \bibfield  {author} {\bibinfo {author} {\bibfnamefont {H.}~\bibnamefont
  {Nakashima}}\ and\ \bibinfo {author} {\bibfnamefont {H.}~\bibnamefont
  {Nakatsuji}},\ }\href@noop {} {\bibfield  {journal} {\bibinfo  {journal} {J.
  Chem. Phys.}\ }\textbf {\bibinfo {volume} {127}},\ \bibinfo {pages} {224104}
  (\bibinfo {year} {2007})}\BibitemShut {NoStop}%
\bibitem [{\citenamefont {Drake}\ and\ \citenamefont {Yan}(1992)}]{Drake_Yan}%
  \BibitemOpen
  \bibfield  {author} {\bibinfo {author} {\bibfnamefont {G.~W.~F.}\
  \bibnamefont {Drake}}\ and\ \bibinfo {author} {\bibfnamefont {Z.-C.}\
  \bibnamefont {Yan}},\ }\href@noop {} {\bibfield  {journal} {\bibinfo
  {journal} {Phys. Rev. A}\ }\textbf {\bibinfo {volume} {46}},\ \bibinfo
  {pages} {2378} (\bibinfo {year} {1992})}\BibitemShut {NoStop}%
\bibitem [{\citenamefont {Korobov}(2000)}]{Korobov}%
  \BibitemOpen
  \bibfield  {author} {\bibinfo {author} {\bibfnamefont {V.~I.}\ \bibnamefont
  {Korobov}},\ }\href@noop {} {\bibfield  {journal} {\bibinfo  {journal}
  {Phys.\ Rev.\ A}\ }\textbf {\bibinfo {volume} {61}},\ \bibinfo {pages}
  {064503} (\bibinfo {year} {2000})}\BibitemShut {NoStop}%
\bibitem [{\citenamefont {Bailey}(1993)}]{Bailey19}%
  \BibitemOpen
  \bibfield  {author} {\bibinfo {author} {\bibfnamefont {D.}~\bibnamefont
  {Bailey}},\ }\href@noop {} {\bibfield  {journal} {\bibinfo  {journal} {ACM
  Trans. Math. Softw.}\ }\textbf {\bibinfo {volume} {19}},\ \bibinfo {pages}
  {288} (\bibinfo {year} {1993})}\BibitemShut {NoStop}%
\bibitem [{\citenamefont {Bailey}(1995)}]{Bailey21}%
  \BibitemOpen
  \bibfield  {author} {\bibinfo {author} {\bibfnamefont {D.}~\bibnamefont
  {Bailey}},\ }\href@noop {} {\bibfield  {journal} {\bibinfo  {journal} {ACM
  Trans. Math. Softw.}\ }\textbf {\bibinfo {volume} {379}},\ \bibinfo {pages}
  {21} (\bibinfo {year} {1995})},\ \bibinfo {note} {see also
  http://crd-legacy.lbl.gov/~dhbailey/mpdist/}\BibitemShut {NoStop}%
\bibitem [{\citenamefont {Drake}\ \emph {et~al.}(2002)\citenamefont {Drake},
  \citenamefont {Cassar},\ and\ \citenamefont {Nistor}}]{Nistor}%
  \BibitemOpen
  \bibfield  {author} {\bibinfo {author} {\bibfnamefont {G.}~\bibnamefont
  {Drake}}, \bibinfo {author} {\bibfnamefont {M.}~\bibnamefont {Cassar}}, \
  and\ \bibinfo {author} {\bibfnamefont {R.}~\bibnamefont {Nistor}},\
  }\href@noop {} {\bibfield  {journal} {\bibinfo  {journal} {Phys. Rev. A}\
  }\textbf {\bibinfo {volume} {65}},\ \bibinfo {pages} {054501} (\bibinfo
  {year} {2002})}\BibitemShut {NoStop}%
\bibitem [{\citenamefont {Wang}\ \emph {et~al.}(2012)\citenamefont {Wang},
  \citenamefont {Yan}, \citenamefont {Qiao},\ and\ \citenamefont
  {Drake}}]{Wang}%
  \BibitemOpen
  \bibfield  {author} {\bibinfo {author} {\bibfnamefont {L.}~\bibnamefont
  {Wang}}, \bibinfo {author} {\bibfnamefont {Z.}~\bibnamefont {Yan}}, \bibinfo
  {author} {\bibfnamefont {H.}~\bibnamefont {Qiao}}, \ and\ \bibinfo {author}
  {\bibfnamefont {G.}~\bibnamefont {Drake}},\ }\href@noop {} {\bibfield
  {journal} {\bibinfo  {journal} {Phys. Rev. A}\ }\textbf {\bibinfo {volume}
  {85}},\ \bibinfo {pages} {052513} (\bibinfo {year} {2012})}\BibitemShut
  {NoStop}%
\bibitem [{\citenamefont {Drake}\ and\ \citenamefont
  {Martin}(1998)}]{DrakeMartin_1998}%
  \BibitemOpen
  \bibfield  {author} {\bibinfo {author} {\bibfnamefont {G.~W.~F.}\
  \bibnamefont {Drake}}\ and\ \bibinfo {author} {\bibfnamefont {W.~C.}\
  \bibnamefont {Martin}},\ }\href {\doibase 10.1139/cjp-76-9-679} {\bibfield
  {journal} {\bibinfo  {journal} {Can. J. Phys.}\ }\textbf {\bibinfo {volume}
  {76}},\ \bibinfo {pages} {679} (\bibinfo {year} {1998})}\BibitemShut
  {NoStop}%
\bibitem [{\citenamefont {Morton}\ \emph {et~al.}(2006)\citenamefont {Morton},
  \citenamefont {Wu},\ and\ \citenamefont {Drake}}]{DrakeMorton_2006}%
  \BibitemOpen
  \bibfield  {author} {\bibinfo {author} {\bibfnamefont {D.~C.}\ \bibnamefont
  {Morton}}, \bibinfo {author} {\bibfnamefont {Q.-X.}\ \bibnamefont {Wu}}, \
  and\ \bibinfo {author} {\bibfnamefont {G.~W.~F.}\ \bibnamefont {Drake}},\
  }\href {\doibase 10.1139/P06-009} {\bibfield  {journal} {\bibinfo  {journal}
  {Can. J. Phys.}\ }\textbf {\bibinfo {volume} {84}},\ \bibinfo {pages} {83}
  (\bibinfo {year} {2006})}\BibitemShut {NoStop}%
\bibitem [{\citenamefont {Yan}\ \emph {et~al.}(2008{\natexlab{a}})\citenamefont
  {Yan}, \citenamefont {Noertershaeuser},\ and\ \citenamefont
  {Drake}}]{DrakeYan_2008}%
  \BibitemOpen
  \bibfield  {author} {\bibinfo {author} {\bibfnamefont {Z.~C.}\ \bibnamefont
  {Yan}}, \bibinfo {author} {\bibfnamefont {W.}~\bibnamefont
  {Noertershaeuser}}, \ and\ \bibinfo {author} {\bibfnamefont {G.~W.~F.}\
  \bibnamefont {Drake}},\ }\href@noop {} {\bibfield  {journal} {\bibinfo
  {journal} {{Phys. Rev. Lett.}}\ }\textbf {\bibinfo {volume} {{100}}},\
  \bibinfo {pages} {{243002}} (\bibinfo {year}
  {{2008}}{\natexlab{a}})}\BibitemShut {NoStop}%
\bibitem [{\citenamefont {Yerokhin}\ and\ \citenamefont
  {Pachucki}(2010)}]{Yerokhin_2010}%
  \BibitemOpen
  \bibfield  {author} {\bibinfo {author} {\bibfnamefont {V.~A.}\ \bibnamefont
  {Yerokhin}}\ and\ \bibinfo {author} {\bibfnamefont {K.}~\bibnamefont
  {Pachucki}},\ }\href@noop {} {\bibfield  {journal} {\bibinfo  {journal}
  {Phys. Rev. A}\ }\textbf {\bibinfo {volume} {81}},\ \bibinfo {pages} {022507}
  (\bibinfo {year} {2010})}\BibitemShut {NoStop}%
\bibitem [{\citenamefont {Drake}(2006)}]{Handbook}%
  \BibitemOpen
  \bibfield  {author} {\bibinfo {author} {\bibfnamefont {G.~W.~F.}\
  \bibnamefont {Drake}},\ }in\ \href@noop {} {\emph {\bibinfo {booktitle}
  {Handbook of Atomic, Molecular, and Optical Physics}}},\ \bibinfo {editor}
  {edited by\ \bibinfo {editor} {\bibfnamefont {G.~W.~F.}\ \bibnamefont
  {Drake}}}\ (\bibinfo  {publisher} {Springer-Verlag, New York},\ \bibinfo
  {year} {2006})\ Chap.~\bibinfo {chapter} {11}\BibitemShut {NoStop}%
\bibitem [{\citenamefont {Drake}(1988)}]{Drake_1987}%
  \BibitemOpen
  \bibfield  {author} {\bibinfo {author} {\bibfnamefont {G.~W.~F.}\
  \bibnamefont {Drake}},\ }\href@noop {} {\bibfield  {journal} {\bibinfo
  {journal} {Can. J. Phys.}\ }\textbf {\bibinfo {volume} {66}},\ \bibinfo
  {pages} {586} (\bibinfo {year} {1988})}\BibitemShut {NoStop}%
\bibitem [{\citenamefont {Artemyev}\ \emph {et~al.}(2005)\citenamefont
  {Artemyev}, \citenamefont {Shabaev}, \citenamefont {Yerokhin}, \citenamefont
  {Plunien},\ and\ \citenamefont {Soff}}]{Artemeyev_2005}%
  \BibitemOpen
  \bibfield  {author} {\bibinfo {author} {\bibfnamefont {A.~N.}\ \bibnamefont
  {Artemyev}}, \bibinfo {author} {\bibfnamefont {V.~M.}\ \bibnamefont
  {Shabaev}}, \bibinfo {author} {\bibfnamefont {V.~A.}\ \bibnamefont
  {Yerokhin}}, \bibinfo {author} {\bibfnamefont {G.}~\bibnamefont {Plunien}}, \
  and\ \bibinfo {author} {\bibfnamefont {G.}~\bibnamefont {Soff}},\ }\href@noop
  {} {\bibfield  {journal} {\bibinfo  {journal} {Phys. Rev. A}\ }\textbf
  {\bibinfo {volume} {71}},\ \bibinfo {pages} {062104} (\bibinfo {year}
  {2005})}\BibitemShut {NoStop}%
\bibitem [{\citenamefont {Chantler}\ \emph {et~al.}(2010)\citenamefont
  {Chantler}, \citenamefont {Lowe},\ and\ \citenamefont
  {Grant}}]{Chantler2010}%
  \BibitemOpen
  \bibfield  {author} {\bibinfo {author} {\bibfnamefont {C.~T.}\ \bibnamefont
  {Chantler}}, \bibinfo {author} {\bibfnamefont {J.}~\bibnamefont {Lowe}}, \
  and\ \bibinfo {author} {\bibfnamefont {I.}~\bibnamefont {Grant}},\
  }\href@noop {} {\bibfield  {journal} {\bibinfo  {journal} {Phys. Rev. A}\
  }\textbf {\bibinfo {volume} {82}},\ \bibinfo {pages} {052505} (\bibinfo
  {year} {2010})}\BibitemShut {NoStop}%
\bibitem [{\citenamefont {Fischer}(2014)}]{atoms2010001}%
  \BibitemOpen
  \bibfield  {author} {\bibinfo {author} {\bibfnamefont {C.~F.}\ \bibnamefont
  {Fischer}},\ }\href {\doibase 10.3390/atoms2010001} {\bibfield  {journal}
  {\bibinfo  {journal} {Atoms}\ }\textbf {\bibinfo {volume} {2}},\ \bibinfo
  {pages} {1} (\bibinfo {year} {2014})}\BibitemShut {NoStop}%
\bibitem [{\citenamefont {Tr{\"a}bert}(2014)}]{atoms2010015}%
  \BibitemOpen
  \bibfield  {author} {\bibinfo {author} {\bibfnamefont {E.}~\bibnamefont
  {Tr{\"a}bert}},\ }\href {\doibase 10.3390/atoms2010015} {\bibfield  {journal}
  {\bibinfo  {journal} {Atoms}\ }\textbf {\bibinfo {volume} {2}},\ \bibinfo
  {pages} {15} (\bibinfo {year} {2014})}\BibitemShut {NoStop}%
\bibitem [{\citenamefont {Kramida}(2014)}]{atoms2020086}%
  \BibitemOpen
  \bibfield  {author} {\bibinfo {author} {\bibfnamefont {A.}~\bibnamefont
  {Kramida}},\ }\href {\doibase 10.3390/atoms2020086} {\bibfield  {journal}
  {\bibinfo  {journal} {Atoms}\ }\textbf {\bibinfo {volume} {2}},\ \bibinfo
  {pages} {86} (\bibinfo {year} {2014})}\BibitemShut {NoStop}%
\bibitem [{\citenamefont {Ekman}\ \emph {et~al.}(2014)\citenamefont {Ekman},
  \citenamefont {Godefroid},\ and\ \citenamefont {Hartman}}]{atoms2020215}%
  \BibitemOpen
  \bibfield  {author} {\bibinfo {author} {\bibfnamefont {J.}~\bibnamefont
  {Ekman}}, \bibinfo {author} {\bibfnamefont {M.~R.}\ \bibnamefont
  {Godefroid}}, \ and\ \bibinfo {author} {\bibfnamefont {H.}~\bibnamefont
  {Hartman}},\ }\href {\doibase 10.3390/atoms2020215} {\bibfield  {journal}
  {\bibinfo  {journal} {Atoms}\ }\textbf {\bibinfo {volume} {2}},\ \bibinfo
  {pages} {215} (\bibinfo {year} {2014})}\BibitemShut {NoStop}%
\bibitem [{\citenamefont {Kelleher}(2014)}]{atoms2040382}%
  \BibitemOpen
  \bibfield  {author} {\bibinfo {author} {\bibfnamefont {D.~E.}\ \bibnamefont
  {Kelleher}},\ }\href {\doibase 10.3390/atoms2040382} {\bibfield  {journal}
  {\bibinfo  {journal} {Atoms}\ }\textbf {\bibinfo {volume} {2}},\ \bibinfo
  {pages} {382} (\bibinfo {year} {2014})}\BibitemShut {NoStop}%
\bibitem [{\citenamefont {Safronova}\ \emph
  {et~al.}(2014{\natexlab{a}})\citenamefont {Safronova}, \citenamefont {Dzuba},
  \citenamefont {Flambaum}, \citenamefont {Safronova}, \citenamefont {Porsev},\
  and\ \citenamefont {Kozlov}}]{Safronova}%
  \BibitemOpen
  \bibfield  {author} {\bibinfo {author} {\bibfnamefont {M.~S.}\ \bibnamefont
  {Safronova}}, \bibinfo {author} {\bibfnamefont {V.~A.}\ \bibnamefont
  {Dzuba}}, \bibinfo {author} {\bibfnamefont {V.~V.}\ \bibnamefont {Flambaum}},
  \bibinfo {author} {\bibfnamefont {U.~I.}\ \bibnamefont {Safronova}}, \bibinfo
  {author} {\bibfnamefont {S.~G.}\ \bibnamefont {Porsev}}, \ and\ \bibinfo
  {author} {\bibfnamefont {M.~G.}\ \bibnamefont {Kozlov}},\ }\href@noop {}
  {\bibfield  {journal} {\bibinfo  {journal} {Phys. Rev. A}\ }\textbf {\bibinfo
  {volume} {90}},\ \bibinfo {pages} {042513} (\bibinfo {year}
  {2014}{\natexlab{a}})}\BibitemShut {NoStop}%
\bibitem [{\citenamefont {Safronova}\ \emph
  {et~al.}(2014{\natexlab{b}})\citenamefont {Safronova}, \citenamefont {Dzuba},
  \citenamefont {Flambaum}, \citenamefont {Safronova}, \citenamefont {Porsev},\
  and\ \citenamefont {Kozlov}}]{Safronova2}%
  \BibitemOpen
  \bibfield  {author} {\bibinfo {author} {\bibfnamefont {M.~S.}\ \bibnamefont
  {Safronova}}, \bibinfo {author} {\bibfnamefont {V.~A.}\ \bibnamefont
  {Dzuba}}, \bibinfo {author} {\bibfnamefont {V.~V.}\ \bibnamefont {Flambaum}},
  \bibinfo {author} {\bibfnamefont {U.~I.}\ \bibnamefont {Safronova}}, \bibinfo
  {author} {\bibfnamefont {S.~G.}\ \bibnamefont {Porsev}}, \ and\ \bibinfo
  {author} {\bibfnamefont {M.~G.}\ \bibnamefont {Kozlov}},\ }\href@noop {}
  {\bibfield  {journal} {\bibinfo  {journal} {Phys. Rev. A}\ }\textbf {\bibinfo
  {volume} {90}},\ \bibinfo {pages} {052509} (\bibinfo {year}
  {2014}{\natexlab{b}})}\BibitemShut {NoStop}%
\bibitem [{\citenamefont {K\'allay}\ \emph {et~al.}(2011)\citenamefont
  {K\'allay}, \citenamefont {Nataraj}, \citenamefont {Sahoo}, \citenamefont
  {Das},\ and\ \citenamefont {Visscher}}]{Sahoo}%
  \BibitemOpen
  \bibfield  {author} {\bibinfo {author} {\bibfnamefont {M.}~\bibnamefont
  {K\'allay}}, \bibinfo {author} {\bibfnamefont {H.~S.}\ \bibnamefont
  {Nataraj}}, \bibinfo {author} {\bibfnamefont {B.~K.}\ \bibnamefont {Sahoo}},
  \bibinfo {author} {\bibfnamefont {B.~P.}\ \bibnamefont {Das}}, \ and\
  \bibinfo {author} {\bibfnamefont {L.}~\bibnamefont {Visscher}},\ }\href@noop
  {} {\bibfield  {journal} {\bibinfo  {journal} {Phys. Rev. A}\ }\textbf
  {\bibinfo {volume} {83}},\ \bibinfo {pages} {030503} (\bibinfo {year}
  {2011})}\BibitemShut {NoStop}%
\bibitem [{\citenamefont {Fischer}(2009)}]{fischer2009evaluating}%
  \BibitemOpen
  \bibfield  {author} {\bibinfo {author} {\bibfnamefont {C.~F.}\ \bibnamefont
  {Fischer}},\ }\href@noop {} {\bibfield  {journal} {\bibinfo  {journal} {Phys.
  Scr.}\ }\textbf {\bibinfo {volume} {2009}},\ \bibinfo {pages} {014019}
  (\bibinfo {year} {2009})}\BibitemShut {NoStop}%
\bibitem [{\citenamefont {Born}\ and\ \citenamefont
  {Oppenheimer}(1927)}]{27BoOp}%
  \BibitemOpen
  \bibfield  {author} {\bibinfo {author} {\bibfnamefont {M.}~\bibnamefont
  {Born}}\ and\ \bibinfo {author} {\bibfnamefont {J.~R.}\ \bibnamefont
  {Oppenheimer}},\ }\href@noop {} {\bibfield  {journal} {\bibinfo  {journal}
  {Ann. Phys.}\ }\textbf {\bibinfo {volume} {84}},\ \bibinfo {pages} {457}
  (\bibinfo {year} {1927})}\BibitemShut {NoStop}%
\bibitem [{\citenamefont {Born}\ and\ \citenamefont {Huang}(1954)}]{1954BoHu}%
  \BibitemOpen
  \bibfield  {author} {\bibinfo {author} {\bibfnamefont {M.}~\bibnamefont
  {Born}}\ and\ \bibinfo {author} {\bibfnamefont {K.}~\bibnamefont {Huang}},\
  }\href@noop {} {\emph {\bibinfo {title} {Dynamical Theory of Crystal
  Lattice}}}\ (\bibinfo  {publisher} {Clarendon Press},\ \bibinfo {year}
  {1954})\BibitemShut {NoStop}%
\bibitem [{\citenamefont {Zobov}\ \emph {et~al.}(1996)\citenamefont {Zobov},
  \citenamefont {Polyansky}, \citenamefont {{Le Sueur}},\ and\ \citenamefont
  {Tennyson}}]{jt186}%
  \BibitemOpen
  \bibfield  {author} {\bibinfo {author} {\bibfnamefont {N.~F.}\ \bibnamefont
  {Zobov}}, \bibinfo {author} {\bibfnamefont {O.~L.}\ \bibnamefont
  {Polyansky}}, \bibinfo {author} {\bibfnamefont {C.~R.}\ \bibnamefont {{Le
  Sueur}}}, \ and\ \bibinfo {author} {\bibfnamefont {J.}~\bibnamefont
  {Tennyson}},\ }\href@noop {} {\bibfield  {journal} {\bibinfo  {journal}
  {Chem. Phys. Lett.}\ }\textbf {\bibinfo {volume} {260}},\ \bibinfo {pages}
  {381} (\bibinfo {year} {1996})}\BibitemShut {NoStop}%
\bibitem [{\citenamefont {Polyansky}\ and\ \citenamefont
  {Tennyson}(1999)}]{jt236}%
  \BibitemOpen
  \bibfield  {author} {\bibinfo {author} {\bibfnamefont {O.~L.}\ \bibnamefont
  {Polyansky}}\ and\ \bibinfo {author} {\bibfnamefont {J.}~\bibnamefont
  {Tennyson}},\ }\href@noop {} {\bibfield  {journal} {\bibinfo  {journal} {J.
  Chem. Phys.}\ }\textbf {\bibinfo {volume} {110}},\ \bibinfo {pages} {5056}
  (\bibinfo {year} {1999})}\BibitemShut {NoStop}%
\bibitem [{\citenamefont {Diniz}\ \emph {et~al.}(2013)\citenamefont {Diniz},
  \citenamefont {Mohallem}, \citenamefont {Alijah}, \citenamefont {Pavanello},
  \citenamefont {Adamowicz}, \citenamefont {Polyansky},\ and\ \citenamefont
  {Tennyson}}]{jt566}%
  \BibitemOpen
  \bibfield  {author} {\bibinfo {author} {\bibfnamefont {L.~G.}\ \bibnamefont
  {Diniz}}, \bibinfo {author} {\bibfnamefont {J.~R.}\ \bibnamefont {Mohallem}},
  \bibinfo {author} {\bibfnamefont {A.}~\bibnamefont {Alijah}}, \bibinfo
  {author} {\bibfnamefont {M.}~\bibnamefont {Pavanello}}, \bibinfo {author}
  {\bibfnamefont {L.}~\bibnamefont {Adamowicz}}, \bibinfo {author}
  {\bibfnamefont {O.~L.}\ \bibnamefont {Polyansky}}, \ and\ \bibinfo {author}
  {\bibfnamefont {J.}~\bibnamefont {Tennyson}},\ }\href {\doibase
  10.1103/PhysRevA.88.032506} {\bibfield  {journal} {\bibinfo  {journal} {Phys.
  Rev. A}\ }\textbf {\bibinfo {volume} {88}},\ \bibinfo {pages} {032506}
  (\bibinfo {year} {2013})}\BibitemShut {NoStop}%
\bibitem [{\citenamefont {E.~M\'tyus}\ and\ \citenamefont
  {Cs\'asz\'ar}(2014)}]{Matyus14}%
  \BibitemOpen
  \bibfield  {author} {\bibinfo {author} {\bibfnamefont {T.~S.}\ \bibnamefont
  {E.~M\'tyus}}\ and\ \bibinfo {author} {\bibfnamefont {A.~G.}\ \bibnamefont
  {Cs\'asz\'ar}},\ }\href {\doibase http://dx.doi.org/10.1063/1.4897566}
  {\bibfield  {journal} {\bibinfo  {journal} {The Journal of Chemical Physics}\
  }\textbf {\bibinfo {volume} {141}},\ \bibinfo {pages} {154111} (\bibinfo
  {year} {2014})}\BibitemShut {NoStop}%
\bibitem [{\citenamefont {Cs\'asz\'ar}\ \emph {et~al.}(2000)\citenamefont
  {Cs\'asz\'ar}, \citenamefont {Allen}, \citenamefont {Yamaguchi},\ and\
  \citenamefont {{Schaefer III}}}]{CMS}%
  \BibitemOpen
  \bibfield  {author} {\bibinfo {author} {\bibfnamefont {A.~G.}\ \bibnamefont
  {Cs\'asz\'ar}}, \bibinfo {author} {\bibfnamefont {W.~D.}\ \bibnamefont
  {Allen}}, \bibinfo {author} {\bibfnamefont {Y.}~\bibnamefont {Yamaguchi}}, \
  and\ \bibinfo {author} {\bibfnamefont {H.~F.}\ \bibnamefont {{Schaefer
  III}}},\ }in\ \href@noop {} {\emph {\bibinfo {booktitle} {Computational
  Molecular Spectroscopy}}}\ (\bibinfo  {publisher} {Wiley, New York},\
  \bibinfo {year} {2000})\ pp.\ \bibinfo {pages} {15--68}\BibitemShut {NoStop}%
\bibitem [{\citenamefont {Cs\'asz\'ar}\ \emph {et~al.}(2001)\citenamefont
  {Cs\'asz\'ar}, \citenamefont {Tarczay}, \citenamefont {Leininger},
  \citenamefont {Polyansky}, \citenamefont {Tennyson},\ and\ \citenamefont
  {Allen}}]{Dream}%
  \BibitemOpen
  \bibfield  {author} {\bibinfo {author} {\bibfnamefont {A.~G.}\ \bibnamefont
  {Cs\'asz\'ar}}, \bibinfo {author} {\bibfnamefont {G.}~\bibnamefont
  {Tarczay}}, \bibinfo {author} {\bibfnamefont {M.~L.}\ \bibnamefont
  {Leininger}}, \bibinfo {author} {\bibfnamefont {O.~L.}\ \bibnamefont
  {Polyansky}}, \bibinfo {author} {\bibfnamefont {J.}~\bibnamefont {Tennyson}},
  \ and\ \bibinfo {author} {\bibfnamefont {W.~D.}\ \bibnamefont {Allen}},\ }in\
  \href@noop {} {\emph {\bibinfo {booktitle} {In: Spectroscopy from space}}}\
  (\bibinfo  {publisher} {Kluwer, Dordrecht},\ \bibinfo {year} {2001})\ pp.\
  \bibinfo {pages} {17--339}\BibitemShut {NoStop}%
\bibitem [{\citenamefont {Yamaguchi}\ \emph {et~al.}(1994)\citenamefont
  {Yamaguchi}, \citenamefont {Osamura}, \citenamefont {Goddard},\ and\
  \citenamefont {Schaefer}}]{94YaOsGoSc}%
  \BibitemOpen
  \bibfield  {author} {\bibinfo {author} {\bibfnamefont {Y.}~\bibnamefont
  {Yamaguchi}}, \bibinfo {author} {\bibfnamefont {Y.}~\bibnamefont {Osamura}},
  \bibinfo {author} {\bibfnamefont {J.~D.}\ \bibnamefont {Goddard}}, \ and\
  \bibinfo {author} {\bibfnamefont {H.~F.}\ \bibnamefont {Schaefer}},\
  }\href@noop {} {\emph {\bibinfo {title} {A New Dimension to Quantum
  Chemistry: Analytic Derivative Methods in Ab Initio Molecular Electronic
  Structure Theory (International Series of Monographs on Chemistry)}}}\
  (\bibinfo  {publisher} {Oxford University Press},\ \bibinfo {year}
  {1994})\BibitemShut {NoStop}%
\bibitem [{\citenamefont {Murrell}\ \emph {et~al.}(1984)\citenamefont
  {Murrell}, \citenamefont {Carter}, \citenamefont {Farantos}, \citenamefont
  {Huxley},\ and\ \citenamefont {Varandas}}]{84MuCaFaHu}%
  \BibitemOpen
  \bibfield  {author} {\bibinfo {author} {\bibfnamefont {J.~N.}\ \bibnamefont
  {Murrell}}, \bibinfo {author} {\bibfnamefont {S.}~\bibnamefont {Carter}},
  \bibinfo {author} {\bibfnamefont {S.~C.}\ \bibnamefont {Farantos}}, \bibinfo
  {author} {\bibfnamefont {P.}~\bibnamefont {Huxley}}, \ and\ \bibinfo {author}
  {\bibfnamefont {A.~J.~C.}\ \bibnamefont {Varandas}},\ }\href@noop {} {\emph
  {\bibinfo {title} {Molecular Potential Energy Surfaces}}}\ (\bibinfo
  {publisher} {Wiley},\ \bibinfo {address} {New York},\ \bibinfo {year}
  {1984})\BibitemShut {NoStop}%
\bibitem [{\citenamefont {Braams}\ and\ \citenamefont
  {Bowman}(2009)}]{braams2009}%
  \BibitemOpen
  \bibfield  {author} {\bibinfo {author} {\bibfnamefont {B.~J.}\ \bibnamefont
  {Braams}}\ and\ \bibinfo {author} {\bibfnamefont {J.~M.}\ \bibnamefont
  {Bowman}},\ }\href@noop {} {\bibfield  {journal} {\bibinfo  {journal} {Int.
  Rev. Phys. Chem.}\ }\textbf {\bibinfo {volume} {28}},\ \bibinfo {pages} {577}
  (\bibinfo {year} {2009})}\BibitemShut {NoStop}%
\bibitem [{\citenamefont {Tarczay}\ \emph {et~al.}(2001)\citenamefont
  {Tarczay}, \citenamefont {Cs\'asz\'ar}, \citenamefont {Klopper},\ and\
  \citenamefont {Quiney}}]{magiccube}%
  \BibitemOpen
  \bibfield  {author} {\bibinfo {author} {\bibfnamefont {G.}~\bibnamefont
  {Tarczay}}, \bibinfo {author} {\bibfnamefont {A.~G.}\ \bibnamefont
  {Cs\'asz\'ar}}, \bibinfo {author} {\bibfnamefont {W.}~\bibnamefont
  {Klopper}}, \ and\ \bibinfo {author} {\bibfnamefont {H.~M.}\ \bibnamefont
  {Quiney}},\ }\href {\doibase 10.1080/00268970110073907} {\bibfield  {journal}
  {\bibinfo  {journal} {Mol. Phys.}\ }\textbf {\bibinfo {volume} {99}},\
  \bibinfo {pages} {1769} (\bibinfo {year} {2001})}\BibitemShut {NoStop}%
\bibitem [{\citenamefont {Dunning}(1989)}]{Dunning1989}%
  \BibitemOpen
  \bibfield  {author} {\bibinfo {author} {\bibfnamefont {T.~H.}\ \bibnamefont
  {Dunning}, \bibfnamefont {Jr.}},\ }\href@noop {} {\bibfield  {journal}
  {\bibinfo  {journal} {J. Chem. Phys.}\ }\textbf {\bibinfo {volume} {90}},\
  \bibinfo {pages} {1007} (\bibinfo {year} {1989})}\BibitemShut {NoStop}%
\bibitem [{\citenamefont {Cs\'asz\'ar}\ and\ \citenamefont
  {Allen}(1996)}]{CsaszarAllen1996}%
  \BibitemOpen
  \bibfield  {author} {\bibinfo {author} {\bibfnamefont {A.~G.}\ \bibnamefont
  {Cs\'asz\'ar}}\ and\ \bibinfo {author} {\bibfnamefont {W.~D.}\ \bibnamefont
  {Allen}},\ }\href@noop {} {\bibfield  {journal} {\bibinfo  {journal} {J.
  Chem. Phys.}\ }\textbf {\bibinfo {volume} {104}},\ \bibinfo {pages} {2746}
  (\bibinfo {year} {1996})}\BibitemShut {NoStop}%
\bibitem [{\citenamefont {{Pyykk\"o}}\ \emph {et~al.}(2001)\citenamefont
  {{Pyykk\"o}}, \citenamefont {Dyall}, \citenamefont {{Cs\'asz\'ar}},
  \citenamefont {Tarczay}, \citenamefont {Polyansky},\ and\ \citenamefont
  {Tennyson}}]{jt265}%
  \BibitemOpen
  \bibfield  {author} {\bibinfo {author} {\bibfnamefont {P.}~\bibnamefont
  {{Pyykk\"o}}}, \bibinfo {author} {\bibfnamefont {K.~G.}\ \bibnamefont
  {Dyall}}, \bibinfo {author} {\bibfnamefont {A.~G.}\ \bibnamefont
  {{Cs\'asz\'ar}}}, \bibinfo {author} {\bibfnamefont {G.}~\bibnamefont
  {Tarczay}}, \bibinfo {author} {\bibfnamefont {O.~L.}\ \bibnamefont
  {Polyansky}}, \ and\ \bibinfo {author} {\bibfnamefont {J.}~\bibnamefont
  {Tennyson}},\ }\href@noop {} {\bibfield  {journal} {\bibinfo  {journal}
  {Phys. Rev. A}\ }\textbf {\bibinfo {volume} {63}},\ \bibinfo {pages} {024502}
  (\bibinfo {year} {2001})}\BibitemShut {NoStop}%
\bibitem [{\citenamefont {Lodi}\ \emph {et~al.}(2008)\citenamefont {Lodi},
  \citenamefont {Tolchenov}, \citenamefont {Tennyson}, \citenamefont
  {Lynas-Gray}, \citenamefont {Shirin}, \citenamefont {Zobov}, \citenamefont
  {Polyansky}, \citenamefont {{Cs\'asz\'ar}}, \citenamefont {{van Stralen}},\
  and\ \citenamefont {Visscher}}]{jt424}%
  \BibitemOpen
  \bibfield  {author} {\bibinfo {author} {\bibfnamefont {L.}~\bibnamefont
  {Lodi}}, \bibinfo {author} {\bibfnamefont {R.~N.}\ \bibnamefont {Tolchenov}},
  \bibinfo {author} {\bibfnamefont {J.}~\bibnamefont {Tennyson}}, \bibinfo
  {author} {\bibfnamefont {A.~E.}\ \bibnamefont {Lynas-Gray}}, \bibinfo
  {author} {\bibfnamefont {S.~V.}\ \bibnamefont {Shirin}}, \bibinfo {author}
  {\bibfnamefont {N.~F.}\ \bibnamefont {Zobov}}, \bibinfo {author}
  {\bibfnamefont {O.~L.}\ \bibnamefont {Polyansky}}, \bibinfo {author}
  {\bibfnamefont {A.~G.}\ \bibnamefont {{Cs\'asz\'ar}}}, \bibinfo {author}
  {\bibfnamefont {J.}~\bibnamefont {{van Stralen}}}, \ and\ \bibinfo {author}
  {\bibfnamefont {L.}~\bibnamefont {Visscher}},\ }\href@noop {} {\bibfield
  {journal} {\bibinfo  {journal} {J. Chem. Phys.}\ }\textbf {\bibinfo {volume}
  {128}},\ \bibinfo {pages} {044304} (\bibinfo {year} {2008})}\BibitemShut
  {NoStop}%
\bibitem [{\citenamefont {Lodi}\ \emph {et~al.}(2011)\citenamefont {Lodi},
  \citenamefont {Tennyson},\ and\ \citenamefont {Polyansky}}]{jt509}%
  \BibitemOpen
  \bibfield  {author} {\bibinfo {author} {\bibfnamefont {L.}~\bibnamefont
  {Lodi}}, \bibinfo {author} {\bibfnamefont {J.}~\bibnamefont {Tennyson}}, \
  and\ \bibinfo {author} {\bibfnamefont {O.~L.}\ \bibnamefont {Polyansky}},\
  }\href@noop {} {\bibfield  {journal} {\bibinfo  {journal} {J. Chem. Phys.}\
  }\textbf {\bibinfo {volume} {135}},\ \bibinfo {pages} {034113} (\bibinfo
  {year} {2011})}\BibitemShut {NoStop}%
\bibitem [{\citenamefont {Jensen}(2006)}]{06Jensen}%
  \BibitemOpen
  \bibfield  {author} {\bibinfo {author} {\bibfnamefont {F.}~\bibnamefont
  {Jensen}},\ }\href@noop {} {\emph {\bibinfo {title} {Introduction to
  Computational Chemistry}}}\ (\bibinfo  {publisher} {Wiley},\ \bibinfo
  {address} {New York},\ \bibinfo {year} {2006})\BibitemShut {NoStop}%
\bibitem [{\citenamefont {Lipi{\'n}ski}(2002)}]{02Lipins.ai}%
  \BibitemOpen
  \bibfield  {author} {\bibinfo {author} {\bibfnamefont {J.}~\bibnamefont
  {Lipi{\'n}ski}},\ }\href {\doibase
  http://dx.doi.org/10.1016/S0009-2614(02)01186-7} {\bibfield  {journal}
  {\bibinfo  {journal} {Chem. Phys. Lett.}\ }\textbf {\bibinfo {volume}
  {363}},\ \bibinfo {pages} {313 } (\bibinfo {year} {2002})}\BibitemShut
  {NoStop}%
\bibitem [{\citenamefont {Werner}\ \emph {et~al.}(1983)\citenamefont {Werner},
  \citenamefont {Rosmus},\ and\ \citenamefont {Reinsch}}]{83WeRoRe.OH}%
  \BibitemOpen
  \bibfield  {author} {\bibinfo {author} {\bibfnamefont {H.~J.}\ \bibnamefont
  {Werner}}, \bibinfo {author} {\bibfnamefont {P.}~\bibnamefont {Rosmus}}, \
  and\ \bibinfo {author} {\bibfnamefont {E.~A.}\ \bibnamefont {Reinsch}},\
  }\href {\doibase 10.1063/1.445867} {\bibfield  {journal} {\bibinfo  {journal}
  {J. Chem. Phys.}\ }\textbf {\bibinfo {volume} {79}},\ \bibinfo {pages} {905}
  (\bibinfo {year} {1983})}\BibitemShut {NoStop}%
\bibitem [{\citenamefont {Ernzerhof}\ \emph {et~al.}(1992)\citenamefont
  {Ernzerhof}, \citenamefont {Marian},\ and\ \citenamefont
  {Peyerimhoff}}]{92ErMaPe.ai}%
  \BibitemOpen
  \bibfield  {author} {\bibinfo {author} {\bibfnamefont {M.}~\bibnamefont
  {Ernzerhof}}, \bibinfo {author} {\bibfnamefont {C.~M.}\ \bibnamefont
  {Marian}}, \ and\ \bibinfo {author} {\bibfnamefont {S.~D.}\ \bibnamefont
  {Peyerimhoff}},\ }\href {\doibase 10.1002/qua.560430505} {\bibfield
  {journal} {\bibinfo  {journal} {Intern. J. Quantum Chem.}\ }\textbf {\bibinfo
  {volume} {43}},\ \bibinfo {pages} {659} (\bibinfo {year} {1992})}\BibitemShut
  {NoStop}%
\bibitem [{\citenamefont {Lodi}\ and\ \citenamefont {Tennyson}(2010)}]{jt475}%
  \BibitemOpen
  \bibfield  {author} {\bibinfo {author} {\bibfnamefont {L.}~\bibnamefont
  {Lodi}}\ and\ \bibinfo {author} {\bibfnamefont {J.}~\bibnamefont
  {Tennyson}},\ }\href@noop {} {\bibfield  {journal} {\bibinfo  {journal} {J.
  Phys. B: At. Mol. Opt. Phys.}\ }\textbf {\bibinfo {volume} {43}},\ \bibinfo
  {pages} {133001} (\bibinfo {year} {2010})}\BibitemShut {NoStop}%
\bibitem [{\citenamefont {Pindzola}\ and\ \citenamefont
  {Kelly}(1975)}]{75PiKe}%
  \BibitemOpen
  \bibfield  {author} {\bibinfo {author} {\bibfnamefont {M.~S.}\ \bibnamefont
  {Pindzola}}\ and\ \bibinfo {author} {\bibfnamefont {H.~P.}\ \bibnamefont
  {Kelly}},\ }\href {\doibase {10.1103/PhysRevA.12.1419}} {\bibfield  {journal}
  {\bibinfo  {journal} {Phys. Rev. A}\ }\textbf {\bibinfo {volume} {{12}}},\
  \bibinfo {pages} {1419} (\bibinfo {year} {{1975}})}\BibitemShut {NoStop}%
\bibitem [{\citenamefont {Griffin}\ \emph {et~al.}(2009)\citenamefont
  {Griffin}, \citenamefont {Ballance},\ and\ \citenamefont
  {Pindzola}}]{09GrBaPi}%
  \BibitemOpen
  \bibfield  {author} {\bibinfo {author} {\bibfnamefont {D.~C.}\ \bibnamefont
  {Griffin}}, \bibinfo {author} {\bibfnamefont {C.~P.}\ \bibnamefont
  {Ballance}}, \ and\ \bibinfo {author} {\bibfnamefont {M.~S.}\ \bibnamefont
  {Pindzola}},\ }\href {\doibase {10.1103/PhysRevA.80.023420}} {\bibfield
  {journal} {\bibinfo  {journal} {Phys. Rev. A}\ }\textbf {\bibinfo {volume}
  {{80}}},\ \bibinfo {pages} {023420} (\bibinfo {year} {{2009}})}\BibitemShut
  {NoStop}%
\bibitem [{\citenamefont {Adamson}\ \emph {et~al.}(1998)\citenamefont
  {Adamson}, \citenamefont {Zaitsevskii},\ and\ \citenamefont
  {Stepanov}}]{98AdZaSt.methods}%
  \BibitemOpen
  \bibfield  {author} {\bibinfo {author} {\bibfnamefont {S.~O.}\ \bibnamefont
  {Adamson}}, \bibinfo {author} {\bibfnamefont {A.}~\bibnamefont
  {Zaitsevskii}}, \ and\ \bibinfo {author} {\bibfnamefont {N.~F.}\ \bibnamefont
  {Stepanov}},\ }\href@noop {} {\bibfield  {journal} {\bibinfo  {journal} {J.
  Phys. B: At. Mol. Opt. Phys.}\ }\textbf {\bibinfo {volume} {31}},\ \bibinfo
  {pages} {5275} (\bibinfo {year} {1998})}\BibitemShut {NoStop}%
\bibitem [{\citenamefont {Lodi}\ \emph {et~al.}(2015)\citenamefont {Lodi},
  \citenamefont {Yurchenko},\ and\ \citenamefont {Tennyson}}]{jt599}%
  \BibitemOpen
  \bibfield  {author} {\bibinfo {author} {\bibfnamefont {L.}~\bibnamefont
  {Lodi}}, \bibinfo {author} {\bibfnamefont {S.~N.}\ \bibnamefont {Yurchenko}},
  \ and\ \bibinfo {author} {\bibfnamefont {J.}~\bibnamefont {Tennyson}},\
  }\href {\doibase 10.1080/00268976.2015.1029996} {\bibfield  {journal}
  {\bibinfo  {journal} {Mol. Phys.}\ }\textbf {\bibinfo {volume} {113}},\
  \bibinfo {pages} {1559} (\bibinfo {year} {2015})}\BibitemShut {NoStop}%
\bibitem [{\citenamefont {McKemmish}\ \emph {et~al.}(2016)\citenamefont
  {McKemmish}, \citenamefont {Yurchenko},\ and\ \citenamefont
  {Tennyson}}]{jt623}%
  \BibitemOpen
  \bibfield  {author} {\bibinfo {author} {\bibfnamefont {L.~K.}\ \bibnamefont
  {McKemmish}}, \bibinfo {author} {\bibfnamefont {S.~N.}\ \bibnamefont
  {Yurchenko}}, \ and\ \bibinfo {author} {\bibfnamefont {J.}~\bibnamefont
  {Tennyson}},\ }\href@noop {} {\bibfield  {journal} {\bibinfo  {journal} {J.
  Chem. Theory Comput.}\ } (\bibinfo {year} {2016})}\BibitemShut {NoStop}%
\bibitem [{\citenamefont {Tennyson}\ \emph {et~al.}(2016)\citenamefont
  {Tennyson}, \citenamefont {Lodi}, \citenamefont {McKemmish},\ and\
  \citenamefont {Yurchenko}}]{jt632}%
  \BibitemOpen
  \bibfield  {author} {\bibinfo {author} {\bibfnamefont {J.}~\bibnamefont
  {Tennyson}}, \bibinfo {author} {\bibfnamefont {L.}~\bibnamefont {Lodi}},
  \bibinfo {author} {\bibfnamefont {L.~K.}\ \bibnamefont {McKemmish}}, \ and\
  \bibinfo {author} {\bibfnamefont {S.~N.}\ \bibnamefont {Yurchenko}},\
  }\href@noop {} {\bibfield  {journal} {\bibinfo  {journal} {J. Phys. B: At.
  Mol. Opt. Phys.}\ } (\bibinfo {year} {2016})},\ \bibinfo {note} {topical
  Review}\BibitemShut {NoStop}%
\bibitem [{\citenamefont {Jain}\ and\ \citenamefont
  {Gianturco}(1991)}]{91JaGi}%
  \BibitemOpen
  \bibfield  {author} {\bibinfo {author} {\bibfnamefont {A.}~\bibnamefont
  {Jain}}\ and\ \bibinfo {author} {\bibfnamefont {F.}~\bibnamefont
  {Gianturco}},\ }\href {\doibase 10.1088/0953-4075/24/9/018} {\bibfield
  {journal} {\bibinfo  {journal} {J. Phys. B: At. Mol. Opt. Phys.}\ }\textbf
  {\bibinfo {volume} {24}},\ \bibinfo {pages} {2387} (\bibinfo {year}
  {1991})}\BibitemShut {NoStop}%
\bibitem [{\citenamefont {Jones}\ and\ \citenamefont {Tennyson}(2010)}]{jt468}%
  \BibitemOpen
  \bibfield  {author} {\bibinfo {author} {\bibfnamefont {M.}~\bibnamefont
  {Jones}}\ and\ \bibinfo {author} {\bibfnamefont {J.}~\bibnamefont
  {Tennyson}},\ }\href@noop {} {\bibfield  {journal} {\bibinfo  {journal} {J.
  Phys. B: At. Mol. Opt. Phys.}\ }\textbf {\bibinfo {volume} {43}},\ \bibinfo
  {pages} {045101} (\bibinfo {year} {2010})}\BibitemShut {NoStop}%
\bibitem [{\citenamefont {Brigg}\ \emph {et~al.}(2014)\citenamefont {Brigg},
  \citenamefont {Tennyson},\ and\ \citenamefont {Plummer}}]{jt585}%
  \BibitemOpen
  \bibfield  {author} {\bibinfo {author} {\bibfnamefont {W.~J.}\ \bibnamefont
  {Brigg}}, \bibinfo {author} {\bibfnamefont {J.}~\bibnamefont {Tennyson}}, \
  and\ \bibinfo {author} {\bibfnamefont {M.}~\bibnamefont {Plummer}},\
  }\href@noop {} {\bibfield  {journal} {\bibinfo  {journal} {J. Phys. B: At.
  Mol. Opt. Phys.}\ }\textbf {\bibinfo {volume} {47}},\ \bibinfo {pages}
  {185203} (\bibinfo {year} {2014})}\BibitemShut {NoStop}%
\bibitem [{\citenamefont {Jonsson}\ \emph {et~al.}(1997)\citenamefont
  {Jonsson}, \citenamefont {Norman},\ and\ \citenamefont {Agren}}]{jna97}%
  \BibitemOpen
  \bibfield  {author} {\bibinfo {author} {\bibfnamefont {D.}~\bibnamefont
  {Jonsson}}, \bibinfo {author} {\bibfnamefont {P.}~\bibnamefont {Norman}}, \
  and\ \bibinfo {author} {\bibfnamefont {H.}~\bibnamefont {Agren}},\
  }\href@noop {} {\bibfield  {journal} {\bibinfo  {journal} {Chem. Phys.}\
  }\textbf {\bibinfo {volume} {224}},\ \bibinfo {pages} {201} (\bibinfo {year}
  {1997})}\BibitemShut {NoStop}%
\bibitem [{\citenamefont {DeYonker}\ \emph {et~al.}(2014)\citenamefont
  {DeYonker}, \citenamefont {Halfen}, \citenamefont {Allen},\ and\
  \citenamefont {Ziurys}}]{DeYonker2014}%
  \BibitemOpen
  \bibfield  {author} {\bibinfo {author} {\bibfnamefont {N.}~\bibnamefont
  {DeYonker}}, \bibinfo {author} {\bibfnamefont {D.}~\bibnamefont {Halfen}},
  \bibinfo {author} {\bibfnamefont {W.}~\bibnamefont {Allen}}, \ and\ \bibinfo
  {author} {\bibfnamefont {L.}~\bibnamefont {Ziurys}},\ }\href@noop {}
  {\bibfield  {journal} {\bibinfo  {journal} {J. Chem. Phys.}\ }\textbf
  {\bibinfo {volume} {141}},\ \bibinfo {pages} {204302} (\bibinfo {year}
  {2014})}\BibitemShut {NoStop}%
\bibitem [{\citenamefont {Valeev}\ \emph {et~al.}(2001)\citenamefont {Valeev},
  \citenamefont {Allen}, \citenamefont {Schaefer}, \citenamefont
  {Cs\'asz\'ar},\ and\ \citenamefont {East}}]{Valeev2001}%
  \BibitemOpen
  \bibfield  {author} {\bibinfo {author} {\bibfnamefont {E.}~\bibnamefont
  {Valeev}}, \bibinfo {author} {\bibfnamefont {W.}~\bibnamefont {Allen}},
  \bibinfo {author} {\bibfnamefont {H.}~\bibnamefont {Schaefer}}, \bibinfo
  {author} {\bibfnamefont {A.}~\bibnamefont {Cs\'asz\'ar}}, \ and\ \bibinfo
  {author} {\bibfnamefont {A.}~\bibnamefont {East}},\ }\href@noop {} {\bibfield
   {journal} {\bibinfo  {journal} {Journal of Physical Chemistry A}\ }\textbf
  {\bibinfo {volume} {105}},\ \bibinfo {pages} {2716} (\bibinfo {year}
  {2001})}\BibitemShut {NoStop}%
\bibitem [{\citenamefont {Bokareva}\ \emph {et~al.}(2009)\citenamefont
  {Bokareva}, \citenamefont {Bataev}, \citenamefont {Pupyshev},\ and\
  \citenamefont {Godunov}}]{Bokareva2009}%
  \BibitemOpen
  \bibfield  {author} {\bibinfo {author} {\bibfnamefont {O.}~\bibnamefont
  {Bokareva}}, \bibinfo {author} {\bibfnamefont {V.}~\bibnamefont {Bataev}},
  \bibinfo {author} {\bibfnamefont {V.}~\bibnamefont {Pupyshev}}, \ and\
  \bibinfo {author} {\bibfnamefont {I.}~\bibnamefont {Godunov}},\ }\href@noop
  {} {\bibfield  {journal} {\bibinfo  {journal} {Spectrochim. Acta Mol. Biomol.
  Spectrosc.}\ }\textbf {\bibinfo {volume} {73}},\ \bibinfo {pages} {654}
  (\bibinfo {year} {2009})}\BibitemShut {NoStop}%
\bibitem [{\citenamefont {Little}\ and\ \citenamefont
  {Tennyson}(2013)}]{jt560}%
  \BibitemOpen
  \bibfield  {author} {\bibinfo {author} {\bibfnamefont {D.~A.}\ \bibnamefont
  {Little}}\ and\ \bibinfo {author} {\bibfnamefont {J.}~\bibnamefont
  {Tennyson}},\ }\href@noop {} {\bibfield  {journal} {\bibinfo  {journal} {J.
  Phys. B: At. Mol. Opt. Phys.}\ }\textbf {\bibinfo {volume} {46}},\ \bibinfo
  {pages} {145102} (\bibinfo {year} {2013})}\BibitemShut {NoStop}%
\bibitem [{\citenamefont {Jungen}(1996)}]{J96}%
  \BibitemOpen
  \bibinfo {editor} {\bibfnamefont {C.}~\bibnamefont {Jungen}},\ ed.,\
  \href@noop {} {\emph {\bibinfo {title} {Molecular Applications of Quantum
  Defect Theory}}}\ (\bibinfo  {publisher} {Taylor and Francis},\ \bibinfo
  {year} {1996})\BibitemShut {NoStop}%
\bibitem [{\citenamefont {Tennyson}(1996)}]{jt189}%
  \BibitemOpen
  \bibfield  {author} {\bibinfo {author} {\bibfnamefont {J.}~\bibnamefont
  {Tennyson}},\ }\href@noop {} {\bibfield  {journal} {\bibinfo  {journal} {J.
  Phys. B: At. Mol. Opt. Phys.}\ }\textbf {\bibinfo {volume} {29}},\ \bibinfo
  {pages} {6185} (\bibinfo {year} {1996})}\BibitemShut {NoStop}%
\bibitem [{\citenamefont {Schneider}\ \emph {et~al.}(2000)\citenamefont
  {Schneider}, \citenamefont {{Rabad\'an}}, \citenamefont {Carata},
  \citenamefont {Tennyson}, \citenamefont {Andersen},\ and\ \citenamefont
  {Suzor-Weiner}}]{jt260}%
  \BibitemOpen
  \bibfield  {author} {\bibinfo {author} {\bibfnamefont {I.~F.}\ \bibnamefont
  {Schneider}}, \bibinfo {author} {\bibfnamefont {I.}~\bibnamefont
  {{Rabad\'an}}}, \bibinfo {author} {\bibfnamefont {L.}~\bibnamefont {Carata}},
  \bibinfo {author} {\bibfnamefont {J.}~\bibnamefont {Tennyson}}, \bibinfo
  {author} {\bibfnamefont {L.~H.}\ \bibnamefont {Andersen}}, \ and\ \bibinfo
  {author} {\bibfnamefont {A.}~\bibnamefont {Suzor-Weiner}},\ }\href@noop {}
  {\bibfield  {journal} {\bibinfo  {journal} {J. Phys. B: At. Mol. Opt. Phys.}\
  }\textbf {\bibinfo {volume} {33}},\ \bibinfo {pages} {4849} (\bibinfo {year}
  {2000})}\BibitemShut {NoStop}%
\bibitem [{\citenamefont {Cocks}\ \emph {et~al.}(2010)\citenamefont {Cocks},
  \citenamefont {Whittingham},\ and\ \citenamefont {Peach}}]{10CoWhPe}%
  \BibitemOpen
  \bibfield  {author} {\bibinfo {author} {\bibfnamefont {D.~G.}\ \bibnamefont
  {Cocks}}, \bibinfo {author} {\bibfnamefont {I.~B.}\ \bibnamefont
  {Whittingham}}, \ and\ \bibinfo {author} {\bibfnamefont {G.}~\bibnamefont
  {Peach}},\ }\href {\doibase {10.1088/0953-4075/43/13/135102}} {\bibfield
  {journal} {\bibinfo  {journal} {J. Phys. B: At. Mol. Opt. Phys.}\ }\textbf
  {\bibinfo {volume} {{43}}},\ \bibinfo {pages} {135102} (\bibinfo {year}
  {{2010}})}\BibitemShut {NoStop}%
\bibitem [{\citenamefont {Bartschat}(2013)}]{0022-3727-46-33-334004}%
  \BibitemOpen
  \bibfield  {author} {\bibinfo {author} {\bibfnamefont {K.}~\bibnamefont
  {Bartschat}},\ }\href {http://stacks.iop.org/0022-3727/46/i=33/a=334004}
  {\bibfield  {journal} {\bibinfo  {journal} {J. Phys. D: Appl. Phys.}\
  }\textbf {\bibinfo {volume} {46}},\ \bibinfo {pages} {334004} (\bibinfo
  {year} {2013})}\BibitemShut {NoStop}%
\bibitem [{\citenamefont {Bartschat}\ and\ \citenamefont
  {Zatsarinny}(2015)}]{1402-4896-90-5-054006}%
  \BibitemOpen
  \bibfield  {author} {\bibinfo {author} {\bibfnamefont {K.}~\bibnamefont
  {Bartschat}}\ and\ \bibinfo {author} {\bibfnamefont {O.}~\bibnamefont
  {Zatsarinny}},\ }\href {http://stacks.iop.org/1402-4896/90/i=5/a=054006}
  {\bibfield  {journal} {\bibinfo  {journal} {Phys. Scr.}\ }\textbf {\bibinfo
  {volume} {90}},\ \bibinfo {pages} {054006} (\bibinfo {year}
  {2015})}\BibitemShut {NoStop}%
\bibitem [{\citenamefont {Burke}(2011)}]{Burke2011}%
  \BibitemOpen
  \bibfield  {author} {\bibinfo {author} {\bibfnamefont {P.~G.}\ \bibnamefont
  {Burke}},\ }\href@noop {} {\emph {\bibinfo {title} {R-Matrix Theory of Atomic
  Collisions: Application to Atomic, Molecular and Optical Processes}}}\
  (\bibinfo  {publisher} {Springer},\ \bibinfo {year} {2011})\BibitemShut
  {NoStop}%
\bibitem [{\citenamefont {Tennyson}(2010)}]{jt474}%
  \BibitemOpen
  \bibfield  {author} {\bibinfo {author} {\bibfnamefont {J.}~\bibnamefont
  {Tennyson}},\ }\href@noop {} {\bibfield  {journal} {\bibinfo  {journal}
  {Phys. Rep.}\ }\textbf {\bibinfo {volume} {491}},\ \bibinfo {pages} {29}
  (\bibinfo {year} {2010})}\BibitemShut {NoStop}%
\bibitem [{\citenamefont {Bray}\ \emph {et~al.}(2012)\citenamefont {Bray},
  \citenamefont {Fursa}, \citenamefont {Kadyrov}, \citenamefont {Stelbovics},
  \citenamefont {Kheifets},\ and\ \citenamefont
  {Mukhamedzhanov}}]{Bray2012135}%
  \BibitemOpen
  \bibfield  {author} {\bibinfo {author} {\bibfnamefont {I.}~\bibnamefont
  {Bray}}, \bibinfo {author} {\bibfnamefont {D.}~\bibnamefont {Fursa}},
  \bibinfo {author} {\bibfnamefont {A.}~\bibnamefont {Kadyrov}}, \bibinfo
  {author} {\bibfnamefont {A.}~\bibnamefont {Stelbovics}}, \bibinfo {author}
  {\bibfnamefont {A.}~\bibnamefont {Kheifets}}, \ and\ \bibinfo {author}
  {\bibfnamefont {A.}~\bibnamefont {Mukhamedzhanov}},\ }\href {\doibase
  http://dx.doi.org/10.1016/j.physrep.2012.07.002} {\bibfield  {journal}
  {\bibinfo  {journal} {Phys. Rep.}\ }\textbf {\bibinfo {volume} {520}},\
  \bibinfo {pages} {135 } (\bibinfo {year} {2012})},\ \bibinfo {note}
  {electron- and photon-impact atomic ionisation}\BibitemShut {NoStop}%
\bibitem [{\citenamefont {Bartschat}\ \emph {et~al.}(1996)\citenamefont
  {Bartschat}, \citenamefont {Hudson}, \citenamefont {Scott}, \citenamefont
  {Burke},\ and\ \citenamefont {Burke}}]{0953-4075-29-1-015}%
  \BibitemOpen
  \bibfield  {author} {\bibinfo {author} {\bibfnamefont {K.}~\bibnamefont
  {Bartschat}}, \bibinfo {author} {\bibfnamefont {E.~T.}\ \bibnamefont
  {Hudson}}, \bibinfo {author} {\bibfnamefont {M.~P.}\ \bibnamefont {Scott}},
  \bibinfo {author} {\bibfnamefont {P.~G.}\ \bibnamefont {Burke}}, \ and\
  \bibinfo {author} {\bibfnamefont {V.~M.}\ \bibnamefont {Burke}},\ }\href
  {http://stacks.iop.org/0953-4075/29/i=1/a=015} {\bibfield  {journal}
  {\bibinfo  {journal} {J. Phys. B: At., Mol. Opt. Phys.}\ }\textbf {\bibinfo
  {volume} {29}},\ \bibinfo {pages} {115} (\bibinfo {year} {1996})}\BibitemShut
  {NoStop}%
\bibitem [{\citenamefont {Gorfinkiel}\ and\ \citenamefont
  {Tennyson}(2004)}]{jt341}%
  \BibitemOpen
  \bibfield  {author} {\bibinfo {author} {\bibfnamefont {J.~D.}\ \bibnamefont
  {Gorfinkiel}}\ and\ \bibinfo {author} {\bibfnamefont {J.}~\bibnamefont
  {Tennyson}},\ }\href@noop {} {\bibfield  {journal} {\bibinfo  {journal} {J.
  Phys. B: At. Mol. Opt. Phys.}\ }\textbf {\bibinfo {volume} {37}},\ \bibinfo
  {pages} {L343} (\bibinfo {year} {2004})}\BibitemShut {NoStop}%
\bibitem [{\citenamefont {Colgan}\ \emph {et~al.}(2002)\citenamefont {Colgan},
  \citenamefont {Pindzola}, \citenamefont {Robicheaux}, \citenamefont
  {Griffin},\ and\ \citenamefont {Baertschy}}]{PhysRevA.65.042721}%
  \BibitemOpen
  \bibfield  {author} {\bibinfo {author} {\bibfnamefont {J.}~\bibnamefont
  {Colgan}}, \bibinfo {author} {\bibfnamefont {M.~S.}\ \bibnamefont
  {Pindzola}}, \bibinfo {author} {\bibfnamefont {F.~J.}\ \bibnamefont
  {Robicheaux}}, \bibinfo {author} {\bibfnamefont {D.~C.}\ \bibnamefont
  {Griffin}}, \ and\ \bibinfo {author} {\bibfnamefont {M.}~\bibnamefont
  {Baertschy}},\ }\href {\doibase 10.1103/PhysRevA.65.042721} {\bibfield
  {journal} {\bibinfo  {journal} {Phys. Rev. A}\ }\textbf {\bibinfo {volume}
  {65}},\ \bibinfo {pages} {042721} (\bibinfo {year} {2002})}\BibitemShut
  {NoStop}%
\bibitem [{\citenamefont {Rescigno}\ \emph {et~al.}(1999)\citenamefont
  {Rescigno}, \citenamefont {Baertschy}, \citenamefont {Isaacs},\ and\
  \citenamefont {McCurdy}}]{Rescigno24121999}%
  \BibitemOpen
  \bibfield  {author} {\bibinfo {author} {\bibfnamefont {T.~N.}\ \bibnamefont
  {Rescigno}}, \bibinfo {author} {\bibfnamefont {M.}~\bibnamefont {Baertschy}},
  \bibinfo {author} {\bibfnamefont {W.~A.}\ \bibnamefont {Isaacs}}, \ and\
  \bibinfo {author} {\bibfnamefont {C.~W.}\ \bibnamefont {McCurdy}},\ }\href
  {\doibase 10.1126/science.286.5449.2474} {\bibfield  {journal} {\bibinfo
  {journal} {Science}\ }\textbf {\bibinfo {volume} {286}},\ \bibinfo {pages}
  {2474} (\bibinfo {year} {1999})}\BibitemShut {NoStop}%
\bibitem [{\citenamefont {Taylor}(1972)}]{taylor1972scattering}%
  \BibitemOpen
  \bibfield  {author} {\bibinfo {author} {\bibfnamefont {J.~R.}\ \bibnamefont
  {Taylor}},\ }\href@noop {} {\emph {\bibinfo {title} {Scattering theory}}}\
  (\bibinfo  {publisher} {Wiley},\ \bibinfo {year} {1972})\BibitemShut
  {NoStop}%
\bibitem [{\citenamefont {Joachain}(1984)}]{joachain1975quantum}%
  \BibitemOpen
  \bibfield  {author} {\bibinfo {author} {\bibfnamefont {C.~J.}\ \bibnamefont
  {Joachain}},\ }\href@noop {} {\emph {\bibinfo {title} {Quantum collision
  theory}}}\ (\bibinfo  {publisher} {Elsevier Science Ltd},\ \bibinfo {year}
  {1984})\BibitemShut {NoStop}%
\bibitem [{\citenamefont {Madison}\ and\ \citenamefont
  {Shelton}(1973)}]{madison1973distorted}%
  \BibitemOpen
  \bibfield  {author} {\bibinfo {author} {\bibfnamefont {D.}~\bibnamefont
  {Madison}}\ and\ \bibinfo {author} {\bibfnamefont {W.}~\bibnamefont
  {Shelton}},\ }\href@noop {} {\bibfield  {journal} {\bibinfo  {journal} {Phys.
  Rev. A}\ }\textbf {\bibinfo {volume} {7}},\ \bibinfo {pages} {499} (\bibinfo
  {year} {1973})}\BibitemShut {NoStop}%
\bibitem [{\citenamefont {Itikawa}(1986)}]{itikawa1986distorted}%
  \BibitemOpen
  \bibfield  {author} {\bibinfo {author} {\bibfnamefont {Y.}~\bibnamefont
  {Itikawa}},\ }\href@noop {} {\bibfield  {journal} {\bibinfo  {journal} {Phys.
  Rep.}\ }\textbf {\bibinfo {volume} {143}},\ \bibinfo {pages} {69} (\bibinfo
  {year} {1986})}\BibitemShut {NoStop}%
\bibitem [{\citenamefont {Bote}\ and\ \citenamefont
  {Salvat}(2008)}]{bote2008calculations}%
  \BibitemOpen
  \bibfield  {author} {\bibinfo {author} {\bibfnamefont {D.}~\bibnamefont
  {Bote}}\ and\ \bibinfo {author} {\bibfnamefont {F.}~\bibnamefont {Salvat}},\
  }\href@noop {} {\bibfield  {journal} {\bibinfo  {journal} {Phys. Rev. A}\
  }\textbf {\bibinfo {volume} {77}},\ \bibinfo {pages} {042701} (\bibinfo
  {year} {2008})}\BibitemShut {NoStop}%
\bibitem [{\citenamefont {Gao}\ \emph {et~al.}(2006)\citenamefont {Gao},
  \citenamefont {Madison},\ and\ \citenamefont {Peacher}}]{Gao06}%
  \BibitemOpen
  \bibfield  {author} {\bibinfo {author} {\bibfnamefont {J.~F.}\ \bibnamefont
  {Gao}}, \bibinfo {author} {\bibfnamefont {D.~H.}\ \bibnamefont {Madison}}, \
  and\ \bibinfo {author} {\bibfnamefont {J.~L.}\ \bibnamefont {Peacher}},\
  }\href {\doibase {10.1088/0953-4075/39/6/002}} {\bibfield  {journal}
  {\bibinfo  {journal} {J. Phys. B: At. Mol. Opt. Phys.}\ }\textbf {\bibinfo
  {volume} {{39}}},\ \bibinfo {pages} {1275} (\bibinfo {year}
  {{2006}})}\BibitemShut {NoStop}%
\bibitem [{\citenamefont {Al-Hagan}\ \emph {et~al.}(2009)\citenamefont
  {Al-Hagan}, \citenamefont {Kaiser}, \citenamefont {Madison},\ and\
  \citenamefont {Murray}}]{AlH09}%
  \BibitemOpen
  \bibfield  {author} {\bibinfo {author} {\bibfnamefont {O.}~\bibnamefont
  {Al-Hagan}}, \bibinfo {author} {\bibfnamefont {C.}~\bibnamefont {Kaiser}},
  \bibinfo {author} {\bibfnamefont {D.}~\bibnamefont {Madison}}, \ and\
  \bibinfo {author} {\bibfnamefont {A.~J.}\ \bibnamefont {Murray}},\ }\href
  {\doibase {10.1038/NPHYS1135}} {\bibfield  {journal} {\bibinfo  {journal}
  {Nature Phys.}\ }\textbf {\bibinfo {volume} {{5}}},\ \bibinfo {pages} {59}
  (\bibinfo {year} {{2009}})}\BibitemShut {NoStop}%
\bibitem [{\citenamefont {Toth}\ and\ \citenamefont {Nagy}(2011)}]{Toth11}%
  \BibitemOpen
  \bibfield  {author} {\bibinfo {author} {\bibfnamefont {I.}~\bibnamefont
  {Toth}}\ and\ \bibinfo {author} {\bibfnamefont {L.}~\bibnamefont {Nagy}},\
  }\href {\doibase {10.1088/0953-4075/44/19/195205}} {\bibfield  {journal}
  {\bibinfo  {journal} {J. Phys. B: At. Mol. Opt. Phys.}\ }\textbf {\bibinfo
  {volume} {{44}}},\ \bibinfo {pages} {195205} (\bibinfo {year}
  {{2011}})}\BibitemShut {NoStop}%
\bibitem [{\citenamefont {Zhang}\ \emph {et~al.}(2014)\citenamefont {Zhang},
  \citenamefont {Li}, \citenamefont {Wang}, \citenamefont {Qu},\ and\
  \citenamefont {Chen}}]{Zhang14}%
  \BibitemOpen
  \bibfield  {author} {\bibinfo {author} {\bibfnamefont {S.~B.}\ \bibnamefont
  {Zhang}}, \bibinfo {author} {\bibfnamefont {X.~Y.}\ \bibnamefont {Li}},
  \bibinfo {author} {\bibfnamefont {J.~G.}\ \bibnamefont {Wang}}, \bibinfo
  {author} {\bibfnamefont {Y.~Z.}\ \bibnamefont {Qu}}, \ and\ \bibinfo {author}
  {\bibfnamefont {X.}~\bibnamefont {Chen}},\ }\href {\doibase
  {10.1103/PhysRevA.89.052711}} {\bibfield  {journal} {\bibinfo  {journal}
  {Phys. Rev. A}\ }\textbf {\bibinfo {volume} {{89}}},\ \bibinfo {pages}
  {052711} (\bibinfo {year} {{2014}})}\BibitemShut {NoStop}%
\bibitem [{\citenamefont {Badnell}(2008)}]{badnell2008dirac}%
  \BibitemOpen
  \bibfield  {author} {\bibinfo {author} {\bibfnamefont {N.}~\bibnamefont
  {Badnell}},\ }\href@noop {} {\bibfield  {journal} {\bibinfo  {journal} {J.
  Phys. B: At., Mol. Opt. Phys.}\ }\textbf {\bibinfo {volume} {41}},\ \bibinfo
  {pages} {175202} (\bibinfo {year} {2008})}\BibitemShut {NoStop}%
\bibitem [{\citenamefont {Fursa}\ and\ \citenamefont
  {Bray}(2008)}]{fursa2008fully}%
  \BibitemOpen
  \bibfield  {author} {\bibinfo {author} {\bibfnamefont {D.~V.}\ \bibnamefont
  {Fursa}}\ and\ \bibinfo {author} {\bibfnamefont {I.}~\bibnamefont {Bray}},\
  }\href@noop {} {\bibfield  {journal} {\bibinfo  {journal} {Phys. Rev. Lett.}\
  }\textbf {\bibinfo {volume} {100}},\ \bibinfo {pages} {113201} (\bibinfo
  {year} {2008})}\BibitemShut {NoStop}%
\bibitem [{\citenamefont {Zuo}\ \emph {et~al.}(1991)\citenamefont {Zuo},
  \citenamefont {McEachran},\ and\ \citenamefont
  {Stauffer}}]{zuo1991relativistic}%
  \BibitemOpen
  \bibfield  {author} {\bibinfo {author} {\bibfnamefont {T.}~\bibnamefont
  {Zuo}}, \bibinfo {author} {\bibfnamefont {R.}~\bibnamefont {McEachran}}, \
  and\ \bibinfo {author} {\bibfnamefont {A.}~\bibnamefont {Stauffer}},\
  }\href@noop {} {\bibfield  {journal} {\bibinfo  {journal} {J. Phys. B: At.,
  Mol. Opt. Phys.}\ }\textbf {\bibinfo {volume} {24}},\ \bibinfo {pages} {2853}
  (\bibinfo {year} {1991})}\BibitemShut {NoStop}%
\bibitem [{\citenamefont {Kim}(2001)}]{PhysRevA.64.032713}%
  \BibitemOpen
  \bibfield  {author} {\bibinfo {author} {\bibfnamefont {Y.-K.}\ \bibnamefont
  {Kim}},\ }\href {\doibase 10.1103/PhysRevA.64.032713} {\bibfield  {journal}
  {\bibinfo  {journal} {Phys. Rev. A}\ }\textbf {\bibinfo {volume} {64}},\
  \bibinfo {pages} {032713} (\bibinfo {year} {2001})}\BibitemShut {NoStop}%
\bibitem [{\citenamefont {Kim}\ and\ \citenamefont
  {Rudd}(1994)}]{PhysRevA.50.3954}%
  \BibitemOpen
  \bibfield  {author} {\bibinfo {author} {\bibfnamefont {Y.-K.}\ \bibnamefont
  {Kim}}\ and\ \bibinfo {author} {\bibfnamefont {M.~E.}\ \bibnamefont {Rudd}},\
  }\href {\doibase 10.1103/PhysRevA.50.3954} {\bibfield  {journal} {\bibinfo
  {journal} {Phys. Rev. A}\ }\textbf {\bibinfo {volume} {50}},\ \bibinfo
  {pages} {3954} (\bibinfo {year} {1994})}\BibitemShut {NoStop}%
\bibitem [{\citenamefont {Laporta}\ \emph {et~al.}(2012)\citenamefont
  {Laporta}, \citenamefont {Cassidy}, \citenamefont {Tennyson},\ and\
  \citenamefont {Celiberto}}]{jt527}%
  \BibitemOpen
  \bibfield  {author} {\bibinfo {author} {\bibfnamefont {V.}~\bibnamefont
  {Laporta}}, \bibinfo {author} {\bibfnamefont {C.~M.}\ \bibnamefont
  {Cassidy}}, \bibinfo {author} {\bibfnamefont {J.}~\bibnamefont {Tennyson}}, \
  and\ \bibinfo {author} {\bibfnamefont {R.}~\bibnamefont {Celiberto}},\
  }\href@noop {} {\bibfield  {journal} {\bibinfo  {journal} {Plasma Sources
  Sci. Technol.}\ }\textbf {\bibinfo {volume} {21}},\ \bibinfo {pages} {045005}
  (\bibinfo {year} {2012})}\BibitemShut {NoStop}%
\bibitem [{\citenamefont {Laporta}\ \emph {et~al.}(2015)\citenamefont
  {Laporta}, \citenamefont {Celiberto},\ and\ \citenamefont
  {Tennyson}}]{jt586}%
  \BibitemOpen
  \bibfield  {author} {\bibinfo {author} {\bibfnamefont {V.}~\bibnamefont
  {Laporta}}, \bibinfo {author} {\bibfnamefont {R.}~\bibnamefont {Celiberto}},
  \ and\ \bibinfo {author} {\bibfnamefont {J.}~\bibnamefont {Tennyson}},\
  }\href {\doibase 10.1103/PhysRevA.91.012701} {\bibfield  {journal} {\bibinfo
  {journal} {Phys. Rev. A}\ }\textbf {\bibinfo {volume} {91}},\ \bibinfo
  {pages} {012701} (\bibinfo {year} {2015})}\BibitemShut {NoStop}%
\bibitem [{\citenamefont {Chang}\ and\ \citenamefont
  {Fano}(1972)}]{chang1972theory}%
  \BibitemOpen
  \bibfield  {author} {\bibinfo {author} {\bibfnamefont {E.}~\bibnamefont
  {Chang}}\ and\ \bibinfo {author} {\bibfnamefont {U.}~\bibnamefont {Fano}},\
  }\href@noop {} {\bibfield  {journal} {\bibinfo  {journal} {Phys. Rev. A}\
  }\textbf {\bibinfo {volume} {6}},\ \bibinfo {pages} {173} (\bibinfo {year}
  {1972})}\BibitemShut {NoStop}%
\bibitem [{\citenamefont {Morrison}\ \emph {et~al.}(1984)\citenamefont
  {Morrison}, \citenamefont {Feldt},\ and\ \citenamefont {Saha}}]{84MoFeSa}%
  \BibitemOpen
  \bibfield  {author} {\bibinfo {author} {\bibfnamefont {M.~A.}\ \bibnamefont
  {Morrison}}, \bibinfo {author} {\bibfnamefont {A.~N.}\ \bibnamefont {Feldt}},
  \ and\ \bibinfo {author} {\bibfnamefont {B.~C.}\ \bibnamefont {Saha}},\
  }\href {\doibase 10.1103/PhysRevA.30.2811} {\bibfield  {journal} {\bibinfo
  {journal} {Phys. Rev. A}\ }\textbf {\bibinfo {volume} {30}},\ \bibinfo
  {pages} {2811} (\bibinfo {year} {1984})}\BibitemShut {NoStop}%
\bibitem [{\citenamefont {Atabek}\ \emph {et~al.}(1974)\citenamefont {Atabek},
  \citenamefont {Dill},\ and\ \citenamefont {Jungen}}]{atabek1974quantum}%
  \BibitemOpen
  \bibfield  {author} {\bibinfo {author} {\bibfnamefont {O.}~\bibnamefont
  {Atabek}}, \bibinfo {author} {\bibfnamefont {D.}~\bibnamefont {Dill}}, \ and\
  \bibinfo {author} {\bibfnamefont {C.}~\bibnamefont {Jungen}},\ }\href@noop {}
  {\bibfield  {journal} {\bibinfo  {journal} {Phys. Rev. Lett.}\ }\textbf
  {\bibinfo {volume} {33}},\ \bibinfo {pages} {123} (\bibinfo {year}
  {1974})}\BibitemShut {NoStop}%
\bibitem [{\citenamefont {Greene}\ and\ \citenamefont
  {Ch.Jungen}(1985)}]{greene85}%
  \BibitemOpen
  \bibfield  {author} {\bibinfo {author} {\bibfnamefont {C.~H.}\ \bibnamefont
  {Greene}}\ and\ \bibinfo {author} {\bibnamefont {Ch.Jungen}},\ }\href@noop {}
  {\bibfield  {journal} {\bibinfo  {journal} {Adv. At. Mol. Phys.}\ }\textbf
  {\bibinfo {volume} {21}},\ \bibinfo {pages} {51} (\bibinfo {year}
  {1985})}\BibitemShut {NoStop}%
\bibitem [{\citenamefont {Seaton}(1966)}]{seaton66}%
  \BibitemOpen
  \bibfield  {author} {\bibinfo {author} {\bibfnamefont {M.~J.}\ \bibnamefont
  {Seaton}},\ }\href@noop {} {\bibfield  {journal} {\bibinfo  {journal} {Proc.
  Phys. Soc. London}\ }\textbf {\bibinfo {volume} {88}},\ \bibinfo {pages}
  {801} (\bibinfo {year} {1966})}\BibitemShut {NoStop}%
\bibitem [{\citenamefont {Aymar}\ \emph {et~al.}(1996)\citenamefont {Aymar},
  \citenamefont {Greene},\ and\ \citenamefont {Luc-Koenig}}]{aymar96}%
  \BibitemOpen
  \bibfield  {author} {\bibinfo {author} {\bibfnamefont {M.}~\bibnamefont
  {Aymar}}, \bibinfo {author} {\bibfnamefont {C.~H.}\ \bibnamefont {Greene}}, \
  and\ \bibinfo {author} {\bibfnamefont {E.}~\bibnamefont {Luc-Koenig}},\
  }\href@noop {} {\bibfield  {journal} {\bibinfo  {journal} {Rev. Mod. Phys.}\
  }\textbf {\bibinfo {volume} {68}},\ \bibinfo {pages} {1015} (\bibinfo {year}
  {1996})}\BibitemShut {NoStop}%
\bibitem [{\citenamefont {Greene}\ \emph {et~al.}(1979)\citenamefont {Greene},
  \citenamefont {Fano},\ and\ \citenamefont {Strinati}}]{greene1979general}%
  \BibitemOpen
  \bibfield  {author} {\bibinfo {author} {\bibfnamefont {C.}~\bibnamefont
  {Greene}}, \bibinfo {author} {\bibfnamefont {U.}~\bibnamefont {Fano}}, \ and\
  \bibinfo {author} {\bibfnamefont {G.}~\bibnamefont {Strinati}},\ }\href@noop
  {} {\bibfield  {journal} {\bibinfo  {journal} {Phys. Rev. A}\ }\textbf
  {\bibinfo {volume} {19}},\ \bibinfo {pages} {1485} (\bibinfo {year}
  {1979})}\BibitemShut {NoStop}%
\bibitem [{\citenamefont {Gao}\ and\ \citenamefont
  {Greene}(1989)}]{gao1989energy}%
  \BibitemOpen
  \bibfield  {author} {\bibinfo {author} {\bibfnamefont {H.}~\bibnamefont
  {Gao}}\ and\ \bibinfo {author} {\bibfnamefont {C.~H.}\ \bibnamefont
  {Greene}},\ }\href@noop {} {\bibfield  {journal} {\bibinfo  {journal} {J.
  Chem. Phys.}\ }\textbf {\bibinfo {volume} {91}},\ \bibinfo {pages} {3988}
  (\bibinfo {year} {1989})}\BibitemShut {NoStop}%
\bibitem [{\citenamefont {Gao}\ and\ \citenamefont
  {Greene}(1990)}]{gao1990alternative}%
  \BibitemOpen
  \bibfield  {author} {\bibinfo {author} {\bibfnamefont {H.}~\bibnamefont
  {Gao}}\ and\ \bibinfo {author} {\bibfnamefont {C.~H.}\ \bibnamefont
  {Greene}},\ }\href@noop {} {\bibfield  {journal} {\bibinfo  {journal} {Phys.
  Rev. A}\ }\textbf {\bibinfo {volume} {42}},\ \bibinfo {pages} {6946}
  (\bibinfo {year} {1990})}\BibitemShut {NoStop}%
\bibitem [{\citenamefont {Sun}\ \emph {et~al.}(1995)\citenamefont {Sun},
  \citenamefont {Morrison}, \citenamefont {Isaacs}, \citenamefont {Trail},
  \citenamefont {Alle}, \citenamefont {Gulley}, \citenamefont {Brennan},\ and\
  \citenamefont {Buckman}}]{95SuMoIs}%
  \BibitemOpen
  \bibfield  {author} {\bibinfo {author} {\bibfnamefont {W.}~\bibnamefont
  {Sun}}, \bibinfo {author} {\bibfnamefont {M.~A.}\ \bibnamefont {Morrison}},
  \bibinfo {author} {\bibfnamefont {W.~A.}\ \bibnamefont {Isaacs}}, \bibinfo
  {author} {\bibfnamefont {W.~K.}\ \bibnamefont {Trail}}, \bibinfo {author}
  {\bibfnamefont {D.~T.}\ \bibnamefont {Alle}}, \bibinfo {author}
  {\bibfnamefont {R.}~\bibnamefont {Gulley}}, \bibinfo {author} {\bibfnamefont
  {M.~J.}\ \bibnamefont {Brennan}}, \ and\ \bibinfo {author} {\bibfnamefont
  {S.~J.}\ \bibnamefont {Buckman}},\ }\href@noop {} {\bibfield  {journal}
  {\bibinfo  {journal} {Phys. Rev. A}\ }\textbf {\bibinfo {volume} {52}},\
  \bibinfo {pages} {1229} (\bibinfo {year} {1995})}\BibitemShut {NoStop}%
\bibitem [{\citenamefont {Chase}(1956)}]{cha56}%
  \BibitemOpen
  \bibfield  {author} {\bibinfo {author} {\bibfnamefont {D.~M.}\ \bibnamefont
  {Chase}},\ }\href@noop {} {\bibfield  {journal} {\bibinfo  {journal} {Phys.
  Rev.}\ }\textbf {\bibinfo {volume} {104}},\ \bibinfo {pages} {838} (\bibinfo
  {year} {1956})}\BibitemShut {NoStop}%
\bibitem [{\citenamefont {Faure}\ \emph
  {et~al.}(2009{\natexlab{a}})\citenamefont {Faure}, \citenamefont
  {Kokoouline}, \citenamefont {Greene},\ and\ \citenamefont
  {Tennyson}}]{jt439}%
  \BibitemOpen
  \bibfield  {author} {\bibinfo {author} {\bibfnamefont {A.}~\bibnamefont
  {Faure}}, \bibinfo {author} {\bibfnamefont {V.}~\bibnamefont {Kokoouline}},
  \bibinfo {author} {\bibfnamefont {C.~H.}\ \bibnamefont {Greene}}, \ and\
  \bibinfo {author} {\bibfnamefont {J.}~\bibnamefont {Tennyson}},\ }\href@noop
  {} {\bibfield  {journal} {\bibinfo  {journal} {J. Phys. Conf. Ser.}\ }\textbf
  {\bibinfo {volume} {192}},\ \bibinfo {pages} {012016} (\bibinfo {year}
  {2009}{\natexlab{a}})}\BibitemShut {NoStop}%
\bibitem [{\citenamefont {Faure}\ \emph
  {et~al.}(2009{\natexlab{b}})\citenamefont {Faure}, \citenamefont {Tennyson},
  \citenamefont {Kokoouline},\ and\ \citenamefont {Greene}}]{faure09}%
  \BibitemOpen
  \bibfield  {author} {\bibinfo {author} {\bibfnamefont {A.}~\bibnamefont
  {Faure}}, \bibinfo {author} {\bibfnamefont {J.}~\bibnamefont {Tennyson}},
  \bibinfo {author} {\bibfnamefont {V.}~\bibnamefont {Kokoouline}}, \ and\
  \bibinfo {author} {\bibfnamefont {C.~H.}\ \bibnamefont {Greene}},\ }\href
  {http://stacks.iop.org/1742-6596/192/012016} {\bibfield  {journal} {\bibinfo
  {journal} {J. Phys. Conf. Ser.}\ }\textbf {\bibinfo {volume} {192}},\
  \bibinfo {pages} {012016} (\bibinfo {year} {2009}{\natexlab{b}})}\BibitemShut
  {NoStop}%
\bibitem [{\citenamefont {Kokoouline}\ \emph {et~al.}(2010)\citenamefont
  {Kokoouline}, \citenamefont {Faure}, \citenamefont {Tennyson},\ and\
  \citenamefont {Greene}}]{kokoouline10a}%
  \BibitemOpen
  \bibfield  {author} {\bibinfo {author} {\bibfnamefont {V.}~\bibnamefont
  {Kokoouline}}, \bibinfo {author} {\bibfnamefont {A.}~\bibnamefont {Faure}},
  \bibinfo {author} {\bibfnamefont {J.}~\bibnamefont {Tennyson}}, \ and\
  \bibinfo {author} {\bibfnamefont {C.~H.}\ \bibnamefont {Greene}},\
  }\href@noop {} {\bibfield  {journal} {\bibinfo  {journal} {Mon. Not. R.
  Astron. Soc.}\ }\textbf {\bibinfo {volume} {405}},\ \bibinfo {pages} {1195}
  (\bibinfo {year} {2010})}\BibitemShut {NoStop}%
\bibitem [{\citenamefont {Norcross}\ and\ \citenamefont {Padial}(1982)}]{np82}%
  \BibitemOpen
  \bibfield  {author} {\bibinfo {author} {\bibfnamefont {D.~W.}\ \bibnamefont
  {Norcross}}\ and\ \bibinfo {author} {\bibfnamefont {N.~T.}\ \bibnamefont
  {Padial}},\ }\href@noop {} {\bibfield  {journal} {\bibinfo  {journal} {Phys.
  Rev. A}\ }\textbf {\bibinfo {volume} {25}},\ \bibinfo {pages} {226} (\bibinfo
  {year} {1982})}\BibitemShut {NoStop}%
\bibitem [{\citenamefont {Sanna}\ and\ \citenamefont
  {Gianturco}(1998)}]{polydcs}%
  \BibitemOpen
  \bibfield  {author} {\bibinfo {author} {\bibfnamefont {N.}~\bibnamefont
  {Sanna}}\ and\ \bibinfo {author} {\bibfnamefont {F.~A.}\ \bibnamefont
  {Gianturco}},\ }\href@noop {} {\bibfield  {journal} {\bibinfo  {journal}
  {Comput. Phys. Commun.}\ }\textbf {\bibinfo {volume} {114}},\ \bibinfo
  {pages} {142} (\bibinfo {year} {1998})}\BibitemShut {NoStop}%
\bibitem [{\citenamefont {O'Malley}(1966)}]{omalley66}%
  \BibitemOpen
  \bibfield  {author} {\bibinfo {author} {\bibfnamefont {T.~F.}\ \bibnamefont
  {O'Malley}},\ }\href@noop {} {\bibfield  {journal} {\bibinfo  {journal}
  {Phys. Rev.}\ }\textbf {\bibinfo {volume} {150}},\ \bibinfo {pages} {14}
  (\bibinfo {year} {1966})}\BibitemShut {NoStop}%
\bibitem [{\citenamefont {Bates}(1991)}]{bates1991relative}%
  \BibitemOpen
  \bibfield  {author} {\bibinfo {author} {\bibfnamefont {D.~R.}\ \bibnamefont
  {Bates}},\ }\href@noop {} {\bibfield  {journal} {\bibinfo  {journal} {J.
  Phys. B: At., Mol. Opt. Phys.}\ }\textbf {\bibinfo {volume} {24}},\ \bibinfo
  {pages} {695} (\bibinfo {year} {1991})}\BibitemShut {NoStop}%
\bibitem [{\citenamefont {Guberman}\ and\ \citenamefont
  {Giusti-Suzor}(1991)}]{guberman1991generation}%
  \BibitemOpen
  \bibfield  {author} {\bibinfo {author} {\bibfnamefont {S.~L.}\ \bibnamefont
  {Guberman}}\ and\ \bibinfo {author} {\bibfnamefont {A.}~\bibnamefont
  {Giusti-Suzor}},\ }\href@noop {} {\bibfield  {journal} {\bibinfo  {journal}
  {J. Chem. Phys.}\ }\textbf {\bibinfo {volume} {95}},\ \bibinfo {pages} {2602}
  (\bibinfo {year} {1991})}\BibitemShut {NoStop}%
\bibitem [{\citenamefont {Morgan}\ and\ \citenamefont {Noble}(1984)}]{mn84}%
  \BibitemOpen
  \bibfield  {author} {\bibinfo {author} {\bibfnamefont {L.~A.}\ \bibnamefont
  {Morgan}}\ and\ \bibinfo {author} {\bibfnamefont {C.~J.}\ \bibnamefont
  {Noble}},\ }\href@noop {} {\bibfield  {journal} {\bibinfo  {journal} {J.
  Phys.B: At. Mol. Phys.}\ }\textbf {\bibinfo {volume} {17}},\ \bibinfo {pages}
  {L369} (\bibinfo {year} {1984})}\BibitemShut {NoStop}%
\bibitem [{\citenamefont {Fujimoto}\ \emph {et~al.}(2012)\citenamefont
  {Fujimoto}, \citenamefont {Brigg},\ and\ \citenamefont {Tennyson}}]{jt533}%
  \BibitemOpen
  \bibfield  {author} {\bibinfo {author} {\bibfnamefont {M.~M.}\ \bibnamefont
  {Fujimoto}}, \bibinfo {author} {\bibfnamefont {W.~J.}\ \bibnamefont {Brigg}},
  \ and\ \bibinfo {author} {\bibfnamefont {J.}~\bibnamefont {Tennyson}},\
  }\href@noop {} {\bibfield  {journal} {\bibinfo  {journal} {Eur. Phys. J. D}\
  }\textbf {\bibinfo {volume} {66}},\ \bibinfo {pages} {204} (\bibinfo {year}
  {2012})}\BibitemShut {NoStop}%
\bibitem [{\citenamefont {Bardsley}(1967)}]{bardsley1967ionization}%
  \BibitemOpen
  \bibfield  {author} {\bibinfo {author} {\bibfnamefont {J.}~\bibnamefont
  {Bardsley}},\ }\href@noop {} {\bibfield  {journal} {\bibinfo  {journal}
  {Chem. Phys. Lett.}\ }\textbf {\bibinfo {volume} {1}},\ \bibinfo {pages}
  {229} (\bibinfo {year} {1967})}\BibitemShut {NoStop}%
\bibitem [{\citenamefont {Kokoouline}\ \emph {et~al.}(2011)\citenamefont
  {Kokoouline}, \citenamefont {Douguet},\ and\ \citenamefont
  {Greene}}]{kokoouline11a}%
  \BibitemOpen
  \bibfield  {author} {\bibinfo {author} {\bibfnamefont {V.}~\bibnamefont
  {Kokoouline}}, \bibinfo {author} {\bibfnamefont {N.}~\bibnamefont {Douguet}},
  \ and\ \bibinfo {author} {\bibfnamefont {C.~H.}\ \bibnamefont {Greene}},\
  }\href@noop {} {\bibfield  {journal} {\bibinfo  {journal} {Chem. Phys.
  Lett.}\ }\textbf {\bibinfo {volume} {507}},\ \bibinfo {pages} {1} (\bibinfo
  {year} {2011})}\BibitemShut {NoStop}%
\bibitem [{\citenamefont {Douguet}\ \emph
  {et~al.}(2012{\natexlab{a}})\citenamefont {Douguet}, \citenamefont {Orel},
  \citenamefont {Greene},\ and\ \citenamefont {Kokoouline}}]{douguet12a}%
  \BibitemOpen
  \bibfield  {author} {\bibinfo {author} {\bibfnamefont {N.}~\bibnamefont
  {Douguet}}, \bibinfo {author} {\bibfnamefont {A.~E.}\ \bibnamefont {Orel}},
  \bibinfo {author} {\bibfnamefont {C.~H.}\ \bibnamefont {Greene}}, \ and\
  \bibinfo {author} {\bibfnamefont {V.}~\bibnamefont {Kokoouline}},\
  }\href@noop {} {\bibfield  {journal} {\bibinfo  {journal} {Phys. Phys.
  Lett.}\ }\textbf {\bibinfo {volume} {108}} (\bibinfo {year}
  {2012}{\natexlab{a}})}\BibitemShut {NoStop}%
\bibitem [{\citenamefont {Douguet}\ \emph
  {et~al.}(2012{\natexlab{b}})\citenamefont {Douguet}, \citenamefont
  {Kokoouline},\ and\ \citenamefont {Orel}}]{douguet12b}%
  \BibitemOpen
  \bibfield  {author} {\bibinfo {author} {\bibfnamefont {N.}~\bibnamefont
  {Douguet}}, \bibinfo {author} {\bibfnamefont {V.}~\bibnamefont {Kokoouline}},
  \ and\ \bibinfo {author} {\bibfnamefont {A.~E.}\ \bibnamefont {Orel}},\
  }\href@noop {} {\bibfield  {journal} {\bibinfo  {journal} {J. Phys. B: At.
  Mol. Opt. Phys.}\ }\textbf {\bibinfo {volume} {45}},\ \bibinfo {pages}
  {051001} (\bibinfo {year} {2012}{\natexlab{b}})}\BibitemShut {NoStop}%
\bibitem [{\citenamefont {{Fonseca dos Santos}}\ \emph
  {et~al.}(2014)\citenamefont {{Fonseca dos Santos}}, \citenamefont {Douguet},
  \citenamefont {Kokoouline},\ and\ \citenamefont {Orel}}]{samantha14}%
  \BibitemOpen
  \bibfield  {author} {\bibinfo {author} {\bibfnamefont {S.}~\bibnamefont
  {{Fonseca dos Santos}}}, \bibinfo {author} {\bibfnamefont {N.}~\bibnamefont
  {Douguet}}, \bibinfo {author} {\bibfnamefont {V.}~\bibnamefont {Kokoouline}},
  \ and\ \bibinfo {author} {\bibfnamefont {A.}~\bibnamefont {Orel}},\
  }\href@noop {} {\bibfield  {journal} {\bibinfo  {journal} {J. Chem. Phys.}\
  }\textbf {\bibinfo {volume} {140}},\ \bibinfo {pages} {164308} (\bibinfo
  {year} {2014})}\BibitemShut {NoStop}%
\bibitem [{\citenamefont {Little}\ \emph {et~al.}(2014)\citenamefont {Little},
  \citenamefont {Chakrabarti}, \citenamefont {Schneider},\ and\ \citenamefont
  {Tennyson}}]{jt591}%
  \BibitemOpen
  \bibfield  {author} {\bibinfo {author} {\bibfnamefont {D.~A.}\ \bibnamefont
  {Little}}, \bibinfo {author} {\bibfnamefont {K.}~\bibnamefont {Chakrabarti}},
  \bibinfo {author} {\bibfnamefont {I.~F.}\ \bibnamefont {Schneider}}, \ and\
  \bibinfo {author} {\bibfnamefont {J.}~\bibnamefont {Tennyson}},\ }\href@noop
  {} {\bibfield  {journal} {\bibinfo  {journal} {Phys. Rev. A}\ }\textbf
  {\bibinfo {volume} {90}},\ \bibinfo {pages} {052705} (\bibinfo {year}
  {2014})}\BibitemShut {NoStop}%
\bibitem [{\citenamefont {Faure}\ and\ \citenamefont {Tennyson}(2002)}]{jt288}%
  \BibitemOpen
  \bibfield  {author} {\bibinfo {author} {\bibfnamefont {A.}~\bibnamefont
  {Faure}}\ and\ \bibinfo {author} {\bibfnamefont {J.}~\bibnamefont
  {Tennyson}},\ }\href@noop {} {\bibfield  {journal} {\bibinfo  {journal} {J.
  Phys. B: At. Mol. Opt. Phys.}\ }\textbf {\bibinfo {volume} {35}},\ \bibinfo
  {pages} {1865} (\bibinfo {year} {2002})}\BibitemShut {NoStop}%
\bibitem [{\citenamefont {Kokoouline}\ and\ \citenamefont
  {Greene}(2003{\natexlab{a}})}]{kokoouline03a}%
  \BibitemOpen
  \bibfield  {author} {\bibinfo {author} {\bibfnamefont {V.}~\bibnamefont
  {Kokoouline}}\ and\ \bibinfo {author} {\bibfnamefont {C.~H.}\ \bibnamefont
  {Greene}},\ }\href@noop {} {\bibfield  {journal} {\bibinfo  {journal} {Phys.
  Rev. Lett.}\ }\textbf {\bibinfo {volume} {90}},\ \bibinfo {pages} {133201}
  (\bibinfo {year} {2003}{\natexlab{a}})}\BibitemShut {NoStop}%
\bibitem [{\citenamefont {Kokoouline}\ and\ \citenamefont
  {Greene}(2003{\natexlab{b}})}]{kokoouline03b}%
  \BibitemOpen
  \bibfield  {author} {\bibinfo {author} {\bibfnamefont {V.}~\bibnamefont
  {Kokoouline}}\ and\ \bibinfo {author} {\bibfnamefont {C.~H.}\ \bibnamefont
  {Greene}},\ }\href@noop {} {\bibfield  {journal} {\bibinfo  {journal} {Phys.
  Rev. A}\ }\textbf {\bibinfo {volume} {68}},\ \bibinfo {pages} {012703}
  (\bibinfo {year} {2003}{\natexlab{b}})}\BibitemShut {NoStop}%
\bibitem [{\citenamefont {Jungen}\ and\ \citenamefont
  {Pratt}(2008{\natexlab{a}})}]{jungen08a}%
  \BibitemOpen
  \bibfield  {author} {\bibinfo {author} {\bibfnamefont {C.}~\bibnamefont
  {Jungen}}\ and\ \bibinfo {author} {\bibfnamefont {S.~T.}\ \bibnamefont
  {Pratt}},\ }\href@noop {} {\bibfield  {journal} {\bibinfo  {journal} {J.
  Chem. Phys.}\ }\textbf {\bibinfo {volume} {129}},\ \bibinfo {pages} {164310}
  (\bibinfo {year} {2008}{\natexlab{a}})}\BibitemShut {NoStop}%
\bibitem [{\citenamefont {Jungen}\ and\ \citenamefont
  {Pratt}(2008{\natexlab{b}})}]{jungen08b}%
  \BibitemOpen
  \bibfield  {author} {\bibinfo {author} {\bibfnamefont {C.}~\bibnamefont
  {Jungen}}\ and\ \bibinfo {author} {\bibfnamefont {S.~T.}\ \bibnamefont
  {Pratt}},\ }\href@noop {} {\bibfield  {journal} {\bibinfo  {journal} {J.
  Chem. Phys.}\ }\textbf {\bibinfo {volume} {129}},\ \bibinfo {pages} {164311}
  (\bibinfo {year} {2008}{\natexlab{b}})}\BibitemShut {NoStop}%
\bibitem [{\citenamefont {Jungen}\ and\ \citenamefont
  {Pratt}(2009)}]{jungen09}%
  \BibitemOpen
  \bibfield  {author} {\bibinfo {author} {\bibfnamefont {C.}~\bibnamefont
  {Jungen}}\ and\ \bibinfo {author} {\bibfnamefont {S.~T.}\ \bibnamefont
  {Pratt}},\ }\href@noop {} {\bibfield  {journal} {\bibinfo  {journal} {Phys.
  Rev. Lett.}\ }\textbf {\bibinfo {volume} {102}},\ \bibinfo {pages} {023201}
  (\bibinfo {year} {2009})}\BibitemShut {NoStop}%
\bibitem [{\citenamefont {Gorfinkiel}\ and\ \citenamefont
  {Tennyson}(2005)}]{jt354}%
  \BibitemOpen
  \bibfield  {author} {\bibinfo {author} {\bibfnamefont {J.~D.}\ \bibnamefont
  {Gorfinkiel}}\ and\ \bibinfo {author} {\bibfnamefont {J.}~\bibnamefont
  {Tennyson}},\ }\href@noop {} {\bibfield  {journal} {\bibinfo  {journal} {J.
  Phys. B: At. Mol. Opt. Phys.}\ }\textbf {\bibinfo {volume} {38}},\ \bibinfo
  {pages} {1607} (\bibinfo {year} {2005})}\BibitemShut {NoStop}%
\bibitem [{\citenamefont {Halmov{\'a}}\ and\ \citenamefont
  {Tennyson}(2008)}]{jt434}%
  \BibitemOpen
  \bibfield  {author} {\bibinfo {author} {\bibfnamefont {G.}~\bibnamefont
  {Halmov{\'a}}}\ and\ \bibinfo {author} {\bibfnamefont {J.}~\bibnamefont
  {Tennyson}},\ }\href@noop {} {\bibfield  {journal} {\bibinfo  {journal}
  {Phys. Rev. Lett.}\ }\textbf {\bibinfo {volume} {100}},\ \bibinfo {pages}
  {213202} (\bibinfo {year} {2008})}\BibitemShut {NoStop}%
\bibitem [{\citenamefont {Pindzola}\ \emph {et~al.}(2012)\citenamefont
  {Pindzola}, \citenamefont {Abdel-Naby}, \citenamefont {Ludlow}, \citenamefont
  {Robicheaux},\ and\ \citenamefont {Colgan}}]{12PiAbLu}%
  \BibitemOpen
  \bibfield  {author} {\bibinfo {author} {\bibfnamefont {M.~S.}\ \bibnamefont
  {Pindzola}}, \bibinfo {author} {\bibfnamefont {S.~A.}\ \bibnamefont
  {Abdel-Naby}}, \bibinfo {author} {\bibfnamefont {J.~A.}\ \bibnamefont
  {Ludlow}}, \bibinfo {author} {\bibfnamefont {F.}~\bibnamefont {Robicheaux}},
  \ and\ \bibinfo {author} {\bibfnamefont {J.}~\bibnamefont {Colgan}},\
  }\href@noop {} {\bibfield  {journal} {\bibinfo  {journal} {Phys. Rev. A}\
  }\textbf {\bibinfo {volume} {85}} (\bibinfo {year} {2012})}\BibitemShut
  {NoStop}%
\bibitem [{\citenamefont {Colgan}\ and\ \citenamefont
  {Pindzola}(2012)}]{12CoPi}%
  \BibitemOpen
  \bibfield  {author} {\bibinfo {author} {\bibfnamefont {J.}~\bibnamefont
  {Colgan}}\ and\ \bibinfo {author} {\bibfnamefont {M.~S.}\ \bibnamefont
  {Pindzola}},\ }\href {\doibase 10.1140/epjd/e2012-30517-2} {\bibfield
  {journal} {\bibinfo  {journal} {Eur. Phys. J. D}\ }\textbf {\bibinfo {volume}
  {66}},\ \bibinfo {pages} {284} (\bibinfo {year} {2012})}\BibitemShut
  {NoStop}%
\bibitem [{\citenamefont {Gao}\ \emph {et~al.}(2005)\citenamefont {Gao},
  \citenamefont {Madison},\ and\ \citenamefont {Peacher}}]{Gao95}%
  \BibitemOpen
  \bibfield  {author} {\bibinfo {author} {\bibfnamefont {J.~F.}\ \bibnamefont
  {Gao}}, \bibinfo {author} {\bibfnamefont {D.~H.}\ \bibnamefont {Madison}}, \
  and\ \bibinfo {author} {\bibfnamefont {J.~L.}\ \bibnamefont {Peacher}},\
  }\href {\doibase {10.1063/1.2126971}} {\bibfield  {journal} {\bibinfo
  {journal} {J. Chem. Phys.}\ }\textbf {\bibinfo {volume} {{123}}},\ \bibinfo
  {pages} {204314} (\bibinfo {year} {{2005}})}\BibitemShut {NoStop}%
\bibitem [{\citenamefont {Kaiser}\ \emph {et~al.}(2007)\citenamefont {Kaiser},
  \citenamefont {Spieker}, \citenamefont {Gao}, \citenamefont {Hussey},
  \citenamefont {Murray},\ and\ \citenamefont {Madison}}]{Kaiser07}%
  \BibitemOpen
  \bibfield  {author} {\bibinfo {author} {\bibfnamefont {C.}~\bibnamefont
  {Kaiser}}, \bibinfo {author} {\bibfnamefont {D.}~\bibnamefont {Spieker}},
  \bibinfo {author} {\bibfnamefont {J.}~\bibnamefont {Gao}}, \bibinfo {author}
  {\bibfnamefont {M.}~\bibnamefont {Hussey}}, \bibinfo {author} {\bibfnamefont
  {A.}~\bibnamefont {Murray}}, \ and\ \bibinfo {author} {\bibfnamefont {D.~H.}\
  \bibnamefont {Madison}},\ }\href {\doibase {10.1088/0953-4075/40/13/003}}
  {\bibfield  {journal} {\bibinfo  {journal} {J. Phys. B: At. Mol. Opt. Phys.}\
  }\textbf {\bibinfo {volume} {{40}}},\ \bibinfo {pages} {2563} (\bibinfo
  {year} {{2007}})}\BibitemShut {NoStop}%
\bibitem [{\citenamefont {Bransden}\ and\ \citenamefont
  {McDowell}(1992)}]{Bransden92}%
  \BibitemOpen
  \bibfield  {author} {\bibinfo {author} {\bibfnamefont {B.~H.}\ \bibnamefont
  {Bransden}}\ and\ \bibinfo {author} {\bibfnamefont {M.~R.~C.}\ \bibnamefont
  {McDowell}},\ }\href@noop {} {\emph {\bibinfo {title} {Charge Exchange and
  the Theory of Ion-Atom Collisions}}}\ (\bibinfo  {publisher} {Clarendon
  Press, Oxford},\ \bibinfo {year} {1992})\BibitemShut {NoStop}%
\bibitem [{\citenamefont {Belki\'{c}}(2008)}]{Belkic08a}%
  \BibitemOpen
  \bibfield  {author} {\bibinfo {author} {\bibfnamefont {D.}~\bibnamefont
  {Belki\'{c}}},\ }\href@noop {} {\emph {\bibinfo {title} {Quantum Theory of
  High-Energy Ion-Atom Collisions}}}\ (\bibinfo  {publisher} {CRC Press, Boca
  Raton},\ \bibinfo {year} {2008})\BibitemShut {NoStop}%
\bibitem [{\citenamefont {Loreau}\ \emph {et~al.}(2014)\citenamefont {Loreau},
  \citenamefont {Ryabchenko},\ and\ \citenamefont {Vaeck}}]{Loreau14}%
  \BibitemOpen
  \bibfield  {author} {\bibinfo {author} {\bibfnamefont {J.}~\bibnamefont
  {Loreau}}, \bibinfo {author} {\bibfnamefont {S.}~\bibnamefont {Ryabchenko}},
  \ and\ \bibinfo {author} {\bibfnamefont {N.}~\bibnamefont {Vaeck}},\
  }\href@noop {} {\bibfield  {journal} {\bibinfo  {journal} {J. Phys. B}\
  }\textbf {\bibinfo {volume} {47}},\ \bibinfo {pages} {135204} (\bibinfo
  {year} {2014})}\BibitemShut {NoStop}%
\bibitem [{\citenamefont {Delos}(1981)}]{Delos81}%
  \BibitemOpen
  \bibfield  {author} {\bibinfo {author} {\bibfnamefont {J.~B.}\ \bibnamefont
  {Delos}},\ }\href@noop {} {\bibfield  {journal} {\bibinfo  {journal} {Rev.
  Mod. Phys.}\ }\textbf {\bibinfo {volume} {53}},\ \bibinfo {pages} {287}
  (\bibinfo {year} {1981})}\BibitemShut {NoStop}%
\bibitem [{\citenamefont {Errea}\ \emph {et~al.}(1994)\citenamefont {Errea},
  \citenamefont {Harel}, \citenamefont {Jouin}, \citenamefont {M\'endez},
  \citenamefont {Pons},\ and\ \citenamefont {R\'{\i}era}}]{Errea94}%
  \BibitemOpen
  \bibfield  {author} {\bibinfo {author} {\bibfnamefont {L.~F.}\ \bibnamefont
  {Errea}}, \bibinfo {author} {\bibfnamefont {C.}~\bibnamefont {Harel}},
  \bibinfo {author} {\bibfnamefont {H.}~\bibnamefont {Jouin}}, \bibinfo
  {author} {\bibfnamefont {L.}~\bibnamefont {M\'endez}}, \bibinfo {author}
  {\bibfnamefont {B.}~\bibnamefont {Pons}}, \ and\ \bibinfo {author}
  {\bibfnamefont {A.}~\bibnamefont {R\'{\i}era}},\ }\href@noop {} {\bibfield
  {journal} {\bibinfo  {journal} {J. Phys. B}\ }\textbf {\bibinfo {volume}
  {27}},\ \bibinfo {pages} {3603} (\bibinfo {year} {1994})}\BibitemShut
  {NoStop}%
\bibitem [{\citenamefont {Liu}\ \emph {et~al.}(2005)\citenamefont {Liu},
  \citenamefont {Cheng}, \citenamefont {Le},\ and\ \citenamefont
  {Lin}}]{Liu05}%
  \BibitemOpen
  \bibfield  {author} {\bibinfo {author} {\bibfnamefont {C.-N.}\ \bibnamefont
  {Liu}}, \bibinfo {author} {\bibfnamefont {S.-C.}\ \bibnamefont {Cheng}},
  \bibinfo {author} {\bibfnamefont {A.-T.}\ \bibnamefont {Le}}, \ and\ \bibinfo
  {author} {\bibfnamefont {C.~D.}\ \bibnamefont {Lin}},\ }\href@noop {}
  {\bibfield  {journal} {\bibinfo  {journal} {Phys. Rev. A}\ }\textbf {\bibinfo
  {volume} {72}},\ \bibinfo {pages} {012717} (\bibinfo {year}
  {2005})}\BibitemShut {NoStop}%
\bibitem [{\citenamefont {Zygelman}\ \emph {et~al.}(1997)\citenamefont
  {Zygelman}, \citenamefont {Stancil}, \citenamefont {Clarke},\ and\
  \citenamefont {Cooper}}]{Zygelman97}%
  \BibitemOpen
  \bibfield  {author} {\bibinfo {author} {\bibfnamefont {B.}~\bibnamefont
  {Zygelman}}, \bibinfo {author} {\bibfnamefont {P.~C.}\ \bibnamefont
  {Stancil}}, \bibinfo {author} {\bibfnamefont {N.~J.}\ \bibnamefont {Clarke}},
  \ and\ \bibinfo {author} {\bibfnamefont {D.~L.}\ \bibnamefont {Cooper}},\
  }\href@noop {} {\bibfield  {journal} {\bibinfo  {journal} {Phys. Rev. A}\
  }\textbf {\bibinfo {volume} {56}},\ \bibinfo {pages} {457} (\bibinfo {year}
  {1997})}\BibitemShut {NoStop}%
\bibitem [{\citenamefont {Li}\ \emph {et~al.}(2015)\citenamefont {Li},
  \citenamefont {Qu}, \citenamefont {Wu}, \citenamefont {Liu}, \citenamefont
  {Wang}, \citenamefont {Liebermann},\ and\ \citenamefont {Buenker}}]{Li15}%
  \BibitemOpen
  \bibfield  {author} {\bibinfo {author} {\bibfnamefont {T.~C.}\ \bibnamefont
  {Li}}, \bibinfo {author} {\bibfnamefont {Y.~Z.}\ \bibnamefont {Qu}}, \bibinfo
  {author} {\bibfnamefont {Y.}~\bibnamefont {Wu}}, \bibinfo {author}
  {\bibfnamefont {L.}~\bibnamefont {Liu}}, \bibinfo {author} {\bibfnamefont
  {J.~G.}\ \bibnamefont {Wang}}, \bibinfo {author} {\bibfnamefont {H.-P.}\
  \bibnamefont {Liebermann}}, \ and\ \bibinfo {author} {\bibfnamefont {R.~J.}\
  \bibnamefont {Buenker}},\ }\href@noop {} {\bibfield  {journal} {\bibinfo
  {journal} {Phys. Rev. A}\ }\textbf {\bibinfo {volume} {91}},\ \bibinfo
  {pages} {052702} (\bibinfo {year} {2015})}\BibitemShut {NoStop}%
\bibitem [{\citenamefont {Abdurakhmanov}\ \emph {et~al.}(2016)\citenamefont
  {Abdurakhmanov}, \citenamefont {Kadyrov},\ and\ \citenamefont
  {Bray}}]{Abdura15}%
  \BibitemOpen
  \bibfield  {author} {\bibinfo {author} {\bibfnamefont {I.~B.}\ \bibnamefont
  {Abdurakhmanov}}, \bibinfo {author} {\bibfnamefont {A.~S.}\ \bibnamefont
  {Kadyrov}}, \ and\ \bibinfo {author} {\bibfnamefont {I.}~\bibnamefont
  {Bray}},\ }\href@noop {} {\bibfield  {journal} {\bibinfo  {journal} {J. Phys.
  B}\ }\textbf {\bibinfo {volume} {49}},\ \bibinfo {pages} {03LT01} (\bibinfo
  {year} {2016})}\BibitemShut {NoStop}%
\bibitem [{\citenamefont {Deumens}\ \emph {et~al.}(1994)\citenamefont
  {Deumens}, \citenamefont {Diz}, \citenamefont {Longo},\ and\ \citenamefont
  {\"Ohrn}}]{Deumens94}%
  \BibitemOpen
  \bibfield  {author} {\bibinfo {author} {\bibfnamefont {E.}~\bibnamefont
  {Deumens}}, \bibinfo {author} {\bibfnamefont {A.}~\bibnamefont {Diz}},
  \bibinfo {author} {\bibfnamefont {R.}~\bibnamefont {Longo}}, \ and\ \bibinfo
  {author} {\bibfnamefont {Y.}~\bibnamefont {\"Ohrn}},\ }\href@noop {}
  {\bibfield  {journal} {\bibinfo  {journal} {Rev. Mod. Phys.}\ }\textbf
  {\bibinfo {volume} {66}},\ \bibinfo {pages} {917} (\bibinfo {year}
  {1994})}\BibitemShut {NoStop}%
\bibitem [{\citenamefont {Cabrera-Trujillo}(2010)}]{Cabrera10}%
  \BibitemOpen
  \bibfield  {author} {\bibinfo {author} {\bibfnamefont {R.}~\bibnamefont
  {Cabrera-Trujillo}},\ }\href@noop {} {\bibfield  {journal} {\bibinfo
  {journal} {Plasma Sources Sci. Technol.}\ }\textbf {\bibinfo {volume} {19}},\
  \bibinfo {pages} {034006} (\bibinfo {year} {2010})}\BibitemShut {NoStop}%
\bibitem [{\citenamefont {Green}\ \emph
  {et~al.}(1982{\natexlab{a}})\citenamefont {Green}, \citenamefont {Shipsey},\
  and\ \citenamefont {Browne}}]{Green82a}%
  \BibitemOpen
  \bibfield  {author} {\bibinfo {author} {\bibfnamefont {T.~A.}\ \bibnamefont
  {Green}}, \bibinfo {author} {\bibfnamefont {E.~J.}\ \bibnamefont {Shipsey}},
  \ and\ \bibinfo {author} {\bibfnamefont {J.~C.}\ \bibnamefont {Browne}},\
  }\href@noop {} {\bibfield  {journal} {\bibinfo  {journal} {Phys. Rev. A}\
  }\textbf {\bibinfo {volume} {25}},\ \bibinfo {pages} {1364} (\bibinfo {year}
  {1982}{\natexlab{a}})}\BibitemShut {NoStop}%
\bibitem [{\citenamefont {Green}\ \emph
  {et~al.}(1982{\natexlab{b}})\citenamefont {Green}, \citenamefont {Riley},
  \citenamefont {Shipsey},\ and\ \citenamefont {Browne}}]{Green82b}%
  \BibitemOpen
  \bibfield  {author} {\bibinfo {author} {\bibfnamefont {T.~A.}\ \bibnamefont
  {Green}}, \bibinfo {author} {\bibfnamefont {M.~E.}\ \bibnamefont {Riley}},
  \bibinfo {author} {\bibfnamefont {E.~J.}\ \bibnamefont {Shipsey}}, \ and\
  \bibinfo {author} {\bibfnamefont {J.~C.}\ \bibnamefont {Browne}},\
  }\href@noop {} {\bibfield  {journal} {\bibinfo  {journal} {Phys. Rev. A}\
  }\textbf {\bibinfo {volume} {26}},\ \bibinfo {pages} {3668} (\bibinfo {year}
  {1982}{\natexlab{b}})}\BibitemShut {NoStop}%
\bibitem [{\citenamefont {Fritsch}\ and\ \citenamefont
  {Lin}(1984)}]{Fritsch84}%
  \BibitemOpen
  \bibfield  {author} {\bibinfo {author} {\bibfnamefont {W.}~\bibnamefont
  {Fritsch}}\ and\ \bibinfo {author} {\bibfnamefont {C.~D.}\ \bibnamefont
  {Lin}},\ }\href@noop {} {\bibfield  {journal} {\bibinfo  {journal} {Phys.
  Rev. A}\ }\textbf {\bibinfo {volume} {29}},\ \bibinfo {pages} {3039}
  (\bibinfo {year} {1984})}\BibitemShut {NoStop}%
\bibitem [{\citenamefont {van Hemert}\ \emph {et~al.}(1985)\citenamefont {van
  Hemert}, \citenamefont {van Dishoeck}, \citenamefont {van~der Hart},\ and\
  \citenamefont {Koike}}]{Hemert85}%
  \BibitemOpen
  \bibfield  {author} {\bibinfo {author} {\bibfnamefont {M.~C.}\ \bibnamefont
  {van Hemert}}, \bibinfo {author} {\bibfnamefont {E.~F.}\ \bibnamefont {van
  Dishoeck}}, \bibinfo {author} {\bibfnamefont {J.~A.}\ \bibnamefont {van~der
  Hart}}, \ and\ \bibinfo {author} {\bibfnamefont {F.}~\bibnamefont {Koike}},\
  }\href@noop {} {\bibfield  {journal} {\bibinfo  {journal} {Phys. Rev. A}\
  }\textbf {\bibinfo {volume} {31}},\ \bibinfo {pages} {2227} (\bibinfo {year}
  {1985})}\BibitemShut {NoStop}%
\bibitem [{\citenamefont {Fritsch}\ and\ \citenamefont
  {Lin}(1991)}]{Fritsch91}%
  \BibitemOpen
  \bibfield  {author} {\bibinfo {author} {\bibfnamefont {W.}~\bibnamefont
  {Fritsch}}\ and\ \bibinfo {author} {\bibfnamefont {C.~D.}\ \bibnamefont
  {Lin}},\ }\href@noop {} {\bibfield  {journal} {\bibinfo  {journal} {Phys.
  Rep.}\ }\textbf {\bibinfo {volume} {202}},\ \bibinfo {pages} {1} (\bibinfo
  {year} {1991})}\BibitemShut {NoStop}%
\bibitem [{\citenamefont {Zapukhlyak}\ \emph {et~al.}(2005)\citenamefont
  {Zapukhlyak}, \citenamefont {Kirchner}, \citenamefont {L\"udde},
  \citenamefont {Knoop}, \citenamefont {Morgenstern},\ and\ \citenamefont
  {Hoekstra}}]{tcbgm}%
  \BibitemOpen
  \bibfield  {author} {\bibinfo {author} {\bibfnamefont {M.}~\bibnamefont
  {Zapukhlyak}}, \bibinfo {author} {\bibfnamefont {T.}~\bibnamefont
  {Kirchner}}, \bibinfo {author} {\bibfnamefont {H.~J.}\ \bibnamefont
  {L\"udde}}, \bibinfo {author} {\bibfnamefont {S.}~\bibnamefont {Knoop}},
  \bibinfo {author} {\bibfnamefont {R.}~\bibnamefont {Morgenstern}}, \ and\
  \bibinfo {author} {\bibfnamefont {R.}~\bibnamefont {Hoekstra}},\ }\href@noop
  {} {\bibfield  {journal} {\bibinfo  {journal} {J. Phys. B}\ }\textbf
  {\bibinfo {volume} {38}},\ \bibinfo {pages} {2353} (\bibinfo {year}
  {2005})}\BibitemShut {NoStop}%
\bibitem [{\citenamefont {Kuang}\ and\ \citenamefont {Lin}(1997)}]{Kuang97}%
  \BibitemOpen
  \bibfield  {author} {\bibinfo {author} {\bibfnamefont {J.}~\bibnamefont
  {Kuang}}\ and\ \bibinfo {author} {\bibfnamefont {C.~D.}\ \bibnamefont
  {Lin}},\ }\href@noop {} {\bibfield  {journal} {\bibinfo  {journal} {J. Phys.
  B}\ }\textbf {\bibinfo {volume} {30}},\ \bibinfo {pages} {101} (\bibinfo
  {year} {1997})}\BibitemShut {NoStop}%
\bibitem [{\citenamefont {Kroneisen}\ \emph {et~al.}(1999)\citenamefont
  {Kroneisen}, \citenamefont {L\"udde}, \citenamefont {Kirchner},\ and\
  \citenamefont {Dreizler}}]{BGM99}%
  \BibitemOpen
  \bibfield  {author} {\bibinfo {author} {\bibfnamefont {O.~J.}\ \bibnamefont
  {Kroneisen}}, \bibinfo {author} {\bibfnamefont {H.~J.}\ \bibnamefont
  {L\"udde}}, \bibinfo {author} {\bibfnamefont {T.}~\bibnamefont {Kirchner}}, \
  and\ \bibinfo {author} {\bibfnamefont {R.~M.}\ \bibnamefont {Dreizler}},\
  }\href@noop {} {\bibfield  {journal} {\bibinfo  {journal} {J. Phys. A}\
  }\textbf {\bibinfo {volume} {32}},\ \bibinfo {pages} {2141} (\bibinfo {year}
  {1999})}\BibitemShut {NoStop}%
\bibitem [{\citenamefont {Minami}\ \emph {et~al.}(2006)\citenamefont {Minami},
  \citenamefont {Pindzola}, \citenamefont {Lee},\ and\ \citenamefont
  {Schultz}}]{Minami06}%
  \BibitemOpen
  \bibfield  {author} {\bibinfo {author} {\bibfnamefont {T.}~\bibnamefont
  {Minami}}, \bibinfo {author} {\bibfnamefont {M.~S.}\ \bibnamefont
  {Pindzola}}, \bibinfo {author} {\bibfnamefont {T.-G.}\ \bibnamefont {Lee}}, \
  and\ \bibinfo {author} {\bibfnamefont {D.~R.}\ \bibnamefont {Schultz}},\
  }\href@noop {} {\bibfield  {journal} {\bibinfo  {journal} {J. Phys. B}\
  }\textbf {\bibinfo {volume} {39}},\ \bibinfo {pages} {2877} (\bibinfo {year}
  {2006})}\BibitemShut {NoStop}%
\bibitem [{\citenamefont {Pindzola}\ and\ \citenamefont
  {Fogle}(2015)}]{Pindzola15}%
  \BibitemOpen
  \bibfield  {author} {\bibinfo {author} {\bibfnamefont {M.~S.}\ \bibnamefont
  {Pindzola}}\ and\ \bibinfo {author} {\bibfnamefont {M.}~\bibnamefont
  {Fogle}},\ }\href@noop {} {\bibfield  {journal} {\bibinfo  {journal} {J.
  Phys. B}\ }\textbf {\bibinfo {volume} {48}},\ \bibinfo {pages} {205203}
  (\bibinfo {year} {2015})}\BibitemShut {NoStop}%
\bibitem [{\citenamefont {Belki\'{c}}\ \emph {et~al.}(2008)\citenamefont
  {Belki\'{c}}, \citenamefont {Man\u{c}ev},\ and\ \citenamefont
  {Hanssen}}]{Belkic08b}%
  \BibitemOpen
  \bibfield  {author} {\bibinfo {author} {\bibfnamefont {D.}~\bibnamefont
  {Belki\'{c}}}, \bibinfo {author} {\bibfnamefont {I.}~\bibnamefont
  {Man\u{c}ev}}, \ and\ \bibinfo {author} {\bibfnamefont {J.}~\bibnamefont
  {Hanssen}},\ }\href@noop {} {\bibfield  {journal} {\bibinfo  {journal} {Rev.
  Mod. Phys.}\ }\textbf {\bibinfo {volume} {80}},\ \bibinfo {pages} {249}
  (\bibinfo {year} {2008})}\BibitemShut {NoStop}%
\bibitem [{\citenamefont {Niehaus}(1986)}]{Niehaus86}%
  \BibitemOpen
  \bibfield  {author} {\bibinfo {author} {\bibfnamefont {A.}~\bibnamefont
  {Niehaus}},\ }\href@noop {} {\bibfield  {journal} {\bibinfo  {journal} {J.
  Phys. B}\ }\textbf {\bibinfo {volume} {19}},\ \bibinfo {pages} {2925}
  (\bibinfo {year} {1986})}\BibitemShut {NoStop}%
\bibitem [{\citenamefont {Olson}\ and\ \citenamefont {Salop}(1977)}]{Olson77}%
  \BibitemOpen
  \bibfield  {author} {\bibinfo {author} {\bibfnamefont {R.~E.}\ \bibnamefont
  {Olson}}\ and\ \bibinfo {author} {\bibfnamefont {A.}~\bibnamefont {Salop}},\
  }\href@noop {} {\bibfield  {journal} {\bibinfo  {journal} {Phys. Rev. A}\
  }\textbf {\bibinfo {volume} {16}},\ \bibinfo {pages} {531} (\bibinfo {year}
  {1977})}\BibitemShut {NoStop}%
\bibitem [{\citenamefont {Otranto}\ \emph {et~al.}(2014)\citenamefont
  {Otranto}, \citenamefont {Cariatore},\ and\ \citenamefont
  {Olson}}]{Otranto14}%
  \BibitemOpen
  \bibfield  {author} {\bibinfo {author} {\bibfnamefont {S.}~\bibnamefont
  {Otranto}}, \bibinfo {author} {\bibfnamefont {N.~D.}\ \bibnamefont
  {Cariatore}}, \ and\ \bibinfo {author} {\bibfnamefont {R.~E.}\ \bibnamefont
  {Olson}},\ }\href@noop {} {\bibfield  {journal} {\bibinfo  {journal} {Phys.
  Rev. A}\ }\textbf {\bibinfo {volume} {90}},\ \bibinfo {pages} {062708}
  (\bibinfo {year} {2014})}\BibitemShut {NoStop}%
\bibitem [{\citenamefont {S\'anchez}\ \emph {et~al.}(2006)\citenamefont
  {S\'anchez}, \citenamefont {N\"ortersh\"auser}, \citenamefont {Ewald},
  \citenamefont {Albers}, \citenamefont {Behr}, \citenamefont {Bricault},
  \citenamefont {Bushaw}, \citenamefont {Dax}, \citenamefont {Dilling},
  \citenamefont {Dombsky}, \citenamefont {Drake}, \citenamefont {G\"otte},
  \citenamefont {Kirchner}, \citenamefont {Kluge}, \citenamefont {K\"uhl},
  \citenamefont {Lassen}, \citenamefont {Levy}, \citenamefont {Pearson},
  \citenamefont {Prime}, \citenamefont {Ryjkov}, \citenamefont {Wojtaszek},
  \citenamefont {Yan},\ and\ \citenamefont {Zimmermann}}]{Sanchez_2006}%
  \BibitemOpen
  \bibfield  {author} {\bibinfo {author} {\bibfnamefont {R.}~\bibnamefont
  {S\'anchez}}, \bibinfo {author} {\bibfnamefont {W.}~\bibnamefont
  {N\"ortersh\"auser}}, \bibinfo {author} {\bibfnamefont {G.}~\bibnamefont
  {Ewald}}, \bibinfo {author} {\bibfnamefont {D.}~\bibnamefont {Albers}},
  \bibinfo {author} {\bibfnamefont {J.}~\bibnamefont {Behr}}, \bibinfo {author}
  {\bibfnamefont {P.}~\bibnamefont {Bricault}}, \bibinfo {author}
  {\bibfnamefont {B.~A.}\ \bibnamefont {Bushaw}}, \bibinfo {author}
  {\bibfnamefont {A.}~\bibnamefont {Dax}}, \bibinfo {author} {\bibfnamefont
  {J.}~\bibnamefont {Dilling}}, \bibinfo {author} {\bibfnamefont
  {M.}~\bibnamefont {Dombsky}}, \bibinfo {author} {\bibfnamefont {G.~W.~F.}\
  \bibnamefont {Drake}}, \bibinfo {author} {\bibfnamefont {S.}~\bibnamefont
  {G\"otte}}, \bibinfo {author} {\bibfnamefont {R.}~\bibnamefont {Kirchner}},
  \bibinfo {author} {\bibfnamefont {H.-J.}\ \bibnamefont {Kluge}}, \bibinfo
  {author} {\bibfnamefont {T.}~\bibnamefont {K\"uhl}}, \bibinfo {author}
  {\bibfnamefont {J.}~\bibnamefont {Lassen}}, \bibinfo {author} {\bibfnamefont
  {C.~D.~P.}\ \bibnamefont {Levy}}, \bibinfo {author} {\bibfnamefont {M.~R.}\
  \bibnamefont {Pearson}}, \bibinfo {author} {\bibfnamefont {E.~J.}\
  \bibnamefont {Prime}}, \bibinfo {author} {\bibfnamefont {V.}~\bibnamefont
  {Ryjkov}}, \bibinfo {author} {\bibfnamefont {A.}~\bibnamefont {Wojtaszek}},
  \bibinfo {author} {\bibfnamefont {Z.-C.}\ \bibnamefont {Yan}}, \ and\
  \bibinfo {author} {\bibfnamefont {C.}~\bibnamefont {Zimmermann}},\ }\href
  {http://link.aps.org/doi/10.1103/PhysRevLett.96.033002} {\bibfield  {journal}
  {\bibinfo  {journal} {Phys. Rev. Lett.}\ }\textbf {\bibinfo {volume} {96}},\
  \bibinfo {pages} {033002} (\bibinfo {year} {2006})}\BibitemShut {NoStop}%
\bibitem [{\citenamefont {Sanchez}\ \emph {et~al.}(2009)\citenamefont
  {Sanchez}, \citenamefont {Zakova}, \citenamefont {Andjelkovic}, \citenamefont
  {Bushaw}, \citenamefont {Dasgupta}, \citenamefont {Ewald}, \citenamefont
  {Geppert}, \citenamefont {Kluge}, \citenamefont {Kraemer}, \citenamefont
  {Nothhelfer}, \citenamefont {Tiedemann}, \citenamefont {Winters},\ and\
  \citenamefont {Noertershaeuser}}]{Sanchez_2009}%
  \BibitemOpen
  \bibfield  {author} {\bibinfo {author} {\bibfnamefont {R.}~\bibnamefont
  {Sanchez}}, \bibinfo {author} {\bibfnamefont {M.}~\bibnamefont {Zakova}},
  \bibinfo {author} {\bibfnamefont {Z.}~\bibnamefont {Andjelkovic}}, \bibinfo
  {author} {\bibfnamefont {B.~A.}\ \bibnamefont {Bushaw}}, \bibinfo {author}
  {\bibfnamefont {K.}~\bibnamefont {Dasgupta}}, \bibinfo {author}
  {\bibfnamefont {G.}~\bibnamefont {Ewald}}, \bibinfo {author} {\bibfnamefont
  {C.}~\bibnamefont {Geppert}}, \bibinfo {author} {\bibfnamefont {H.-J.}\
  \bibnamefont {Kluge}}, \bibinfo {author} {\bibfnamefont {J.}~\bibnamefont
  {Kraemer}}, \bibinfo {author} {\bibfnamefont {M.}~\bibnamefont {Nothhelfer}},
  \bibinfo {author} {\bibfnamefont {D.}~\bibnamefont {Tiedemann}}, \bibinfo
  {author} {\bibfnamefont {D.~F.~A.}\ \bibnamefont {Winters}}, \ and\ \bibinfo
  {author} {\bibfnamefont {W.}~\bibnamefont {Noertershaeuser}},\ }\href@noop {}
  {\bibfield  {journal} {\bibinfo  {journal} {New J. Phys.}\ }\textbf {\bibinfo
  {volume} {11}} (\bibinfo {year} {2009})}\BibitemShut {NoStop}%
\bibitem [{\citenamefont {Yan}\ \emph {et~al.}(2008{\natexlab{b}})\citenamefont
  {Yan}, \citenamefont {N\"ortersh\"auser},\ and\ \citenamefont
  {Drake}}]{YanDrake_2008}%
  \BibitemOpen
  \bibfield  {author} {\bibinfo {author} {\bibfnamefont {Z.-C.}\ \bibnamefont
  {Yan}}, \bibinfo {author} {\bibfnamefont {W.}~\bibnamefont
  {N\"ortersh\"auser}}, \ and\ \bibinfo {author} {\bibfnamefont {G.~W.~F.}\
  \bibnamefont {Drake}},\ }\href {\doibase 10.1103/PhysRevLett.100.243002}
  {\bibfield  {journal} {\bibinfo  {journal} {Phys. Rev. Lett.}\ }\textbf
  {\bibinfo {volume} {100}},\ \bibinfo {pages} {243002} (\bibinfo {year}
  {2008}{\natexlab{b}})},\ \bibinfo {note} {erratum: Phys.\ Rev.\ Lett.\ {\bf
  102}, 249903 (2009)}\BibitemShut {NoStop}%
\bibitem [{\citenamefont {Yan}\ \emph {et~al.}(2009)\citenamefont {Yan},
  \citenamefont {N\"ortersh\"auser},\ and\ \citenamefont
  {Drake}}]{YanDrake_2008E}%
  \BibitemOpen
  \bibfield  {author} {\bibinfo {author} {\bibfnamefont {Z.-C.}\ \bibnamefont
  {Yan}}, \bibinfo {author} {\bibfnamefont {W.}~\bibnamefont
  {N\"ortersh\"auser}}, \ and\ \bibinfo {author} {\bibfnamefont {G.~W.~F.}\
  \bibnamefont {Drake}},\ }\href {\doibase 10.1103/PhysRevLett.102.249903}
  {\bibfield  {journal} {\bibinfo  {journal} {Phys.\ Rev.\ Lett.}\ }\textbf
  {\bibinfo {volume} {102}},\ \bibinfo {pages} {249903} (\bibinfo {year}
  {2009})}\BibitemShut {NoStop}%
\bibitem [{\citenamefont {Puchalski}\ \emph {et~al.}(2010)\citenamefont
  {Puchalski}, \citenamefont {Kedziera},\ and\ \citenamefont
  {Pachucki}}]{Puchalski_2010}%
  \BibitemOpen
  \bibfield  {author} {\bibinfo {author} {\bibfnamefont {M.}~\bibnamefont
  {Puchalski}}, \bibinfo {author} {\bibfnamefont {D.}~\bibnamefont {Kedziera}},
  \ and\ \bibinfo {author} {\bibfnamefont {K.}~\bibnamefont {Pachucki}},\
  }\href {\doibase 10.1103/PhysRevA.82.062509} {\bibfield  {journal} {\bibinfo
  {journal} {Phys. Rev. A}\ }\textbf {\bibinfo {volume} {82}},\ \bibinfo
  {pages} {062509} (\bibinfo {year} {2010})}\BibitemShut {NoStop}%
\bibitem [{\citenamefont {Yan}\ and\ \citenamefont
  {Drake}(2003)}]{YanDrake_QED}%
  \BibitemOpen
  \bibfield  {author} {\bibinfo {author} {\bibfnamefont {Z.-C.}\ \bibnamefont
  {Yan}}\ and\ \bibinfo {author} {\bibfnamefont {G.~W.~F.}\ \bibnamefont
  {Drake}},\ }\href {\doibase 10.1103/PhysRevLett.91.113004} {\bibfield
  {journal} {\bibinfo  {journal} {Phys. Rev. Lett.}\ }\textbf {\bibinfo
  {volume} {91}},\ \bibinfo {pages} {113004} (\bibinfo {year}
  {2003})}\BibitemShut {NoStop}%
\bibitem [{\citenamefont {N\"ortersh\"auser}\ \emph {et~al.}(2011)\citenamefont
  {N\"ortersh\"auser}, \citenamefont {S\'anchez}, \citenamefont {Ewald},
  \citenamefont {Dax}, \citenamefont {Behr}, \citenamefont {Bricault},
  \citenamefont {Bushaw}, \citenamefont {Dilling}, \citenamefont {Dombsky},
  \citenamefont {Drake}, \citenamefont {G\"otte}, \citenamefont {Kluge},
  \citenamefont {K\"uhl}, \citenamefont {Lassen}, \citenamefont {Levy},
  \citenamefont {Pachucki}, \citenamefont {Pearson}, \citenamefont {Puchalski},
  \citenamefont {Wojtaszek}, \citenamefont {Yan},\ and\ \citenamefont
  {Zimmermann}}]{Drake_LiShift}%
  \BibitemOpen
  \bibfield  {author} {\bibinfo {author} {\bibfnamefont {W.}~\bibnamefont
  {N\"ortersh\"auser}}, \bibinfo {author} {\bibfnamefont {R.}~\bibnamefont
  {S\'anchez}}, \bibinfo {author} {\bibfnamefont {G.}~\bibnamefont {Ewald}},
  \bibinfo {author} {\bibfnamefont {A.}~\bibnamefont {Dax}}, \bibinfo {author}
  {\bibfnamefont {J.}~\bibnamefont {Behr}}, \bibinfo {author} {\bibfnamefont
  {P.}~\bibnamefont {Bricault}}, \bibinfo {author} {\bibfnamefont {B.~A.}\
  \bibnamefont {Bushaw}}, \bibinfo {author} {\bibfnamefont {J.}~\bibnamefont
  {Dilling}}, \bibinfo {author} {\bibfnamefont {M.}~\bibnamefont {Dombsky}},
  \bibinfo {author} {\bibfnamefont {G.~W.~F.}\ \bibnamefont {Drake}}, \bibinfo
  {author} {\bibfnamefont {S.}~\bibnamefont {G\"otte}}, \bibinfo {author}
  {\bibfnamefont {H.-J.}\ \bibnamefont {Kluge}}, \bibinfo {author}
  {\bibfnamefont {T.}~\bibnamefont {K\"uhl}}, \bibinfo {author} {\bibfnamefont
  {J.}~\bibnamefont {Lassen}}, \bibinfo {author} {\bibfnamefont {C.~D.~P.}\
  \bibnamefont {Levy}}, \bibinfo {author} {\bibfnamefont {K.}~\bibnamefont
  {Pachucki}}, \bibinfo {author} {\bibfnamefont {M.}~\bibnamefont {Pearson}},
  \bibinfo {author} {\bibfnamefont {M.}~\bibnamefont {Puchalski}}, \bibinfo
  {author} {\bibfnamefont {A.}~\bibnamefont {Wojtaszek}}, \bibinfo {author}
  {\bibfnamefont {Z.-C.}\ \bibnamefont {Yan}}, \ and\ \bibinfo {author}
  {\bibfnamefont {C.}~\bibnamefont {Zimmermann}},\ }\href {\doibase
  10.1103/PhysRevA.83.012516} {\bibfield  {journal} {\bibinfo  {journal} {Phys.
  Rev. A}\ }\textbf {\bibinfo {volume} {83}},\ \bibinfo {pages} {012516}
  (\bibinfo {year} {2011})}\BibitemShut {NoStop}%
\bibitem [{\citenamefont {Lu}\ \emph {et~al.}(2013)\citenamefont {Lu},
  \citenamefont {Mueller}, \citenamefont {Drake}, \citenamefont
  {N\"ortersh\"auser}, \citenamefont {Pieper},\ and\ \citenamefont
  {Yan}}]{Drake_RMP}%
  \BibitemOpen
  \bibfield  {author} {\bibinfo {author} {\bibfnamefont {Z.-T.}\ \bibnamefont
  {Lu}}, \bibinfo {author} {\bibfnamefont {P.}~\bibnamefont {Mueller}},
  \bibinfo {author} {\bibfnamefont {G.~W.~F.}\ \bibnamefont {Drake}}, \bibinfo
  {author} {\bibfnamefont {W.}~\bibnamefont {N\"ortersh\"auser}}, \bibinfo
  {author} {\bibfnamefont {S.~C.}\ \bibnamefont {Pieper}}, \ and\ \bibinfo
  {author} {\bibfnamefont {Z.-C.}\ \bibnamefont {Yan}},\ }\href {\doibase
  10.1103/RevModPhys.85.1383} {\bibfield  {journal} {\bibinfo  {journal} {Rev.
  Mod. Phys.}\ }\textbf {\bibinfo {volume} {85}},\ \bibinfo {pages} {1383}
  (\bibinfo {year} {2013})}\BibitemShut {NoStop}%
\bibitem [{\citenamefont {Puzzarini}\ \emph {et~al.}(2008)\citenamefont
  {Puzzarini}, \citenamefont {Heckert},\ and\ \citenamefont {Gauss}}]{phg08}%
  \BibitemOpen
  \bibfield  {author} {\bibinfo {author} {\bibfnamefont {C.}~\bibnamefont
  {Puzzarini}}, \bibinfo {author} {\bibfnamefont {M.}~\bibnamefont {Heckert}},
  \ and\ \bibinfo {author} {\bibfnamefont {J.}~\bibnamefont {Gauss}},\ }\href
  {\doibase {10.1063/1.2912941}} {\bibfield  {journal} {\bibinfo  {journal} {J.
  Chem. Phys.}\ }\textbf {\bibinfo {volume} {{128}}},\ \bibinfo {pages}
  {194108} (\bibinfo {year} {{2008}})}\BibitemShut {NoStop}%
\bibitem [{\citenamefont {Demaison}\ \emph {et~al.}(2010)\citenamefont
  {Demaison}, \citenamefont {Boggs},\ and\ \citenamefont
  {Csaszar}}]{Demaison2010}%
  \BibitemOpen
  \bibfield  {author} {\bibinfo {author} {\bibfnamefont {J.}~\bibnamefont
  {Demaison}}, \bibinfo {author} {\bibfnamefont {J.~E.}\ \bibnamefont {Boggs}},
  \ and\ \bibinfo {author} {\bibfnamefont {A.~G.}\ \bibnamefont {Csaszar}},\
  }\href@noop {} {\emph {\bibinfo {title} {{Equilibrium Molecular Structures:
  From Spectroscopy to Quantum Chemistry}}}}\ (\bibinfo  {publisher} {CRC
  Press, Boca Raton},\ \bibinfo {year} {2010})\BibitemShut {NoStop}%
\bibitem [{\citenamefont {Makarov}\ \emph {et~al.}(2015)\citenamefont
  {Makarov}, \citenamefont {Koshelev}, \citenamefont {Zobov},\ and\
  \citenamefont {Boyarkin}}]{15MaKoZoBo}%
  \BibitemOpen
  \bibfield  {author} {\bibinfo {author} {\bibfnamefont {D.~S.}\ \bibnamefont
  {Makarov}}, \bibinfo {author} {\bibfnamefont {M.~A.}\ \bibnamefont
  {Koshelev}}, \bibinfo {author} {\bibfnamefont {N.~F.}\ \bibnamefont {Zobov}},
  \ and\ \bibinfo {author} {\bibfnamefont {O.~V.}\ \bibnamefont {Boyarkin}},\
  }\href {\doibase 10.1016/j.cplett.2015.03.036} {\bibfield  {journal}
  {\bibinfo  {journal} {Chem. Phys. Lett.}\ }\textbf {\bibinfo {volume}
  {627}},\ \bibinfo {pages} {73} (\bibinfo {year} {2015})}\BibitemShut
  {NoStop}%
\bibitem [{\citenamefont {Polyansky}\ \emph {et~al.}(2015)\citenamefont
  {Polyansky}, \citenamefont {Bielska}, \citenamefont {Ghysels}, \citenamefont
  {Lodi}, \citenamefont {Zobov}, \citenamefont {Hodges},\ and\ \citenamefont
  {Tennyson}}]{jt613}%
  \BibitemOpen
  \bibfield  {author} {\bibinfo {author} {\bibfnamefont {O.~L.}\ \bibnamefont
  {Polyansky}}, \bibinfo {author} {\bibfnamefont {K.}~\bibnamefont {Bielska}},
  \bibinfo {author} {\bibfnamefont {M.}~\bibnamefont {Ghysels}}, \bibinfo
  {author} {\bibfnamefont {L.}~\bibnamefont {Lodi}}, \bibinfo {author}
  {\bibfnamefont {N.~F.}\ \bibnamefont {Zobov}}, \bibinfo {author}
  {\bibfnamefont {J.~T.}\ \bibnamefont {Hodges}}, \ and\ \bibinfo {author}
  {\bibfnamefont {J.}~\bibnamefont {Tennyson}},\ }\href {\doibase
  10.1103/PhysRevLett.114.243001} {\bibfield  {journal} {\bibinfo  {journal}
  {Phys. Rev. Lett.}\ }\textbf {\bibinfo {volume} {114}},\ \bibinfo {pages}
  {243001} (\bibinfo {year} {2015})}\BibitemShut {NoStop}%
\bibitem [{\citenamefont {Lodi}\ and\ \citenamefont {Tennyson}(2012)}]{jt522}%
  \BibitemOpen
  \bibfield  {author} {\bibinfo {author} {\bibfnamefont {L.}~\bibnamefont
  {Lodi}}\ and\ \bibinfo {author} {\bibfnamefont {J.}~\bibnamefont
  {Tennyson}},\ }\href@noop {} {\bibfield  {journal} {\bibinfo  {journal} {J.
  Quant. Spectrosc. Radiat. Transf.}\ }\textbf {\bibinfo {volume} {113}},\
  \bibinfo {pages} {850} (\bibinfo {year} {2012})}\BibitemShut {NoStop}%
\bibitem [{\citenamefont {Zak}\ \emph {et~al.}(2016)\citenamefont {Zak},
  \citenamefont {Tennyson}, \citenamefont {Polyansky}, \citenamefont {Lodi},
  \citenamefont {Tashkun},\ and\ \citenamefont {Perevalov}}]{jt625}%
  \BibitemOpen
  \bibfield  {author} {\bibinfo {author} {\bibfnamefont {E.}~\bibnamefont
  {Zak}}, \bibinfo {author} {\bibfnamefont {J.}~\bibnamefont {Tennyson}},
  \bibinfo {author} {\bibfnamefont {O.~L.}\ \bibnamefont {Polyansky}}, \bibinfo
  {author} {\bibfnamefont {L.}~\bibnamefont {Lodi}}, \bibinfo {author}
  {\bibfnamefont {S.~A.}\ \bibnamefont {Tashkun}}, \ and\ \bibinfo {author}
  {\bibfnamefont {V.~I.}\ \bibnamefont {Perevalov}},\ }\href@noop {} {\bibfield
   {journal} {\bibinfo  {journal} {J. Quant. Spectrosc. Radiat. Transf.}\ }
  (\bibinfo {year} {2016})}\BibitemShut {NoStop}%
\bibitem [{\citenamefont {Elford}(2000)}]{Landolt}%
  \BibitemOpen
  \bibfield  {author} {\bibinfo {author} {\bibfnamefont {M.~T.}\ \bibnamefont
  {Elford}},\ }\enquote {\bibinfo {title} {{Photon and Electron Interactions
  with Atoms, Molecules, and Ions (Landolt-Börnstein: Numerical Data and
  Functional Relationships in Science and Technology / Elementary Particles,
  Nuclei and Atoms)}},}\ \ (\bibinfo  {publisher} {Springer, New York},\
  \bibinfo {year} {2000})\BibitemShut {NoStop}%
\bibitem [{\citenamefont {Trajmar}(1973)}]{PhysRevA.8.191}%
  \BibitemOpen
  \bibfield  {author} {\bibinfo {author} {\bibfnamefont {S.}~\bibnamefont
  {Trajmar}},\ }\href {\doibase 10.1103/PhysRevA.8.191} {\bibfield  {journal}
  {\bibinfo  {journal} {Phys. Rev. A}\ }\textbf {\bibinfo {volume} {8}},\
  \bibinfo {pages} {191} (\bibinfo {year} {1973})}\BibitemShut {NoStop}%
\bibitem [{\citenamefont {Hall}\ \emph {et~al.}(1972)\citenamefont {Hall},
  \citenamefont {Joyez}, \citenamefont {Mazeau}, \citenamefont {Reinhard},\
  and\ \citenamefont {Schermann}}]{Hall1973}%
  \BibitemOpen
  \bibfield  {author} {\bibinfo {author} {\bibfnamefont {R.~I.}\ \bibnamefont
  {Hall}}, \bibinfo {author} {\bibfnamefont {G.}~\bibnamefont {Joyez}},
  \bibinfo {author} {\bibfnamefont {Y.}~\bibnamefont {Mazeau}}, \bibinfo
  {author} {\bibfnamefont {J.}~\bibnamefont {Reinhard}}, \ and\ \bibinfo
  {author} {\bibfnamefont {C.}~\bibnamefont {Schermann}},\ }\href@noop {}
  {\bibfield  {journal} {\bibinfo  {journal} {J.~Physique}\ }\textbf {\bibinfo
  {volume} {34}},\ \bibinfo {pages} {827} (\bibinfo {year} {1972})}\BibitemShut
  {NoStop}%
\bibitem [{\citenamefont {Donaldson}\ \emph {et~al.}(1972)\citenamefont
  {Donaldson}, \citenamefont {Hender},\ and\ \citenamefont
  {McConkey}}]{0022-3700-5-6-022}%
  \BibitemOpen
  \bibfield  {author} {\bibinfo {author} {\bibfnamefont {F.~G.}\ \bibnamefont
  {Donaldson}}, \bibinfo {author} {\bibfnamefont {M.~A.}\ \bibnamefont
  {Hender}}, \ and\ \bibinfo {author} {\bibfnamefont {J.~W.}\ \bibnamefont
  {McConkey}},\ }\href {http://stacks.iop.org/0022-3700/5/i=6/a=022} {\bibfield
   {journal} {\bibinfo  {journal} {J. Phys. B: At., Mol. Phys.}\ }\textbf
  {\bibinfo {volume} {5}},\ \bibinfo {pages} {1192} (\bibinfo {year}
  {1972})}\BibitemShut {NoStop}%
\bibitem [{\citenamefont {Fursa}\ and\ \citenamefont
  {Bray}(1995)}]{PhysRevA.52.1279}%
  \BibitemOpen
  \bibfield  {author} {\bibinfo {author} {\bibfnamefont {D.~V.}\ \bibnamefont
  {Fursa}}\ and\ \bibinfo {author} {\bibfnamefont {I.}~\bibnamefont {Bray}},\
  }\href {\doibase 10.1103/PhysRevA.52.1279} {\bibfield  {journal} {\bibinfo
  {journal} {Phys. Rev. A}\ }\textbf {\bibinfo {volume} {52}},\ \bibinfo
  {pages} {1279} (\bibinfo {year} {1995})}\BibitemShut {NoStop}%
\bibitem [{\citenamefont {Bartschat}(1998)}]{0953-4075-31-10-005}%
  \BibitemOpen
  \bibfield  {author} {\bibinfo {author} {\bibfnamefont {K.}~\bibnamefont
  {Bartschat}},\ }\href@noop {} {\bibfield  {journal} {\bibinfo  {journal} {J.
  Phys. B: At., Mol. Opt. Phys.}\ }\textbf {\bibinfo {volume} {31}},\ \bibinfo
  {pages} {L469} (\bibinfo {year} {1998})}\BibitemShut {NoStop}%
\bibitem [{\citenamefont {Zatsarinny}\ \emph {et~al.}(2014)\citenamefont
  {Zatsarinny}, \citenamefont {Wang},\ and\ \citenamefont
  {Bartschat}}]{PhysRevA.89.022706}%
  \BibitemOpen
  \bibfield  {author} {\bibinfo {author} {\bibfnamefont {O.}~\bibnamefont
  {Zatsarinny}}, \bibinfo {author} {\bibfnamefont {Y.}~\bibnamefont {Wang}}, \
  and\ \bibinfo {author} {\bibfnamefont {K.}~\bibnamefont {Bartschat}},\ }\href
  {\doibase 10.1103/PhysRevA.89.022706} {\bibfield  {journal} {\bibinfo
  {journal} {Phys. Rev. A}\ }\textbf {\bibinfo {volume} {89}},\ \bibinfo
  {pages} {022706} (\bibinfo {year} {2014})}\BibitemShut {NoStop}%
\bibitem [{\citenamefont {Boffard}\ \emph {et~al.}(1999)\citenamefont
  {Boffard}, \citenamefont {Piech}, \citenamefont {Gehrke}, \citenamefont
  {Anderson},\ and\ \citenamefont {Lin}}]{PhysRevA.59.2749}%
  \BibitemOpen
  \bibfield  {author} {\bibinfo {author} {\bibfnamefont {J.~B.}\ \bibnamefont
  {Boffard}}, \bibinfo {author} {\bibfnamefont {G.~A.}\ \bibnamefont {Piech}},
  \bibinfo {author} {\bibfnamefont {M.~F.}\ \bibnamefont {Gehrke}}, \bibinfo
  {author} {\bibfnamefont {L.~W.}\ \bibnamefont {Anderson}}, \ and\ \bibinfo
  {author} {\bibfnamefont {C.~C.}\ \bibnamefont {Lin}},\ }\href {\doibase
  10.1103/PhysRevA.59.2749} {\bibfield  {journal} {\bibinfo  {journal} {Phys.
  Rev. A}\ }\textbf {\bibinfo {volume} {59}},\ \bibinfo {pages} {2749}
  (\bibinfo {year} {1999})}\BibitemShut {NoStop}%
\bibitem [{\citenamefont {Hibbert}(1975)}]{CIV3}%
  \BibitemOpen
  \bibfield  {author} {\bibinfo {author} {\bibfnamefont {A.}~\bibnamefont
  {Hibbert}},\ }\href@noop {} {\bibfield  {journal} {\bibinfo  {journal} {Comp.
  Phys. Commun.}\ }\textbf {\bibinfo {volume} {9}},\ \bibinfo {pages} {141}
  (\bibinfo {year} {1975})}\BibitemShut {NoStop}%
\bibitem [{\citenamefont {Eissner}\ \emph {et~al.}(1974)\citenamefont
  {Eissner}, \citenamefont {Jones},\ and\ \citenamefont {Nussbaumer}}]{SS}%
  \BibitemOpen
  \bibfield  {author} {\bibinfo {author} {\bibfnamefont {W.}~\bibnamefont
  {Eissner}}, \bibinfo {author} {\bibfnamefont {M.}~\bibnamefont {Jones}}, \
  and\ \bibinfo {author} {\bibfnamefont {H.}~\bibnamefont {Nussbaumer}},\
  }\href@noop {} {\bibfield  {journal} {\bibinfo  {journal} {Comp. Phys.
  Commun.}\ }\textbf {\bibinfo {volume} {8}},\ \bibinfo {pages} {270} (\bibinfo
  {year} {1974})}\BibitemShut {NoStop}%
\bibitem [{\citenamefont {Zatsarinny}\ and\ \citenamefont
  {Bartschat}(2005)}]{ISI:000231564200009}%
  \BibitemOpen
  \bibfield  {author} {\bibinfo {author} {\bibfnamefont {O.}~\bibnamefont
  {Zatsarinny}}\ and\ \bibinfo {author} {\bibfnamefont {K.}~\bibnamefont
  {Bartschat}},\ }\href {\doibase {10.1103/PhysRevA.72.020702}} {\bibfield
  {journal} {\bibinfo  {journal} {Phys. Rev. A}\ }\textbf {\bibinfo {volume}
  {{72}}},\ \bibinfo {pages} {{020702}} (\bibinfo {year} {{2005}})}\BibitemShut
  {NoStop}%
\bibitem [{\citenamefont {Ramsbottom}\ \emph {et~al.}(2002)\citenamefont
  {Ramsbottom}, \citenamefont {Scott}, \citenamefont {Bell}, \citenamefont
  {Keenan}, \citenamefont {McLaughlin}, \citenamefont {Sunderland},
  \citenamefont {Burke}, \citenamefont {Noble},\ and\ \citenamefont
  {Burke}}]{0953-4075-35-16-308}%
  \BibitemOpen
  \bibfield  {author} {\bibinfo {author} {\bibfnamefont {C.~A.}\ \bibnamefont
  {Ramsbottom}}, \bibinfo {author} {\bibfnamefont {M.~P.}\ \bibnamefont
  {Scott}}, \bibinfo {author} {\bibfnamefont {K.~L.}\ \bibnamefont {Bell}},
  \bibinfo {author} {\bibfnamefont {F.~P.}\ \bibnamefont {Keenan}}, \bibinfo
  {author} {\bibfnamefont {B.~M.}\ \bibnamefont {McLaughlin}}, \bibinfo
  {author} {\bibfnamefont {A.~G.}\ \bibnamefont {Sunderland}}, \bibinfo
  {author} {\bibfnamefont {V.~M.}\ \bibnamefont {Burke}}, \bibinfo {author}
  {\bibfnamefont {C.~J.}\ \bibnamefont {Noble}}, \ and\ \bibinfo {author}
  {\bibfnamefont {P.~G.}\ \bibnamefont {Burke}},\ }\href
  {http://stacks.iop.org/0953-4075/35/i=16/a=308} {\bibfield  {journal}
  {\bibinfo  {journal} {J. Phys. B: At., Mol. Opt. Phys.}\ }\textbf {\bibinfo
  {volume} {35}},\ \bibinfo {pages} {3451} (\bibinfo {year}
  {2002})}\BibitemShut {NoStop}%
\bibitem [{\citenamefont {Nussbaumer}\ and\ \citenamefont
  {Storey}(1980)}]{NS80}%
  \BibitemOpen
  \bibfield  {author} {\bibinfo {author} {\bibfnamefont {H.}~\bibnamefont
  {Nussbaumer}}\ and\ \bibinfo {author} {\bibfnamefont {P.~J.}\ \bibnamefont
  {Storey}},\ }\href@noop {} {\bibfield  {journal} {\bibinfo  {journal}
  {Astron. Astrophys.}\ }\textbf {\bibinfo {volume} {89}},\ \bibinfo {pages}
  {308} (\bibinfo {year} {1980})}\BibitemShut {NoStop}%
\bibitem [{\citenamefont {Pradhan}\ and\ \citenamefont
  {Berrington}(1993)}]{PB93}%
  \BibitemOpen
  \bibfield  {author} {\bibinfo {author} {\bibfnamefont {A.~K.}\ \bibnamefont
  {Pradhan}}\ and\ \bibinfo {author} {\bibfnamefont {K.~A.}\ \bibnamefont
  {Berrington}},\ }\href@noop {} {\bibfield  {journal} {\bibinfo  {journal} {J.
  Phys. B}\ }\textbf {\bibinfo {volume} {26}},\ \bibinfo {pages} {157}
  (\bibinfo {year} {1993})}\BibitemShut {NoStop}%
\bibitem [{\citenamefont {Falk}\ and\ \citenamefont
  {Dunn}(1983)}]{PhysRevA.27.754}%
  \BibitemOpen
  \bibfield  {author} {\bibinfo {author} {\bibfnamefont {R.~A.}\ \bibnamefont
  {Falk}}\ and\ \bibinfo {author} {\bibfnamefont {G.~H.}\ \bibnamefont
  {Dunn}},\ }\href {\doibase 10.1103/PhysRevA.27.754} {\bibfield  {journal}
  {\bibinfo  {journal} {Phys. Rev. A}\ }\textbf {\bibinfo {volume} {27}},\
  \bibinfo {pages} {754} (\bibinfo {year} {1983})}\BibitemShut {NoStop}%
\bibitem [{\citenamefont {Colgan}\ \emph {et~al.}(2003)\citenamefont {Colgan},
  \citenamefont {Loch}, \citenamefont {Pindzola}, \citenamefont {Ballance},\
  and\ \citenamefont {Griffin}}]{PhysRevA.68.032712}%
  \BibitemOpen
  \bibfield  {author} {\bibinfo {author} {\bibfnamefont {J.}~\bibnamefont
  {Colgan}}, \bibinfo {author} {\bibfnamefont {S.~D.}\ \bibnamefont {Loch}},
  \bibinfo {author} {\bibfnamefont {M.~S.}\ \bibnamefont {Pindzola}}, \bibinfo
  {author} {\bibfnamefont {C.~P.}\ \bibnamefont {Ballance}}, \ and\ \bibinfo
  {author} {\bibfnamefont {D.~C.}\ \bibnamefont {Griffin}},\ }\href {\doibase
  10.1103/PhysRevA.68.032712} {\bibfield  {journal} {\bibinfo  {journal} {Phys.
  Rev. A}\ }\textbf {\bibinfo {volume} {68}},\ \bibinfo {pages} {032712}
  (\bibinfo {year} {2003})}\BibitemShut {NoStop}%
\bibitem [{\citenamefont {Mitnik}\ \emph {et~al.}(1999)\citenamefont {Mitnik},
  \citenamefont {Pindzol}, , \citenamefont {Griffin},\ and\ \citenamefont
  {Badnell}}]{Mitnik1999}%
  \BibitemOpen
  \bibfield  {author} {\bibinfo {author} {\bibfnamefont {D.~M.}\ \bibnamefont
  {Mitnik}}, \bibinfo {author} {\bibfnamefont {M.~S.}\ \bibnamefont {Pindzol}},
  , \bibinfo {author} {\bibfnamefont {D.~C.}\ \bibnamefont {Griffin}}, \ and\
  \bibinfo {author} {\bibfnamefont {N.~R.}\ \bibnamefont {Badnell}},\
  }\href@noop {} {\ \textbf {\bibinfo {volume} {32}},\ \bibinfo {pages} {L479}
  (\bibinfo {year} {1999})}\BibitemShut {NoStop}%
\bibitem [{\citenamefont {Davey}\ \emph {et~al.}(2000)\citenamefont {Davey},
  \citenamefont {Storey},\ and\ \citenamefont {Kisielius}}]{CII}%
  \BibitemOpen
  \bibfield  {author} {\bibinfo {author} {\bibfnamefont {A.~R.}\ \bibnamefont
  {Davey}}, \bibinfo {author} {\bibfnamefont {P.~J.}\ \bibnamefont {Storey}}, \
  and\ \bibinfo {author} {\bibfnamefont {R.}~\bibnamefont {Kisielius}},\
  }\href@noop {} {\bibfield  {journal} {\bibinfo  {journal} {Astron. Astrophys.
  Suppl. Ser.}\ }\textbf {\bibinfo {volume} {{142}}},\ \bibinfo {pages} {85}
  (\bibinfo {year} {{2000}})}\BibitemShut {NoStop}%
\bibitem [{\citenamefont {Spruck}\ \emph {et~al.}(2014)\citenamefont {Spruck},
  \citenamefont {Badnell}, \citenamefont {Krantz}, \citenamefont {Novotny},
  \citenamefont {Becker}, \citenamefont {Bernhardt},\ and\ \citenamefont
  {et~al.}}]{Spruck2014}%
  \BibitemOpen
  \bibfield  {author} {\bibinfo {author} {\bibfnamefont {K.}~\bibnamefont
  {Spruck}}, \bibinfo {author} {\bibfnamefont {N.~R.}\ \bibnamefont {Badnell}},
  \bibinfo {author} {\bibfnamefont {C.}~\bibnamefont {Krantz}}, \bibinfo
  {author} {\bibfnamefont {O.}~\bibnamefont {Novotny}}, \bibinfo {author}
  {\bibfnamefont {A.}~\bibnamefont {Becker}}, \bibinfo {author} {\bibfnamefont
  {D.}~\bibnamefont {Bernhardt}}, \ and\ \bibinfo {author} {\bibfnamefont
  {M.~G.}\ \bibnamefont {et~al.}},\ }\href@noop {} {\bibfield  {journal}
  {\bibinfo  {journal} {Phys. Rev. A}\ }\textbf {\bibinfo {volume} {90}},\
  \bibinfo {pages} {032715} (\bibinfo {year} {2014})}\BibitemShut {NoStop}%
\bibitem [{\citenamefont {Badnell}\ \emph {et~al.}(2016)\citenamefont
  {Badnell}, \citenamefont {Spruck}, \citenamefont {Krantz}, \citenamefont
  {Novotny}, \citenamefont {Becker}, \citenamefont {Bernhardt},\ and\
  \citenamefont {et~al}}]{Badnell2016}%
  \BibitemOpen
  \bibfield  {author} {\bibinfo {author} {\bibfnamefont {N.~R.}\ \bibnamefont
  {Badnell}}, \bibinfo {author} {\bibfnamefont {K.}~\bibnamefont {Spruck}},
  \bibinfo {author} {\bibfnamefont {C.}~\bibnamefont {Krantz}}, \bibinfo
  {author} {\bibfnamefont {O.}~\bibnamefont {Novotny}}, \bibinfo {author}
  {\bibfnamefont {A.}~\bibnamefont {Becker}}, \bibinfo {author} {\bibfnamefont
  {D.}~\bibnamefont {Bernhardt}}, \ and\ \bibinfo {author} {\bibfnamefont
  {M.~G.}\ \bibnamefont {et~al}},\ }\href@noop {} {\bibfield  {journal}
  {\bibinfo  {journal} {Phys. Rev. A}\ }\textbf {\bibinfo {volume} {93}},\
  \bibinfo {pages} {052703} (\bibinfo {year} {2016})}\BibitemShut {NoStop}%
\bibitem [{\citenamefont {Badnell}\ \emph {et~al.}(2015)\citenamefont
  {Badnell}, \citenamefont {Ferland}, \citenamefont {Gorczyca}, \citenamefont
  {Nikoli.},\ and\ \citenamefont {Wagle}}]{Badnell2015}%
  \BibitemOpen
  \bibfield  {author} {\bibinfo {author} {\bibfnamefont {N.~R.}\ \bibnamefont
  {Badnell}}, \bibinfo {author} {\bibfnamefont {G.~J.}\ \bibnamefont
  {Ferland}}, \bibinfo {author} {\bibfnamefont {T.~W.}\ \bibnamefont
  {Gorczyca}}, \bibinfo {author} {\bibfnamefont {D.}~\bibnamefont {Nikoli.}}, \
  and\ \bibinfo {author} {\bibfnamefont {G.~A.}\ \bibnamefont {Wagle}},\
  }\href@noop {} {\bibfield  {journal} {\bibinfo  {journal} {The Astrophysical
  Journal}\ }\textbf {\bibinfo {volume} {804}},\ \bibinfo {pages} {100}
  (\bibinfo {year} {2015})}\BibitemShut {NoStop}%
\bibitem [{\citenamefont {Zammit}\ \emph {et~al.}(2014)\citenamefont {Zammit},
  \citenamefont {Fursa},\ and\ \citenamefont {Bray}}]{14ZaFuBr}%
  \BibitemOpen
  \bibfield  {author} {\bibinfo {author} {\bibfnamefont {M.~C.}\ \bibnamefont
  {Zammit}}, \bibinfo {author} {\bibfnamefont {D.~V.}\ \bibnamefont {Fursa}}, \
  and\ \bibinfo {author} {\bibfnamefont {I.}~\bibnamefont {Bray}},\ }\href@noop
  {} {\bibfield  {journal} {\bibinfo  {journal} {Phys. Rev. A}\ }\textbf
  {\bibinfo {volume} {90}},\ \bibinfo {pages} {022711} (\bibinfo {year}
  {2014})}\BibitemShut {NoStop}%
\bibitem [{\citenamefont {Douguet}\ \emph {et~al.}(2014)\citenamefont
  {Douguet}, \citenamefont {Kokoouline},\ and\ \citenamefont
  {Orel}}]{douguet14}%
  \BibitemOpen
  \bibfield  {author} {\bibinfo {author} {\bibfnamefont {N.}~\bibnamefont
  {Douguet}}, \bibinfo {author} {\bibfnamefont {V.}~\bibnamefont {Kokoouline}},
  \ and\ \bibinfo {author} {\bibfnamefont {A.~E.}\ \bibnamefont {Orel}},\
  }\href@noop {} {\bibfield  {journal} {\bibinfo  {journal} {Phys. Rev. A}\
  }\textbf {\bibinfo {volume} {90}},\ \bibinfo {pages} {063410} (\bibinfo
  {year} {2014})}\BibitemShut {NoStop}%
\bibitem [{\citenamefont {Best}\ \emph {et~al.}(2011)\citenamefont {Best},
  \citenamefont {Otto}, \citenamefont {Trippel}, \citenamefont {Hlavenka},
  \citenamefont {von Zastrow}, \citenamefont {Eisenbach}, \citenamefont
  {Jezouin}, \citenamefont {Wester}, \citenamefont {Vigren}, \citenamefont
  {Hamberg},\ and\ \citenamefont {Geppert}}]{best11}%
  \BibitemOpen
  \bibfield  {author} {\bibinfo {author} {\bibfnamefont {T.}~\bibnamefont
  {Best}}, \bibinfo {author} {\bibfnamefont {R.}~\bibnamefont {Otto}}, \bibinfo
  {author} {\bibfnamefont {S.}~\bibnamefont {Trippel}}, \bibinfo {author}
  {\bibfnamefont {P.}~\bibnamefont {Hlavenka}}, \bibinfo {author}
  {\bibfnamefont {A.}~\bibnamefont {von Zastrow}}, \bibinfo {author}
  {\bibfnamefont {S.}~\bibnamefont {Eisenbach}}, \bibinfo {author}
  {\bibfnamefont {S.}~\bibnamefont {Jezouin}}, \bibinfo {author} {\bibfnamefont
  {R.}~\bibnamefont {Wester}}, \bibinfo {author} {\bibfnamefont
  {E.}~\bibnamefont {Vigren}}, \bibinfo {author} {\bibfnamefont
  {M.}~\bibnamefont {Hamberg}}, \ and\ \bibinfo {author} {\bibfnamefont
  {W.~D.}\ \bibnamefont {Geppert}},\ }\href@noop {} {\bibfield  {journal}
  {\bibinfo  {journal} {Astrophys. J.}\ }\textbf {\bibinfo {volume} {742}},\
  \bibinfo {pages} {63} (\bibinfo {year} {2011})}\BibitemShut {NoStop}%
\bibitem [{\citenamefont {Tennyson}\ \emph {et~al.}(2007)\citenamefont
  {Tennyson}, \citenamefont {Brown}, \citenamefont {Munro}, \citenamefont
  {Rozum}, \citenamefont {Varambhia},\ and\ \citenamefont {Vinci}}]{jt416}%
  \BibitemOpen
  \bibfield  {author} {\bibinfo {author} {\bibfnamefont {J.}~\bibnamefont
  {Tennyson}}, \bibinfo {author} {\bibfnamefont {D.~B.}\ \bibnamefont {Brown}},
  \bibinfo {author} {\bibfnamefont {J.~J.}\ \bibnamefont {Munro}}, \bibinfo
  {author} {\bibfnamefont {I.}~\bibnamefont {Rozum}}, \bibinfo {author}
  {\bibfnamefont {H.~N.}\ \bibnamefont {Varambhia}}, \ and\ \bibinfo {author}
  {\bibfnamefont {N.}~\bibnamefont {Vinci}},\ }\href@noop {} {\bibfield
  {journal} {\bibinfo  {journal} {J. Phys. Conf. Ser.}\ }\textbf {\bibinfo
  {volume} {86}},\ \bibinfo {pages} {012001} (\bibinfo {year}
  {2007})}\BibitemShut {NoStop}%
\bibitem [{\citenamefont {Carr}\ \emph {et~al.}(2012)\citenamefont {Carr},
  \citenamefont {Galiatsatos}, \citenamefont {Gorfinkiel}, \citenamefont
  {Harvey}, \citenamefont {Lysaght}, \citenamefont {Madden}, \citenamefont
  {Masin}, \citenamefont {Plummer},\ and\ \citenamefont {Tennyson}}]{jt518}%
  \BibitemOpen
  \bibfield  {author} {\bibinfo {author} {\bibfnamefont {J.~M.}\ \bibnamefont
  {Carr}}, \bibinfo {author} {\bibfnamefont {P.~G.}\ \bibnamefont
  {Galiatsatos}}, \bibinfo {author} {\bibfnamefont {J.~D.}\ \bibnamefont
  {Gorfinkiel}}, \bibinfo {author} {\bibfnamefont {A.~G.}\ \bibnamefont
  {Harvey}}, \bibinfo {author} {\bibfnamefont {M.~A.}\ \bibnamefont {Lysaght}},
  \bibinfo {author} {\bibfnamefont {D.}~\bibnamefont {Madden}}, \bibinfo
  {author} {\bibfnamefont {Z.}~\bibnamefont {Masin}}, \bibinfo {author}
  {\bibfnamefont {M.}~\bibnamefont {Plummer}}, \ and\ \bibinfo {author}
  {\bibfnamefont {J.}~\bibnamefont {Tennyson}},\ }\href@noop {} {\bibfield
  {journal} {\bibinfo  {journal} {Eur. Phys. J. D}\ }\textbf {\bibinfo {volume}
  {66}},\ \bibinfo {pages} {58} (\bibinfo {year} {2012})}\BibitemShut {NoStop}%
\bibitem [{\citenamefont {Khamesian}\ \emph {et~al.}(2016)\citenamefont
  {Khamesian}, \citenamefont {Douguet}, \citenamefont {{Fonseca dos Santos}},
  \citenamefont {Dulieu}, \citenamefont {Raoult},\ and\ \citenamefont
  {Kokoouline}}]{khamesyan16}%
  \BibitemOpen
  \bibfield  {author} {\bibinfo {author} {\bibfnamefont {M.}~\bibnamefont
  {Khamesian}}, \bibinfo {author} {\bibfnamefont {N.}~\bibnamefont {Douguet}},
  \bibinfo {author} {\bibfnamefont {S.}~\bibnamefont {{Fonseca dos Santos}}},
  \bibinfo {author} {\bibfnamefont {O.}~\bibnamefont {Dulieu}}, \bibinfo
  {author} {\bibfnamefont {M.}~\bibnamefont {Raoult}}, \ and\ \bibinfo {author}
  {\bibfnamefont {V.}~\bibnamefont {Kokoouline}},\ }\href@noop {} {\bibfield
  {journal} {\bibinfo  {journal} {Eur. Phys. J. D}\ ,\ \bibinfo {pages}
  {Submitted}} (\bibinfo {year} {2016})}\BibitemShut {NoStop}%
\bibitem [{\citenamefont {Vejby-Christensen}\ \emph {et~al.}(1997)\citenamefont
  {Vejby-Christensen}, \citenamefont {Andersen}, \citenamefont {Heber},
  \citenamefont {Kella}, \citenamefont {Pedersen}, \citenamefont {Schmidt},\
  and\ \citenamefont {Zajfman}}]{vejby97}%
  \BibitemOpen
  \bibfield  {author} {\bibinfo {author} {\bibfnamefont {L.}~\bibnamefont
  {Vejby-Christensen}}, \bibinfo {author} {\bibfnamefont {L.}~\bibnamefont
  {Andersen}}, \bibinfo {author} {\bibfnamefont {O.}~\bibnamefont {Heber}},
  \bibinfo {author} {\bibfnamefont {D.}~\bibnamefont {Kella}}, \bibinfo
  {author} {\bibfnamefont {H.}~\bibnamefont {Pedersen}}, \bibinfo {author}
  {\bibfnamefont {H.}~\bibnamefont {Schmidt}}, \ and\ \bibinfo {author}
  {\bibfnamefont {D.}~\bibnamefont {Zajfman}},\ }\href@noop {} {\bibfield
  {journal} {\bibinfo  {journal} {Astrophys. J.}\ }\textbf {\bibinfo {volume}
  {483}},\ \bibinfo {pages} {531} (\bibinfo {year} {1997})}\BibitemShut
  {NoStop}%
\bibitem [{\citenamefont {Neau}\ \emph {et~al.}(2000)\citenamefont {Neau},
  \citenamefont {Khalili}, \citenamefont {Ros{\'e}n}, \citenamefont {Padellec},
  \citenamefont {Derkatch}, \citenamefont {Shi}, \citenamefont {Vikor},
  \citenamefont {Larsson}, \citenamefont {Semaniak}, \citenamefont {Thomas},
  \citenamefont {N{\aa}{g{\aa}rd}}, \citenamefont {Andersson}, \citenamefont
  {Danared},\ and\ \citenamefont {{af Ugglas}}}]{neau00}%
  \BibitemOpen
  \bibfield  {author} {\bibinfo {author} {\bibfnamefont {A.}~\bibnamefont
  {Neau}}, \bibinfo {author} {\bibfnamefont {A.~A.}\ \bibnamefont {Khalili}},
  \bibinfo {author} {\bibfnamefont {S.}~\bibnamefont {Ros{\'e}n}}, \bibinfo
  {author} {\bibfnamefont {A.~L.}\ \bibnamefont {Padellec}}, \bibinfo {author}
  {\bibfnamefont {A.~M.}\ \bibnamefont {Derkatch}}, \bibinfo {author}
  {\bibfnamefont {W.}~\bibnamefont {Shi}}, \bibinfo {author} {\bibfnamefont
  {L.}~\bibnamefont {Vikor}}, \bibinfo {author} {\bibfnamefont
  {M.}~\bibnamefont {Larsson}}, \bibinfo {author} {\bibfnamefont
  {J.}~\bibnamefont {Semaniak}}, \bibinfo {author} {\bibfnamefont
  {R.}~\bibnamefont {Thomas}}, \bibinfo {author} {\bibfnamefont {M.~B.}\
  \bibnamefont {N{\aa}{g{\aa}rd}}}, \bibinfo {author} {\bibfnamefont
  {K.}~\bibnamefont {Andersson}}, \bibinfo {author} {\bibfnamefont
  {H.}~\bibnamefont {Danared}}, \ and\ \bibinfo {author} {\bibfnamefont
  {M.}~\bibnamefont {{af Ugglas}}},\ }\href {\doibase 10.1063/1.481979}
  {\bibfield  {journal} {\bibinfo  {journal} {J. Chem. Phys.}\ }\textbf
  {\bibinfo {volume} {113}},\ \bibinfo {pages} {1762} (\bibinfo {year}
  {2000})}\BibitemShut {NoStop}%
\bibitem [{\citenamefont {Jensen}\ \emph {et~al.}(2000)\citenamefont {Jensen},
  \citenamefont {Bilodeau}, \citenamefont {Safvan}, \citenamefont {Seiersen},
  \citenamefont {Andersen}, \citenamefont {Pedersen},\ and\ \citenamefont
  {Heber}}]{jensen00}%
  \BibitemOpen
  \bibfield  {author} {\bibinfo {author} {\bibfnamefont {M.~J.}\ \bibnamefont
  {Jensen}}, \bibinfo {author} {\bibfnamefont {R.~C.}\ \bibnamefont
  {Bilodeau}}, \bibinfo {author} {\bibfnamefont {C.~P.}\ \bibnamefont
  {Safvan}}, \bibinfo {author} {\bibfnamefont {K.}~\bibnamefont {Seiersen}},
  \bibinfo {author} {\bibfnamefont {L.~H.}\ \bibnamefont {Andersen}}, \bibinfo
  {author} {\bibfnamefont {H.~B.}\ \bibnamefont {Pedersen}}, \ and\ \bibinfo
  {author} {\bibfnamefont {O.}~\bibnamefont {Heber}},\ }\href@noop {}
  {\bibfield  {journal} {\bibinfo  {journal} {Astrophys. J.}\ }\textbf
  {\bibinfo {volume} {543}},\ \bibinfo {pages} {764} (\bibinfo {year}
  {2000})}\BibitemShut {NoStop}%
\bibitem [{\citenamefont {Shepard}\ \emph {et~al.}(2008)\citenamefont
  {Shepard}, \citenamefont {Shavitt}, \citenamefont {Pitzer}, \citenamefont
  {Dallos}, \citenamefont {M{\"u}ller}, \citenamefont {Szalay}, \citenamefont
  {Brown}, \citenamefont {Ahlrichs}, \citenamefont {B{\"o}hm}, \citenamefont
  {Chang}, \citenamefont {Comeau}, \citenamefont {Gdanitz}, \citenamefont
  {Dachsel}, \citenamefont {Ehrhardt}, \citenamefont {Ernzerhof}, \citenamefont
  {H{\"o}chtl}, \citenamefont {Irle}, \citenamefont {Kedziora}, \citenamefont
  {Kovar}, \citenamefont {Parasuk}, \citenamefont {Pepper}, \citenamefont
  {Scharf}, \citenamefont {Schiffer}, \citenamefont {Schindler}, \citenamefont
  {Sch{\"u}ler}, \citenamefont {Seth}, \citenamefont {Stahlberg}, \citenamefont
  {Zhao}, \citenamefont {Yabushita}, \citenamefont {Zhang}, \citenamefont
  {Barbatti}, \citenamefont {Matsika}, \citenamefont {Schuurmann},
  \citenamefont {Brozell}, \citenamefont {Beck}, \citenamefont {{H. Lischka}},
  \citenamefont {Ruckenbauer}, \citenamefont {Sellner}, \citenamefont
  {Plasser}, \citenamefont {{J.-P. Blaudeau}},\ and\ \citenamefont
  {Szymczak}}]{columbus08}%
  \BibitemOpen
  \bibfield  {author} {\bibinfo {author} {\bibfnamefont {R.}~\bibnamefont
  {Shepard}}, \bibinfo {author} {\bibfnamefont {I.}~\bibnamefont {Shavitt}},
  \bibinfo {author} {\bibfnamefont {R.~M.}\ \bibnamefont {Pitzer}}, \bibinfo
  {author} {\bibfnamefont {M.}~\bibnamefont {Dallos}}, \bibinfo {author}
  {\bibfnamefont {T.}~\bibnamefont {M{\"u}ller}}, \bibinfo {author}
  {\bibfnamefont {P.~G.}\ \bibnamefont {Szalay}}, \bibinfo {author}
  {\bibfnamefont {F.~B.}\ \bibnamefont {Brown}}, \bibinfo {author}
  {\bibfnamefont {R.}~\bibnamefont {Ahlrichs}}, \bibinfo {author}
  {\bibfnamefont {H.~J.}\ \bibnamefont {B{\"o}hm}}, \bibinfo {author}
  {\bibfnamefont {A.}~\bibnamefont {Chang}}, \bibinfo {author} {\bibfnamefont
  {D.~C.}\ \bibnamefont {Comeau}}, \bibinfo {author} {\bibfnamefont
  {R.}~\bibnamefont {Gdanitz}}, \bibinfo {author} {\bibfnamefont
  {H.}~\bibnamefont {Dachsel}}, \bibinfo {author} {\bibfnamefont
  {C.}~\bibnamefont {Ehrhardt}}, \bibinfo {author} {\bibfnamefont
  {M.}~\bibnamefont {Ernzerhof}}, \bibinfo {author} {\bibfnamefont
  {P.}~\bibnamefont {H{\"o}chtl}}, \bibinfo {author} {\bibfnamefont
  {S.}~\bibnamefont {Irle}}, \bibinfo {author} {\bibfnamefont {G.}~\bibnamefont
  {Kedziora}}, \bibinfo {author} {\bibfnamefont {T.}~\bibnamefont {Kovar}},
  \bibinfo {author} {\bibfnamefont {V.}~\bibnamefont {Parasuk}}, \bibinfo
  {author} {\bibfnamefont {M.~J.~M.}\ \bibnamefont {Pepper}}, \bibinfo {author}
  {\bibfnamefont {P.}~\bibnamefont {Scharf}}, \bibinfo {author} {\bibfnamefont
  {H.}~\bibnamefont {Schiffer}}, \bibinfo {author} {\bibfnamefont
  {M.}~\bibnamefont {Schindler}}, \bibinfo {author} {\bibfnamefont
  {M.}~\bibnamefont {Sch{\"u}ler}}, \bibinfo {author} {\bibfnamefont
  {M.}~\bibnamefont {Seth}}, \bibinfo {author} {\bibfnamefont {E.~A.}\
  \bibnamefont {Stahlberg}}, \bibinfo {author} {\bibfnamefont {J.-G.}\
  \bibnamefont {Zhao}}, \bibinfo {author} {\bibfnamefont {S.}~\bibnamefont
  {Yabushita}}, \bibinfo {author} {\bibfnamefont {Z.}~\bibnamefont {Zhang}},
  \bibinfo {author} {\bibfnamefont {M.}~\bibnamefont {Barbatti}}, \bibinfo
  {author} {\bibfnamefont {S.}~\bibnamefont {Matsika}}, \bibinfo {author}
  {\bibfnamefont {M.}~\bibnamefont {Schuurmann}}, \bibinfo {author}
  {\bibfnamefont {D.~R. Y. S.~R.}\ \bibnamefont {Brozell}}, \bibinfo {author}
  {\bibfnamefont {E.~V.}\ \bibnamefont {Beck}}, \bibinfo {author} {\bibnamefont
  {{H. Lischka}}}, \bibinfo {author} {\bibfnamefont {M.}~\bibnamefont
  {Ruckenbauer}}, \bibinfo {author} {\bibfnamefont {B.}~\bibnamefont
  {Sellner}}, \bibinfo {author} {\bibfnamefont {F.}~\bibnamefont {Plasser}},
  \bibinfo {author} {\bibnamefont {{J.-P. Blaudeau}}}, \ and\ \bibinfo {author}
  {\bibfnamefont {J.~J.}\ \bibnamefont {Szymczak}},\ }\href@noop {} {\enquote
  {\bibinfo {title} {Columbus},}\ } (\bibinfo {year} {2008}),\ \bibinfo {note}
  {columbus, an {\it ab initio} electronic structure program, release 5.9.2
  (2008)}\BibitemShut {NoStop}%
\bibitem [{\citenamefont {Igenbergs}\ \emph {et~al.}(2012)\citenamefont
  {Igenbergs}, \citenamefont {Schweinzer}, \citenamefont {Veiter},
  \citenamefont {Perneczky}, \citenamefont {Fr\"uhwirth}, \citenamefont
  {Wallerberger}, \citenamefont {Olson},\ and\ \citenamefont
  {Aumayr}}]{Igenbergs12}%
  \BibitemOpen
  \bibfield  {author} {\bibinfo {author} {\bibfnamefont {K.}~\bibnamefont
  {Igenbergs}}, \bibinfo {author} {\bibfnamefont {J.}~\bibnamefont
  {Schweinzer}}, \bibinfo {author} {\bibfnamefont {A.}~\bibnamefont {Veiter}},
  \bibinfo {author} {\bibfnamefont {L.}~\bibnamefont {Perneczky}}, \bibinfo
  {author} {\bibfnamefont {E.}~\bibnamefont {Fr\"uhwirth}}, \bibinfo {author}
  {\bibfnamefont {M.}~\bibnamefont {Wallerberger}}, \bibinfo {author}
  {\bibfnamefont {R.~E.}\ \bibnamefont {Olson}}, \ and\ \bibinfo {author}
  {\bibfnamefont {F.}~\bibnamefont {Aumayr}},\ }\href@noop {} {\bibfield
  {journal} {\bibinfo  {journal} {J. Phys. B}\ }\textbf {\bibinfo {volume}
  {45}},\ \bibinfo {pages} {065203} (\bibinfo {year} {2012})}\BibitemShut
  {NoStop}%
\bibitem [{\citenamefont {Jorge}\ \emph {et~al.}(2014)\citenamefont {Jorge},
  \citenamefont {Errea}, \citenamefont {Illescas},\ and\ \citenamefont
  {M\'{e}ndez}}]{Jorge14}%
  \BibitemOpen
  \bibfield  {author} {\bibinfo {author} {\bibfnamefont {A.}~\bibnamefont
  {Jorge}}, \bibinfo {author} {\bibfnamefont {L.~F.}\ \bibnamefont {Errea}},
  \bibinfo {author} {\bibfnamefont {C.}~\bibnamefont {Illescas}}, \ and\
  \bibinfo {author} {\bibfnamefont {L.}~\bibnamefont {M\'{e}ndez}},\
  }\href@noop {} {\bibfield  {journal} {\bibinfo  {journal} {Eur. Phys. J. D}\
  }\textbf {\bibinfo {volume} {68}},\ \bibinfo {pages} {227} (\bibinfo {year}
  {2014})}\BibitemShut {NoStop}%
\bibitem [{\citenamefont {Suno}\ and\ \citenamefont {Kato}(2006)}]{Suno06}%
  \BibitemOpen
  \bibfield  {author} {\bibinfo {author} {\bibfnamefont {H.}~\bibnamefont
  {Suno}}\ and\ \bibinfo {author} {\bibfnamefont {T.}~\bibnamefont {Kato}},\
  }\href@noop {} {\bibfield  {journal} {\bibinfo  {journal} {At. Data Nucl.
  Data Tables}\ }\textbf {\bibinfo {volume} {92}},\ \bibinfo {pages} {407}
  (\bibinfo {year} {2006})}\BibitemShut {NoStop}%
\bibitem [{\citenamefont {Harel}\ \emph {et~al.}(1998)\citenamefont {Harel},
  \citenamefont {Jouin},\ and\ \citenamefont {Pons}}]{Harel98}%
  \BibitemOpen
  \bibfield  {author} {\bibinfo {author} {\bibfnamefont {C.}~\bibnamefont
  {Harel}}, \bibinfo {author} {\bibfnamefont {H.}~\bibnamefont {Jouin}}, \ and\
  \bibinfo {author} {\bibfnamefont {B.}~\bibnamefont {Pons}},\ }\href@noop {}
  {\bibfield  {journal} {\bibinfo  {journal} {At. Data Nucl. Data Tables}\
  }\textbf {\bibinfo {volume} {68}},\ \bibinfo {pages} {279} (\bibinfo {year}
  {1998})}\BibitemShut {NoStop}%
\bibitem [{\citenamefont {Cariatore}\ \emph {et~al.}(2015)\citenamefont
  {Cariatore}, \citenamefont {Otranto},\ and\ \citenamefont
  {Olson}}]{Cariatore15}%
  \BibitemOpen
  \bibfield  {author} {\bibinfo {author} {\bibfnamefont {N.~D.}\ \bibnamefont
  {Cariatore}}, \bibinfo {author} {\bibfnamefont {S.}~\bibnamefont {Otranto}},
  \ and\ \bibinfo {author} {\bibfnamefont {R.~E.}\ \bibnamefont {Olson}},\
  }\href@noop {} {\bibfield  {journal} {\bibinfo  {journal} {Phys. Rev. A}\
  }\textbf {\bibinfo {volume} {91}},\ \bibinfo {pages} {042709} (\bibinfo
  {year} {2015})}\BibitemShut {NoStop}%
\bibitem [{\citenamefont {Kimura}\ and\ \citenamefont {Lin}(1985)}]{Kimura85}%
  \BibitemOpen
  \bibfield  {author} {\bibinfo {author} {\bibfnamefont {M.}~\bibnamefont
  {Kimura}}\ and\ \bibinfo {author} {\bibfnamefont {C.~D.}\ \bibnamefont
  {Lin}},\ }\href@noop {} {\bibfield  {journal} {\bibinfo  {journal} {Phys.
  Rev. A}\ }\textbf {\bibinfo {volume} {32}},\ \bibinfo {pages} {1357}
  (\bibinfo {year} {1985})}\BibitemShut {NoStop}%
\bibitem [{\citenamefont {Caillat}\ \emph {et~al.}(2000)\citenamefont
  {Caillat}, \citenamefont {Dubois},\ and\ \citenamefont {Hansen}}]{Caillat00}%
  \BibitemOpen
  \bibfield  {author} {\bibinfo {author} {\bibfnamefont {J.}~\bibnamefont
  {Caillat}}, \bibinfo {author} {\bibfnamefont {A.}~\bibnamefont {Dubois}}, \
  and\ \bibinfo {author} {\bibfnamefont {J.~P.}\ \bibnamefont {Hansen}},\
  }\href@noop {} {\bibfield  {journal} {\bibinfo  {journal} {J. Phys. B}\
  }\textbf {\bibinfo {volume} {33}},\ \bibinfo {pages} {L715} (\bibinfo {year}
  {2000})}\BibitemShut {NoStop}%
\bibitem [{\citenamefont {Errea}\ \emph {et~al.}(2015)\citenamefont {Errea},
  \citenamefont {Illescas}, \citenamefont {Jorge}, \citenamefont {M\'endez},
  \citenamefont {Rabad\'an},\ and\ \citenamefont {Su\'arez}}]{Errea15}%
  \BibitemOpen
  \bibfield  {author} {\bibinfo {author} {\bibfnamefont {L.~F.}\ \bibnamefont
  {Errea}}, \bibinfo {author} {\bibfnamefont {C.}~\bibnamefont {Illescas}},
  \bibinfo {author} {\bibfnamefont {A.}~\bibnamefont {Jorge}}, \bibinfo
  {author} {\bibfnamefont {L.}~\bibnamefont {M\'endez}}, \bibinfo {author}
  {\bibfnamefont {I.}~\bibnamefont {Rabad\'an}}, \ and\ \bibinfo {author}
  {\bibfnamefont {J.}~\bibnamefont {Su\'arez}},\ }\href@noop {} {\bibfield
  {journal} {\bibinfo  {journal} {J. Phys.: Conf. Ser.}\ }\textbf {\bibinfo
  {volume} {576}},\ \bibinfo {pages} {012002} (\bibinfo {year}
  {2015})}\BibitemShut {NoStop}%
\bibitem [{\citenamefont {Guevara}\ \emph {et~al.}(2011)\citenamefont
  {Guevara}, \citenamefont {Teixeira}, \citenamefont {Hall}, \citenamefont
  {\"Ohrn}, \citenamefont {Deumens},\ and\ \citenamefont {Sabin}}]{Guevara11}%
  \BibitemOpen
  \bibfield  {author} {\bibinfo {author} {\bibfnamefont {N.~L.}\ \bibnamefont
  {Guevara}}, \bibinfo {author} {\bibfnamefont {E.}~\bibnamefont {Teixeira}},
  \bibinfo {author} {\bibfnamefont {B.}~\bibnamefont {Hall}}, \bibinfo {author}
  {\bibfnamefont {Y.}~\bibnamefont {\"Ohrn}}, \bibinfo {author} {\bibfnamefont
  {E.}~\bibnamefont {Deumens}}, \ and\ \bibinfo {author} {\bibfnamefont
  {J.~R.}\ \bibnamefont {Sabin}},\ }\href@noop {} {\bibfield  {journal}
  {\bibinfo  {journal} {Phys. Rev. A}\ }\textbf {\bibinfo {volume} {83}},\
  \bibinfo {pages} {052709} (\bibinfo {year} {2011})}\BibitemShut {NoStop}%
\bibitem [{\citenamefont {Herrero}\ \emph {et~al.}(1995)\citenamefont
  {Herrero}, \citenamefont {Cooper}, \citenamefont {Dickinson},\ and\
  \citenamefont {Flower}}]{Herrero95}%
  \BibitemOpen
  \bibfield  {author} {\bibinfo {author} {\bibfnamefont {B.}~\bibnamefont
  {Herrero}}, \bibinfo {author} {\bibfnamefont {I.~L.}\ \bibnamefont {Cooper}},
  \bibinfo {author} {\bibfnamefont {A.~S.}\ \bibnamefont {Dickinson}}, \ and\
  \bibinfo {author} {\bibfnamefont {D.~R.}\ \bibnamefont {Flower}},\
  }\href@noop {} {\bibfield  {journal} {\bibinfo  {journal} {J. Phys. B}\
  }\textbf {\bibinfo {volume} {28}},\ \bibinfo {pages} {711} (\bibinfo {year}
  {1995})}\BibitemShut {NoStop}%
\bibitem [{\citenamefont {Tseng}\ and\ \citenamefont {Lin}(1999)}]{Tseng99}%
  \BibitemOpen
  \bibfield  {author} {\bibinfo {author} {\bibfnamefont {H.~C.}\ \bibnamefont
  {Tseng}}\ and\ \bibinfo {author} {\bibfnamefont {C.~D.}\ \bibnamefont
  {Lin}},\ }\href@noop {} {\bibfield  {journal} {\bibinfo  {journal} {J. Phys.
  B}\ }\textbf {\bibinfo {volume} {32}},\ \bibinfo {pages} {5271} (\bibinfo
  {year} {1999})}\BibitemShut {NoStop}%
\bibitem [{\citenamefont {Heil}\ \emph {et~al.}(1981)\citenamefont {Heil},
  \citenamefont {Butler},\ and\ \citenamefont {Dalgarno}}]{Heil81}%
  \BibitemOpen
  \bibfield  {author} {\bibinfo {author} {\bibfnamefont {T.~G.}\ \bibnamefont
  {Heil}}, \bibinfo {author} {\bibfnamefont {S.~E.}\ \bibnamefont {Butler}}, \
  and\ \bibinfo {author} {\bibfnamefont {A.}~\bibnamefont {Dalgarno}},\
  }\href@noop {} {\bibfield  {journal} {\bibinfo  {journal} {Phys. Rev. A}\
  }\textbf {\bibinfo {volume} {23}},\ \bibinfo {pages} {1100} (\bibinfo {year}
  {1981})}\BibitemShut {NoStop}%
\bibitem [{\citenamefont {Errea}\ \emph {et~al.}(1991)\citenamefont {Errea},
  \citenamefont {Herrero}, \citenamefont {M\'endez},\ and\ \citenamefont
  {R\'{\i}era}}]{Errea91}%
  \BibitemOpen
  \bibfield  {author} {\bibinfo {author} {\bibfnamefont {L.~F.}\ \bibnamefont
  {Errea}}, \bibinfo {author} {\bibfnamefont {B.}~\bibnamefont {Herrero}},
  \bibinfo {author} {\bibfnamefont {L.}~\bibnamefont {M\'endez}}, \ and\
  \bibinfo {author} {\bibfnamefont {A.}~\bibnamefont {R\'{\i}era}},\
  }\href@noop {} {\bibfield  {journal} {\bibinfo  {journal} {J. Phys. B}\
  }\textbf {\bibinfo {volume} {24}},\ \bibinfo {pages} {4061} (\bibinfo {year}
  {1991})}\BibitemShut {NoStop}%
\bibitem [{\citenamefont {Bienstock}\ \emph {et~al.}(1982)\citenamefont
  {Bienstock}, \citenamefont {Heil}, \citenamefont {Bottcher},\ and\
  \citenamefont {Dalgarno}}]{Bienstock82}%
  \BibitemOpen
  \bibfield  {author} {\bibinfo {author} {\bibfnamefont {S.}~\bibnamefont
  {Bienstock}}, \bibinfo {author} {\bibfnamefont {T.~G.}\ \bibnamefont {Heil}},
  \bibinfo {author} {\bibfnamefont {C.}~\bibnamefont {Bottcher}}, \ and\
  \bibinfo {author} {\bibfnamefont {A.}~\bibnamefont {Dalgarno}},\ }\href@noop
  {} {\bibfield  {journal} {\bibinfo  {journal} {Phys. Rev. A}\ }\textbf
  {\bibinfo {volume} {25}},\ \bibinfo {pages} {2850} (\bibinfo {year}
  {1982})}\BibitemShut {NoStop}%
\bibitem [{\citenamefont {Havener}\ \emph {et~al.}(1995)\citenamefont
  {Havener}, \citenamefont {M\"uller}, \citenamefont {Zeijlmans~van
  Emmichoven},\ and\ \citenamefont {Phaneuf}}]{Havener95}%
  \BibitemOpen
  \bibfield  {author} {\bibinfo {author} {\bibfnamefont {C.~C.}\ \bibnamefont
  {Havener}}, \bibinfo {author} {\bibfnamefont {A.}~\bibnamefont {M\"uller}},
  \bibinfo {author} {\bibfnamefont {P.~A.}\ \bibnamefont {Zeijlmans~van
  Emmichoven}}, \ and\ \bibinfo {author} {\bibfnamefont {R.~A.}\ \bibnamefont
  {Phaneuf}},\ }\href@noop {} {\bibfield  {journal} {\bibinfo  {journal} {Phys.
  Rev. A}\ }\textbf {\bibinfo {volume} {51}},\ \bibinfo {pages} {2982}
  (\bibinfo {year} {1995})}\BibitemShut {NoStop}%
\bibitem [{\citenamefont {Liu}\ \emph {et~al.}(2014)\citenamefont {Liu},
  \citenamefont {Wang},\ and\ \citenamefont {Janev}}]{Liu14}%
  \BibitemOpen
  \bibfield  {author} {\bibinfo {author} {\bibfnamefont {L.}~\bibnamefont
  {Liu}}, \bibinfo {author} {\bibfnamefont {J.~G.}\ \bibnamefont {Wang}}, \
  and\ \bibinfo {author} {\bibfnamefont {R.~K.}\ \bibnamefont {Janev}},\
  }\href@noop {} {\bibfield  {journal} {\bibinfo  {journal} {Phys. Rev. A}\
  }\textbf {\bibinfo {volume} {89}},\ \bibinfo {pages} {012710} (\bibinfo
  {year} {2014})}\BibitemShut {NoStop}%
\bibitem [{\citenamefont {Leung}\ and\ \citenamefont
  {Kirchner}(2015)}]{Leung15}%
  \BibitemOpen
  \bibfield  {author} {\bibinfo {author} {\bibfnamefont {A.~C.~K.}\
  \bibnamefont {Leung}}\ and\ \bibinfo {author} {\bibfnamefont
  {T.}~\bibnamefont {Kirchner}},\ }\href@noop {} {\bibfield  {journal}
  {\bibinfo  {journal} {Phys. Rev. A}\ }\textbf {\bibinfo {volume} {92}},\
  \bibinfo {pages} {032712} (\bibinfo {year} {2015})}\BibitemShut {NoStop}%
\bibitem [{\citenamefont {Ali}\ \emph {et~al.}(2010)\citenamefont {Ali},
  \citenamefont {Neill}, \citenamefont {Beiersd\"orfer}, \citenamefont
  {Harris}, \citenamefont {Schultz},\ and\ \citenamefont {Stancil}}]{Ali10}%
  \BibitemOpen
  \bibfield  {author} {\bibinfo {author} {\bibfnamefont {R.}~\bibnamefont
  {Ali}}, \bibinfo {author} {\bibfnamefont {P.~A.}\ \bibnamefont {Neill}},
  \bibinfo {author} {\bibfnamefont {P.}~\bibnamefont {Beiersd\"orfer}},
  \bibinfo {author} {\bibfnamefont {C.~L.}\ \bibnamefont {Harris}}, \bibinfo
  {author} {\bibfnamefont {D.~R.}\ \bibnamefont {Schultz}}, \ and\ \bibinfo
  {author} {\bibfnamefont {P.~C.}\ \bibnamefont {Stancil}},\ }\href@noop {}
  {\bibfield  {journal} {\bibinfo  {journal} {Astrophys. J. Lett.}\ }\textbf
  {\bibinfo {volume} {716}},\ \bibinfo {pages} {L95} (\bibinfo {year}
  {2010})}\BibitemShut {NoStop}%
\bibitem [{\citenamefont {Trassinelli}\ \emph {et~al.}(2012)\citenamefont
  {Trassinelli}, \citenamefont {Prigent}, \citenamefont {Lamour}, \citenamefont
  {Mezdari}, \citenamefont {M\'erot}, \citenamefont {Reuschl}, \citenamefont
  {Rozet}, \citenamefont {Steydli},\ and\ \citenamefont
  {Vernhet}}]{Trassinelli12}%
  \BibitemOpen
  \bibfield  {author} {\bibinfo {author} {\bibfnamefont {M.}~\bibnamefont
  {Trassinelli}}, \bibinfo {author} {\bibfnamefont {C.}~\bibnamefont
  {Prigent}}, \bibinfo {author} {\bibfnamefont {E.}~\bibnamefont {Lamour}},
  \bibinfo {author} {\bibfnamefont {F.}~\bibnamefont {Mezdari}}, \bibinfo
  {author} {\bibfnamefont {J.}~\bibnamefont {M\'erot}}, \bibinfo {author}
  {\bibfnamefont {R.}~\bibnamefont {Reuschl}}, \bibinfo {author} {\bibfnamefont
  {J.-P.}\ \bibnamefont {Rozet}}, \bibinfo {author} {\bibfnamefont
  {S.}~\bibnamefont {Steydli}}, \ and\ \bibinfo {author} {\bibfnamefont
  {D.}~\bibnamefont {Vernhet}},\ }\href@noop {} {\bibfield  {journal} {\bibinfo
   {journal} {J. Phys. B}\ }\textbf {\bibinfo {volume} {45}},\ \bibinfo {pages}
  {085202} (\bibinfo {year} {2012})}\BibitemShut {NoStop}%
\bibitem [{\citenamefont {Salehzadeh}\ and\ \citenamefont
  {Kirchner}(2013)}]{Arash13}%
  \BibitemOpen
  \bibfield  {author} {\bibinfo {author} {\bibfnamefont {A.}~\bibnamefont
  {Salehzadeh}}\ and\ \bibinfo {author} {\bibfnamefont {T.}~\bibnamefont
  {Kirchner}},\ }\href@noop {} {\bibfield  {journal} {\bibinfo  {journal} {J.
  Phys. B}\ }\textbf {\bibinfo {volume} {46}},\ \bibinfo {pages} {025201}
  (\bibinfo {year} {2013})}\BibitemShut {NoStop}%
\bibitem [{\citenamefont {O'Hagan}(2013)}]{OHagan2013}%
  \BibitemOpen
  \bibfield  {author} {\bibinfo {author} {\bibfnamefont {A.}~\bibnamefont
  {O'Hagan}},\ }\href@noop {} {\emph {\bibinfo {title} {Polynomial chaos: A
  tutorial and critique from a statistician.s perspective}}}\ (\bibinfo
  {publisher} {SIAM/ASA J. Uncertainty Quantification},\ \bibinfo {year}
  {2013})\ \bibinfo {note}
  {http://www.tonyohagan.co.uk/academic/pdf/Polynomial-chaos.pdf}\BibitemShut
  {NoStop}%
\bibitem [{\citenamefont {LeMaitre}\ and\ \citenamefont
  {Omar}(2010)}]{LeMaitre2010}%
  \BibitemOpen
  \bibfield  {author} {\bibinfo {author} {\bibfnamefont {O.~P.}\ \bibnamefont
  {LeMaitre}}\ and\ \bibinfo {author} {\bibfnamefont {M.~K.}\ \bibnamefont
  {Omar}},\ }\href@noop {} {\emph {\bibinfo {title} {Spectral methods for
  uncertainty quantification}}}\ (\bibinfo  {publisher} {Springer, New York},\
  \bibinfo {year} {2010})\BibitemShut {NoStop}%
\end{thebibliography}

%
\end{document}